\documentclass[aps,onecolumn,superscriptaddress,prx,preprint,longbibliography]{revtex4-2}
\usepackage{graphicx}
\usepackage{epsfig}
\usepackage{bm}
\usepackage{dcolumn}
\usepackage{amsmath}
\usepackage{amssymb}
\usepackage{amsfonts}
\usepackage{bbm}
\usepackage{color}
\usepackage{soul}
\usepackage[bookmarks=false,linkcolor=blue,urlcolor=blue,colorlinks,citecolor=blue]{hyperref}
\usepackage{multirow}
\usepackage[tight]{subfigure}

\usepackage{tikz}
\usepackage{tikz-feynman}
\tikzfeynmanset{compat=1.1.0}

\newcommand{\beg}{\begin{equation}}
\newcommand{\en}{\end{equation}}
\newcommand{\bp}{\mathbf p}
\newcommand{\bq}{\mathbf q}

\newcommand{\bk}{\mathbf k}
\newcommand{\br}{\mathbf r}

\newcommand{\bn}{\mathbf n}

\newcommand{\bl}{\boldsymbol{\ell} }

\newcommand \bel  {\begin{align}}
\newcommand \enl  {\end{align}}

\newcommand{\eps}{\epsilon}
\newcommand{\veps}{\varepsilon}

\def \nn{\nonumber \\}
\def\XXint#1#2#3{{\setbox0=\hbox{$#1{#2#3}{\int}$}
     \vcenter{\hbox{$#2#3$}}\kern-.5\wd0}}

\begin{document}

\title{Spatially resolved collective modes in $d$-wave superconductors}

\author{Kazi Ranjibul Islam}
\affiliation{Department of Physics, University of Wisconsin-Milwaukee, Milwaukee, Wisconsin 53201, USA}

\author{Samuel Awelewa}
\affiliation{Department of Physics, Kent State University, Kent, Ohio 44242, USA}

\author{Andrey V. Chubukov}
\affiliation{School of Physics and Astronomy and William I. Fine Theoretical Physics Institute, University of Minnesota, Minneapolis, MN 55455, USA}

\author{Maxim Dzero}
\affiliation{Department of Physics, Kent State University, Kent, Ohio 44242, USA}

\date{\today}

\begin{abstract}
We 
 analyze the dispersion of 
 collective modes in a 
 superconductor
  with 
  $d-$wave 
  symmetry of the order parameter 
  in the presence of long-range Coulomb interaction.
  We use 
  diagrammatic
  technique 
   and quasiclassical theory in Keldysh-Nambu
   formalism to  
    compute  
    longitudinal and transverse pair susceptibilities
     and extract from them the dispersion of the 
      longitudinal 
       and transverse 
        collective mode. 
         We show that at $T=0$, the  dispersion of the transverse (plasma) mode  
        is the same as in an  $s$-wave 
        superconductor,   but at a finite $T$ it is softer 
         and 
           has a much larger decay rate 
             due to the partial screening of the Coulomb potential by nodal quasiparticles.  We show that 
             the dispersion of the longitudinal mode depends on the direction of momentum with respect to
               the positions of the nodes of the $d-$wave gap,
               while the decay rate of this mode    
        does not depend on momentum.  We discuss experimental implications of our results.
\end{abstract}
\maketitle
\tableofcontents
\section{Introduction}
The studies of collective excitations in correlated electronic systems
 is one of 
  the major research activities in condensed matter physics where significant progress has been made
  through the years.
 It has been realized early on~\cite{GL,BCS1957,Anderson1958,Anderson1958b,NNB1958} that  
   because an order parameter in a superconductor has an 
     amplitude and a
      phase,
      there exist two types of collective excitations: longitudinal (amplitude) and transverse (phase) modes. 
       For an $s-$wave gap function, multiple earlier studies have demonstrated
          \cite{AndersonGauge,CarlsonGoldman73,Volkov1975,SchmidSchon1975,Volkov1979,SchmidSchon1979,Kulik1981,TakadaSwave1997,
  TakadaPlasma1998} 
        that  
      in  a charge-neutral superfluid, the 
    transverse phase mode is 
    a gapless Goldstone mode, often called 
     Anderson-Bogolubov (AB) mode 
    \cite{Anderson1958b,NNB1958}, while in a superconductor with charge $2e$ bound pairs,  
  the inclusion of a long-ranged Coulomb interaction converts AB mode into a plasma mode, which has a finite gap in three dimensions (3D). 
   This conversion has been argued 
 to be  
 the realization of the Higgs mechanism in 3D in which  
  a Goldstone mode acquires a gap due to 
  coupling to an external electromagnetic field.  At a finite $T$, the 
     gap decreases due to 
      screening of  long-range Coulomb potential by thermally excited quasiparticles \cite{CarlsonGoldman73,Kamenev2011} and eventually vanishes at a superconducting  $T_c$.  A 
       transverse mode with a small gap  near $T_c$ 
       has  been experimentally detected~~\cite{CarlsonGoldman73} and 
        is frequently called  
    Carlson-Goldman (CG) mode.

Longitudinal excitations  have also been extensively studied theoretically
\cite{ASchmid,VolkovKogan1973,Galaiko1972,Galperin1981,Shumeiko1990,Spivak2004,Andreev2004,Enolski2005a,
Levitov2006,Gapless2006,Levitov2007,Yuzbashyan2006,Yuzbashyan2008,Yuzbashyan2015}.
 These excitations are gapped and are often called  Schmid-Higgs (SH) mode \cite{Kulik1981}. {In an 
    $s-$wave superconductor with gap $\Delta$, longitudinal excitations develop at frequencies above $2\Delta$ and manifest as resonances in a continuum rather than sharp modes}. 
    There have been numerous attempts
      over the years 
       to find conditions under which  a longitudinal mode would emerge 
    below $2\Delta$   as an
     undamped  Higgs mode (see e,g., Refs. \cite{VarmaLit1,VarmaLit2,AndersonAllThat2015,Varma2014}). 
Recent interest in longitudinal collective excitations in superconductors  
has been  driven by experimental advances in optical  and Raman measurements \cite{Pioneers2012,Shimano2013,Shimano2014,THz3,Sherman2015-Disorder,KatoKatsumi2024,Measson2014,
 Behrle2018,Measson2019,Shimano2020,NonReciprocal2020,Katsumi2018,Katsumi2020}.
 Some recent theoretical studies~\cite{Papenkort2007,Axt2009,Assa2011,Sachdev2012,Assa2013,Volovik2014,Dupuis2014,Manske2014,Cea2016,Moore2017,Joerg2018,
 Basov2020,Barresi2023,Eremin2024,Lorenzana2023,Phan2023,Li2025,Uhrig2016,DzeroIFE2024,Dzero2024,Uhrig2025,Kamenev2025,
 Pasha2025}  analyzed the 
manifestation of the longitudinal mode in nonlinear response functions, others  
    discussed the possibilities of observing longitudinal modes in realistic experimental settings by
     taking into account the effects of potential and pair-breaking disorder \cite{Silaev2019-Disorder,Seibold2021-Disorder,Haenel2021-Disorder,Yang2022-Disorder,
     Yang2020-Disorder2,IFE-Dzero2024}, including the cases of parametrically induced spatial inhomogeneities \cite{Dzero2009,dzero2018comment,Chern2019,Grankin2025}.

 In this communication, we analyze 
  collective excitations in a $d-$wave superconductor. We compute
 longitudinal and transverse  particle-particle susceptibilities and 
  extract from them 
 the dispersion of the longitudinal and transverse collective modes.
    We chiefly focus on the  2D case, but also present some results for a 3D superconductor. 
We use two complementary theoretical approaches: the diagrammatic one  and 
 the one
  based on the quasiclassical equation for  a single-particle propagator in Keldysh-Nambu 
   formalism.
 We find the same results using both 
 methods.

Our study
 has been inspired by recent experimental studies of collective excitations in 
   $d$-wave
    cuprate
     superconductors  
     at various dopings  
  (see e.g.,  \cite{KotaHighTc1,KotaHighTc2,Cuprates2020}).
   On the theory side, the longitudinal
 mode 
    in 
 a  $d$-wave 
 superconductor has 
  been 
    analyzed by several groups \cite{BarlasVarma2013,PRL2015,Kirmani2019,Wu2020,Awelewa2025}. These studies, however, 
    focused on zero momentum and did not analyze the dispersion of the longitudinal mode.
    We show that  this dispersion is rather peculiar as the mode frequency 
  (the position of the maximum in the longitudinal susceptibility) 
   depends 
     on the direction of momentum, i.e., is different for  momenta 
     along nodal and antinodal directions. 
 The transverse mode in a d-wave superconductor has been analyzed 
  in Refs.~\cite{ArtemenkoSN1997,ArtemenkoSN2001,Paramekanti2000,TakadaDwave,Sharapov2002,karuzin2025plasmon}   
    at $T=0$ and was  shown to be the same 
       as in an $s-$wave superconductor, both with and without Coulomb interaction. 
        We extend these studies to finite $T$ and show  
       that at a finite $T$, 
        the longitudinal mode in a $d-$wave superconductor (CG mode at $T \leq T_c$) gets softer and much broader than in an $s-$wave case due to  
       partial screening of Coulomb interaction by nodal quasiparticles. 
       
The structure of the paper is the following. In the next section, we formulate the model and outline two theoretical frameworks that we employ to analyze
collective modes. In Section \ref{CGQA}, we present the 
expressions for the transverse susceptibility and determine the energy and dispersion of the transverse mode. 
In Section \ref{SHQA}, we analyze the longitudinal susceptibility.
 In Section \ref{Discussion}, we discuss our results in the context of recent optical experiments in $d$-wave superconductors.
   We present a summary of our work and the 
  conclusions 
  in Section \ref{Conclusions}. 
   Technical details of our calculations 
    are presented 
    in Appendices \ref{AppendixA}-\ref{AppendixD}. We adopt the units $\hbar=k_B=1$ throughout the paper. 

\section{Preliminaries}\label{Prelim}
In this section, we introduce the model and 
discuss the
two
 methods that we will use to analyze the collective modes in a $d$-wave superconductor. 

\subsection{Model}
We consider a one-band model of 2D
 fermions with $d-$wave attractive interaction in the 
  Cooper 
  channel: 
\begin{align}
\hat{\cal H}=\sum_{\bk,\sigma} \xi_\bk \hat{c}^\dagger_{\bk,\sigma}\hat{c}_{\bk,\sigma}+
    \sum_{\bk,\bp,\bq} V_{\mathrm{d}}(\bk,\bp) \hat{c}^\dagger_{\bk+\bq/2,\uparrow} \hat{c}^\dagger_{-\bk+\bq/2 ,\downarrow}\hat{c}_{-\bp+\bq/2 ,\downarrow}\hat{c}_{\bp+\bq/2,\uparrow},
    \label{Hamiltonian for s-d}
\end{align}
 where 
 $\hat{c}^\dagger$ ($\hat{c}$) are the creation (annihilation) fermionic operators,
 $V_{\mathrm{d}}$ is the pairing interaction, and 
 $\xi_\bk= 
  \epsilon_k - \veps_F$, where $\epsilon_k$ is 
  a  single particle dispersion,
   which, for simplicity, we set to be parabolic, and 
  $\veps_F$ is the Fermi energy. 
     We project $V_{\mathrm{d}}(\bk,\bp)$ into the $d-$wave channel  and approximate it as  
\begin{align}
    V_{\mathrm{d}}(\bk,\bp)=- g\gamma(\theta_\bk)\, \gamma(\theta_\bp),
   \label{Pairing Interaction in s-d}
\end{align}
 where 
  $g>0$ is the coupling constant, $\gamma(\theta_\bk)=\sqrt{2}\cos2\theta_\bk$ is the normalized $d$-wave form factor and $\theta_\bk$ defines the direction of the momentum on the Fermi surface 
  with respect to $\hat x$:  $\bk_F=k_F(\cos\theta_\bk,\sin\theta_\bk)$.

In addition 
to  $V_{\mathrm{d}}$, 
 we include 
 the long-range Coulomb interaction between fermions
\begin{align}
    \hat{H}_{C} = \dfrac{1}{2}\sum_\bq V_\bq \rho_\bq \rho_{-\bq},
\end{align}
 where $\rho_\bq=\sum_{\bk,\sigma} c^\dagger_{\bk+\bq,\sigma}c_{\bk,\sigma}$ is the particle density operator and  $V_\bq$ is the Coulomb potential $V_\bq=\pi e^2(2/|\bq|)$ 
($V_\bq=\pi e^2(2/|\bq|)^2$ in 3D). The total Hamiltonian is $\hat{H} =   \hat{H}_{C} + \hat{\cal H}$.

  At $T=0$, Eq.~\eqref{Hamiltonian for s-d} describes a superconductor with a $d-$wave pairing gap
  $\Delta_\bk=\Delta\, \gamma(\theta_\bk)$. The gap magnitude is obtained  
 by solving the non-linear 
  BCS-like 
  gap equation at $T=0$:
  \begin{align}
     1=\frac{g\nu_F}{2}\int_{-\Lambda}^\Lambda d\xi_\bk \int_0^{2\pi}\dfrac{d\theta_\bk}{2\pi}\dfrac{\gamma^2(\theta_\bk)}{ \sqrt{\xi^2_\bk+\Delta_\bk^2}}
     \label{Gap equation s-d}
     \end{align}
where 
 $\nu_F = m/(2\pi)$ is the density of states at the Fermi level and $\Lambda$ is an ultraviolet cutoff,
 which we introduce symmetrically with respect to $\xi_\bk$. 

\subsection{Theoretical methods}

We will use two independent approaches 
to study collective excitations in a $d$-wave superconductor.
One approach is based on deriving the expression for the pair susceptibility by solving the quasiclassical equation for the single-particle propagator $\check{g}$ 
in Keldysh-Nambu formalism~\cite{Larkin1965,LarkinVertex,Eilenberger1968,Usadel1970},
 another is based on  direct diagrammatic calculation of a pair susceptibility
   as a two-point correlation function 
 \cite{AGD,Kulik1981,Silaev2019-Disorder,Phan2023,Pasha2025}.

\subsubsection {Quasiclassical theory.}
The 
 quasiclassical equation for the single-particle propagator $\check{g}$ is \cite{Larkin1965,LarkinVertex,Eilenberger1968,Usadel1970}
\beg\label{EilenMain}
[\eps\check{\tau}_3-\check\Delta_\bn(\br,t)\stackrel{\circ},\check{g}_{\bn\eps}]+\frac{i}{2}\left\{\check{\tau}_3,\partial_t\check{g}_{\bn\eps}\right\}+i{v}_F(\bn\cdot\mbox{\boldmath $\nabla$}_\br)\check{g}_{\bn\eps}=0.
\en
Here $\check{\tau}_3=\hat{\tau}_3\otimes\hat{\sigma}_0$ is a $4\times4$ matrix defined in Keldysh and Nambu spaces
(note that $\check{\tau}_3$ is diagonal in Keldysh space) and $\bn=\bk/k$. Quasiclassical propagator 
$\check{g}$ is an 4$\times$4 matrix in Keldysh and Nambu spaces
\beg\label{KeldyshProps}
\check{g}=\left[\begin{matrix} \hat{g}^R & \hat{g}^K \\ 0 & \hat{g}^A\end{matrix}\right]
\en
and it must satisfy the normalization condition
\beg\label{Norm}
\check{g}\circ\check{g}=\check{{\mathbbm{1}}},
\en
where symbol $\circ$ denotes the usual convolution 
\beg\label{BasicConvolution}
(F\circ G)(x,x')=\int F(x,y)G(y,x')dy,
\en
and we use $x=(\br,t)$. Upon the Wigner transformation \eqref{BasicConvolution} acquires the following form 
\beg\label{Conv}
F\circ G=\hat{F}_{\bp\eps}(\br,t)e^{\frac{i}{2}\left(\stackrel{\leftarrow}\partial_\br\stackrel{\rightarrow}\partial_\bp-\stackrel{\leftarrow}\partial_t\stackrel{\rightarrow}\partial_\eps-\stackrel{\leftarrow}\partial_\bp\stackrel{\rightarrow}\partial_\br+\stackrel{\leftarrow}\partial_\eps\stackrel{\rightarrow}\partial_t\right)}\hat{G}_{\bp\eps}(\br,t).
\en
Here $\br$ and $t$ refer to the 'center-of-mass' coordinate $X=(x+x')/2$ and the Fourier transform has been performed with respect to the relative coordinates $x-x'$.

In the ground state, we choose the order parameter to be particle-hole symmetric, $\hat{\Delta}_\bn(\br,t)= i\hat{\tau}_2\Delta_\bn$. It is
 then straightforward to show that the retarded and advanced blocks of the quasiclassical function \eqref{KeldyshProps} are given by
\beg\label{ggsRA}
\hat{g}_{\bn\eps}^{R(A)}= \frac{\eps}{\eta_{\bn\eps}^{R(A)}}\hat{\tau}_3-\frac{\Delta_{\bn}}{\eta_{\bn\eps}^{R(A)}}i\hat{\tau}_2,
\en
where $\hat{\tau}_j$ are the Pauli matrices which act in Nambu space and
functions $\eta_{\bn\eps}^{R(A)}$ are defined according to  
\beg\label{etaRA}
\eta_{\bn\eps}^{R(A)}=\left\{\begin{aligned} &\pm\mathrm{sign}(\eps)\sqrt{(\eps\pm i\delta)^2-\Delta_{\bn}^2}, \quad |\eps|\geq|\Delta_\bn|, \\
&i\sqrt{\Delta_\bn^2-\eps^2}, \quad |\eps|<|\Delta_\bn|.
\end{aligned}
\right.
\en
Given the normalization condition (\ref{Norm}), the Keldysh component is a simple parametrization
$\hat{g}_{\bn\eps}^{K}=\left(\hat{g}_{\bn\eps}^{R}-\hat{g}_{\bn\eps}^{A}\right)\tanh\left({\eps}/{2T}\right)$ (here $T$ is temperature).  Lastly, the value of the pairing amplitude in the ground state is determined by the solution of the self-consistency equation
\beg\label{Coupling}
{\Delta}=\frac{g\nu_F}{4}\int\limits_0^{2\pi}\gamma(\theta_\bn)\frac{d\theta_\bn}{2\pi}\int\limits_{-\Lambda}^{\Lambda}d\eps\left(f_{\bn\eps}^R-f_{\bn\eps}^A\right)t_\eps,
\en
which is of course equivalent to \eqref{Gap equation s-d}.

We present the  explicit expressions for the transverse and longitudinal parts of the pair susceptibility, 
 obtained 
 using this 
  method, 
   in Secs. \ref{CGQA} and 
   \ref{SHQA} respectively.  

\subsubsection{Diagrammatic approach.}
 \label{Diagrammatic method}
  
Within the diagrammatic technique, we introduce normal and anomalous Green's functions in the superconducting state 
   along the Matsubara axis, 
    \beg\label{Green's function s-d}
G(\bk,\omega_m)=\dfrac{i\, \omega_m+\xi_\bk}{(i\, \omega_m)^2-E^2_{\bk}},\quad  F(\bk,\omega_m)=\dfrac{\Delta_\bk}{(i\, \omega_m)^2-E^2_{\bk}},
\en
where $E_\bk=\sqrt{\xi^2_\bk+\Delta^2_\bk}$ is the quasi-particle excitation energy, and 
 $\omega_m = \pi T (2m+1)$ is the fermionic Matsubara frequency, $T$ is the  temperature, and 
 compute diagrammatically
  the two-point pair-pair correlation function \cite{Kulik1981,Phan2023,Pasha2025}
 \begin{align}
     \chi
     (\bq,\Omega_m)&=\int_{-\infty}^\infty d\tau \, e^{i \Omega_m\tau} \int \dfrac{d^2\bk}{(2\pi)^2} \int \dfrac{d^2\bp}{(2\pi)^2} \gamma(\theta_\bk) \gamma(\theta_\bp)\nn &\times\langle T_\tau \, c_{\bk+\bq/2,\uparrow}(\tau) c_{-\bk+ \bq/2,\downarrow}(\tau) c^\dagger_{-\bp+\bq/2,\downarrow}(0) c^\dagger_{\bp+\bq/2,\uparrow}(0) \rangle,
\label{chi expression}
 \end{align}
where 
$\bq$ and $\Omega_m=2\pi mT$  are bosonic momentum and Matsubara frequency.
The retarded pair susceptibility $\chi_R(\bq,\Omega)$, whose poles (branch cuts)  
 determine the spectrum of  collective modes, is obtained by 
  a simple rotation $\Omega_m \rightarrow \Omega+i\, \delta$,
    as we explicitly verified.
    %AC Kazi - pls  verify by Cauchy relation that a rotation is a safe procedure here 
   %AC_l   Pls verify
   
\begin{figure}
     \centering
     \includegraphics[width=\linewidth]{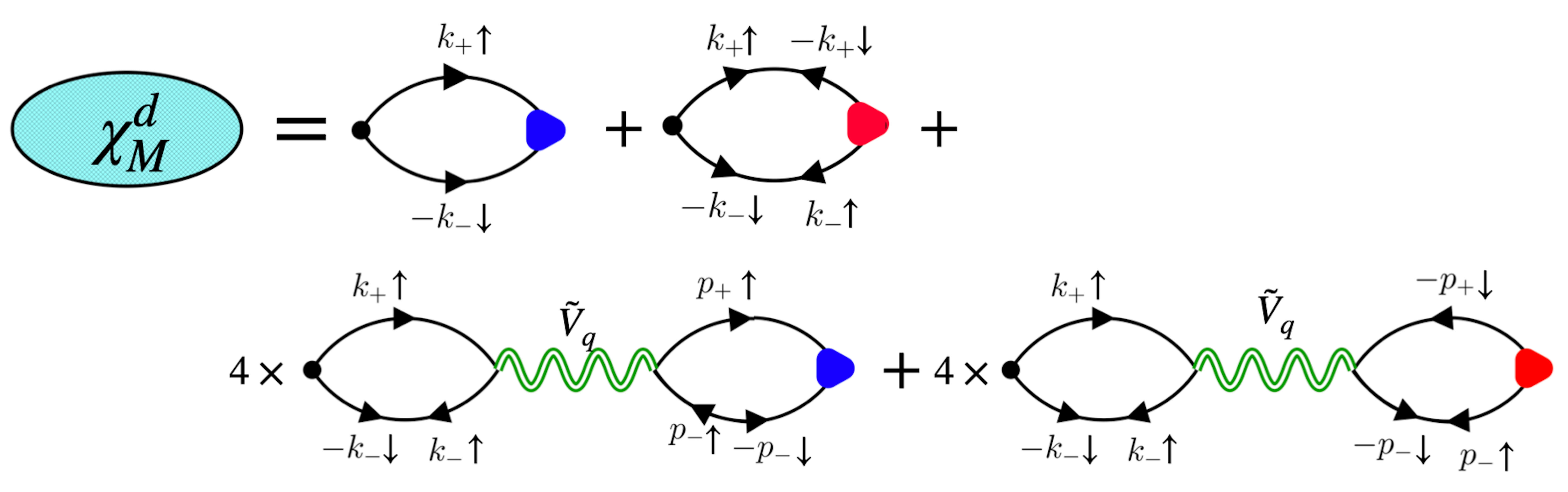}
     \caption{Dyson equation for the pair-pair susceptibility $\chi_M(\bq,\Omega_m)$ with momentum $\bq$ and Matsubara frequency $\Omega_m$. The single arrow solid line represents normal Green's function $G$, and the double-headed arrow solid lines represent anomalous Green's function $F$ and $F^+$.
      The red and blue triangular vertices represent renormalized particle-particle vertices $\Gamma_\bk$ with two incoming fermion momenta $k_+=k+q/2$ and $-k_-=-k+q/2$ and $\bar{\Gamma}_\bk$ with two outgoing fermion momenta $k_-=k-q/2$ and $-k_+=-k-q/2$ respectively. The double wavy green line represents fully renormalized Coulomb interaction, $\tilde{V}_q$. The intermediate momenta are labeled by $k_\pm=k\pm q/2$ and $p_\pm=p\pm q/2$ where $k=(\bk,\omega_k),p=(\bp,\omega_p)$, and $q=(\bq,\Omega_m)$. The black dot on the right vertex represents the d-wave form factor $\gamma(\theta_\bk)=  \sqrt{2}\cos2\theta_{\bk}$. $\uparrow$ and $\downarrow$ represent up and down spin components. }
     \label{fig:susceptibility_figure}
\end{figure}

\begin{figure}
     \centering
     \includegraphics[width=\linewidth]{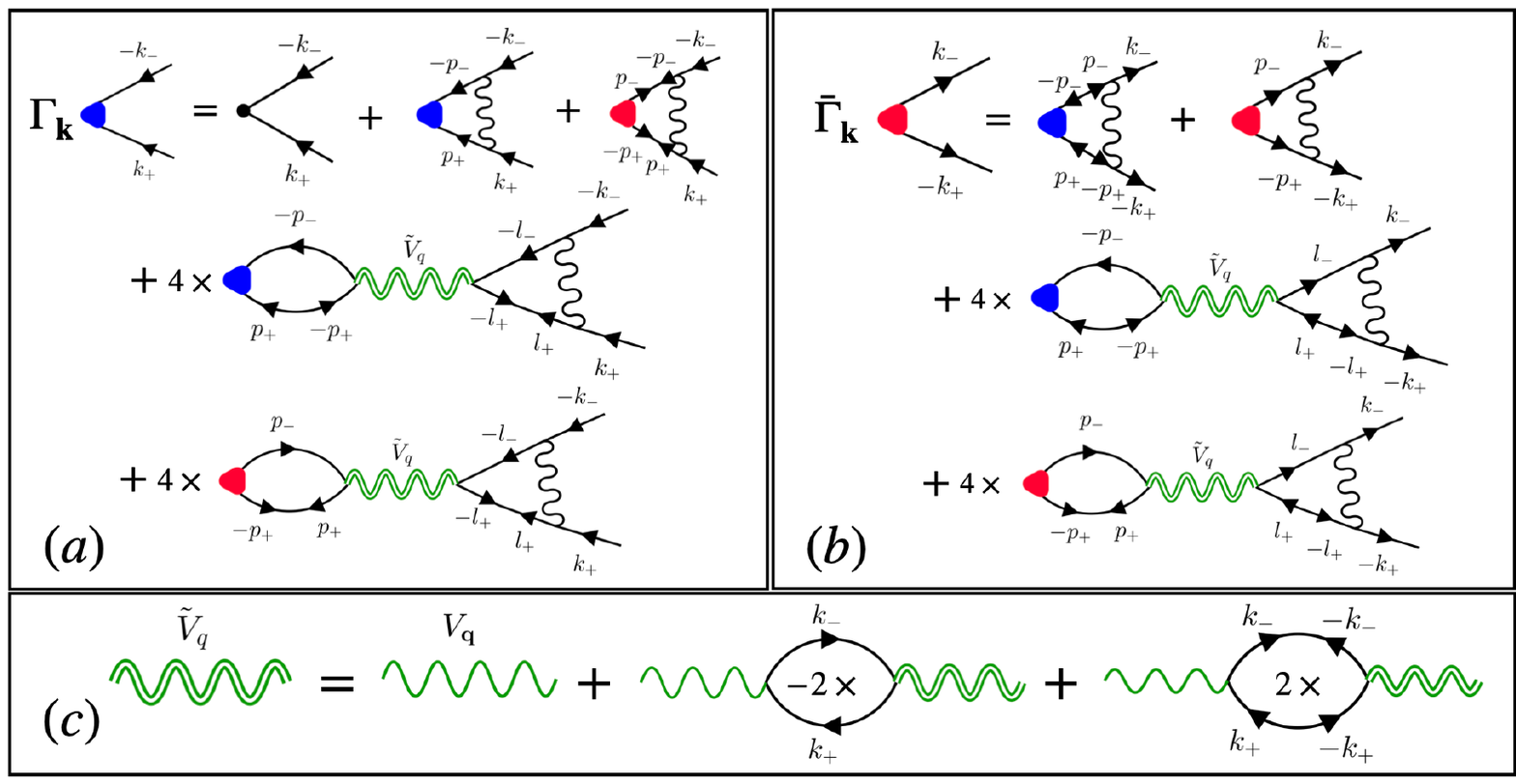}
     \caption{Dyson equation for the renormalized particle-particle vertices (a) $\Gamma_\bk$ (blue triangle with two incoming fermion momenta $k_+=k+q/2$ and $-k_-=-k+q/2$), (b) $\bar{\Gamma}_\bk$ (red triangle with two outgoing fermion momenta $k_-=k-q/2$ and $-k_+=-k-q/2$), and (c) renormalized Coulomb interaction, $\tilde{V}_q$. In the diagrams shown here, the black single wavy line represents the pairing interaction, the green single wavy line represents the long-range Coulomb interaction $V_\bq$, and the double wavy green line stands for the renormalized Coulomb potential $\tilde{V}_q$. (c) Diagrammatic form of the Dyson equation for the renormalized Coulomb potential. The single arrow solid lines represent normal Green's function $G$, and the double-headed arrow solid lines represent anomalous Green's function $F$ and $F^+$.
       The intermediate momenta are labeled by $k_\pm=k\pm q/2$,  $p_\pm=p\pm q/2$, $\ell_\pm=\ell\pm q/2$ where $k=(\bk,\omega_k),p=(\bp,\omega_p)$, $\ell=(\bl,\omega_\ell)$ and $q=(\bq,\Omega_m)$. The black dot  represent the d-wave form factor $\gamma(\theta_\bk)=  \sqrt{2}\cos2\theta_{\bk}$.}
     \label{fig:vertex_figure}
 \end{figure}

    We compute the pair-pair susceptibility $\chi (\bq,\Omega_m)$ \eqref{chi expression} within the 
  standard ladder/bubble approximation. 
   Namely, we sum up a series of ladder diagrams for the two-fermion vertices (see below) and a series of bubble diagrams for the renormalization (screening) of the Coulomb interaction.  The 
    corresponding 
    diagrams are shown in Figs.
 \ref{fig:susceptibility_figure} and \ref{fig:vertex_figure}. 
 
  The expression for $\chi
     (\bq,\Omega_m)$ in terms of single particle propagators 
      $G$ and $F$ 
      is
\begin{align}
    \chi (q)&=\int_k \gamma_\bk\, G(k+\dfrac{q}{2})\, G(-k+\dfrac{q}{2})\, \Gamma_\bk(q)- \int_k \gamma_\bk\, F(k+\dfrac{q}{2})\, F(-k+\dfrac{q}{2})\, \bar{\Gamma}_\bk(q) \nn & 
    -4 \tilde{V}_q\, \int_k \gamma_\bk \, G(k+\dfrac{q}{2})F(-k+\dfrac{q}{2}) \int_p  G(p+\dfrac{q}{2})F(-p+\dfrac{q}{2})\, \Gamma_\bp(q) \nn & -4 \tilde{V}_q\, \int_k \gamma_\bk \, G(k+\dfrac{q}{2})F(-k+\dfrac{q}{2}) \int_p  G(-p-\dfrac{q}{2})F(p-\dfrac{q}{2})\, \bar{\Gamma}_\bp(q),
    \label{chi_ecpression_2}
\end{align}
 where $q=(\bq,i\Omega_m)$, $\gamma_\bk=\gamma(\theta_\bk)$, and the integration stands for $\int_k= T\sum_{\omega_m} \, \int d^2\bk/(2\pi)^2$,
  and  $\tilde{V}_q$ is the  dressed Coulomb interaction:
 \begin{align}
    \tilde{V}_q&=\dfrac{V_\bq}{1-2V_\bq\Pi_0(q)},
    \label{tilde_v_expression}
\end{align}
where the density polarization bubble $\Pi_0(q)$ is 
\begin{align}
    \Pi_0(q)=\int_k \left[G(k+\dfrac{q}{2})\,  G(k-\dfrac{q}{2})-F(k+\dfrac{q}{2})\, F(k-\dfrac{q}{2})\right].
    \label{pi_0_def}
\end{align}
The 
 equations for the two-particle 
vertices $\Gamma_\bk(q)$ and $\bar{\Gamma}_\bk(q)$ 
 are shown graphically in 
 Fig.~\ref{fig:vertex_figure}a,b.  Both vertices have  $d-$wave structure:
 $\Gamma_\bk(q)=\gamma_\bk\, \Gamma(q)$,  $\bar{\Gamma}_\bk(q)=
 \gamma_\bk \bar{\Gamma}(q)$. 
 The set of coupled equations for 
   $\Gamma(q)$ and $\bar{\Gamma}(q)$ is  
\beg\label{bar_gamma_0gamma_0}
\begin{aligned}
    \Gamma(q)&=1+g\, \Gamma(q)\, \Pi_{GG}(q) -g\, \bar{\Gamma}(q)\, \Pi_{FF}(q) 
    -4\, g \tilde{V}_q\, \left(\Gamma(q)\, \Pi^2_{GF}(q)
     + \bar{\Gamma}(q)\, \Pi_{GF}(q)\, \Pi_{GF}(-q)\right),\\
  \bar{\Gamma}(q)&=  -g\Gamma(q)\,  \Pi_{FF}(q)+g\, \bar{\Gamma}(q)\, \Pi_{GG}(-q)-4 g\,\tilde{V}_q\, \left(\Gamma(q)\,  \Pi_{GF}(q)\, \Pi_{GF}(-q)
   + \, \bar{\Gamma}(q)\,  \Pi^2_{GF}(-q)\right),
\end{aligned}
\en
%AC_l Kazi - pls check my changes (already in the previous version) 
%\textcolor{green}{KRI: Done}
The polarization bubbles in these equations are
\begin{align}
\label{Pi_G}
    \Pi_{GG}(q)&=\int_p \gamma^2_\bp\, G(p+\dfrac{q}{2})\, G(-p+\dfrac{q}{2}),\quad
   \Pi_{FF}(q)= \int_p \, \gamma^2_\bp\, F(p+\dfrac{q}{2})\, F(-p+\dfrac{q}{2}),\\
  & \hspace{2.5 cm} \Pi_{\mathrm{GF}}(q)= \int_p \gamma_\bp\, G(p+\dfrac{q}{2})\, F(-p+\dfrac{q}{2}).
   \label{Pi_m}
\end{align}
Solving these equations using 
 $\Pi_{GG}(q)=\Pi_{GG}(-q)$, $\Pi_{FF}(q)=\Pi_{FF}(-q)$ and $\Pi_{GF}(q)=-\Pi_{GF}(-q)$, 
  we obtain (see Appendix~\ref{Bubble_calculation} for details)
\begin{align}
\label{gamma_0_full_expression}
    \Gamma(q)&= \dfrac{1-g \, \Pi_{GG}(q)+4\, \tilde{V}_q\, g\, \Pi^2_{GF}(q)}{\left[1-g \,\left(\Pi_{GG}(q)-\Pi_{FF}(q)\right) \right]\, \left[1-g\, \left(\Pi_{GG}(q)+\Pi_{FF}(q)-8\, \tilde{V}_q\, \Pi^2_{GF}(q)\right)\right]}\\
    \bar{\Gamma}(q)&= - \dfrac{g \Pi_{FF}(q)-4\, \tilde{V}_q\, g\, \Pi^2_{GF}(q)}{\left[1-g \,\left(\Pi_{GG}(q)-\Pi_{FF}(q)\right) \right]\, \left[1-g\, \left(\Pi_{GG}(q)+\Pi_{FF}(q)-8\, \tilde{V}_q\, \Pi^2_{GF}(q)\right)\right]}.
    \label{bar_gamma_0_full_expression}
\end{align}
 Substituting into 
\eqref{tilde_v_expression}, we obtain after 
  some  algebra
  the particle-particle susceptibility $\chi (\bq,\Omega_m)$  as the 
   sum of two terms, 
    which we identify as transverse and longitudinal susceptibilities 
  \begin{align} 
   \chi (q) = \chi_T (q) + \chi_L (q),
   \end{align}
    where
 \begin{align}
    \chi_T (q)=\dfrac{1}{2}
    \dfrac{\bar{\Pi}_T(q)}{1-g \bar{\Pi}_T(q)},~~ \chi_L (q)=\dfrac{1}{2} \dfrac{\Pi_L(q)}{1-g\, \Pi_L(q)}.
    \label{susceptibility_equation}
\end{align}
 and the transverse ($\bar{\Pi}_T$) and longitudinal ($\Pi_L$) polarization bubbles are
\begin{align}
\label{Pi_T_expression}
    \bar{\Pi}_T(q)&={\Pi}_T(q)- 8\Pi^2_{\mathrm{GF}}(q) {\tilde V}_\bq \\
    \Pi_{T,L}(q)&=\Pi_{GG}(q)\pm\Pi_{FF}(q).
    \label{Pi_L_expression}
\end{align}
%AC pls check
%\textcolor{green}{KRI: Done}
 Note that the Coulomb potential appears in the transverse part of the pair-pair susceptibility, while the longitudinal part does not depend on it. This is consistent with Ref. \cite{Kulik1981}.
  We analyze the transverse and longitudinal susceptibilities  
   separately in Secs. \ref{CGQA} and 
   \ref{SHQA}, respectively. 

\section{ Transverse susceptibility and dispersion of the transverse mode}\label{CGQA}
\subsection{Quasiclassical approach}
For the case when an external electromagnetic field is applied, the  main idea behind the calculation of the pair susceptibility consists in evaluating perturbative corrections to $\check{g}$ up to the second order in powers of the vector potential \cite{Moore2017,Li2025,Eremin2024,Awelewa2025}. The second-order correction is required due to the fact that the order parameter is a scalar quantity and therefore the changes to the pairing amplitude may only appear at the order $O({\mathbf A}^2)$ unless the supercurrent is flowing in a superconductor \cite{Moore2017}. Thus, generally one needs to be looking for the perturbative solution of the equation \eqref{EilenMain} in the form
$\check{g}(\bn\eps;\br t)=\check{g}_{\bn\eps}+\check{g}_1(\bn\eps;\br t)+\check{g}_2(\bn\eps;\br t)$. From the normalization condition (\ref{Norm}) it follows that functions $\check{g}_1$ and $\check{g}_2$ must satisfy 
\beg\label{norm12}
\check{g}_0\circ\check{g}_1+\check{g}_1\circ\check{g}_0=0, \quad \check{g}_0\circ\check{g}_2+\check{g}_2\circ\check{g}_0+\check{g}_1\circ\check{g}_1=0.
\en 

With this comment in mind, let us assume that the system has been subjected to an external perturbation which in turn produced a change in the order parameter corresponding to the fluctuation in the transverse (phase) $d$-wave channel. We will describe this by a function
\beg\label{dDeltaT}
\delta\hat{\Delta}_\bn(\br,t)=\gamma(\theta_\bn)\delta\hat{\Delta}_{\bk\omega}^Te^{2i(\bk\br-\omega t)},
\en
where $\delta\hat{\Delta}_{\bk\omega}^T=(i\hat{\tau}_1)\delta\Delta_{\bk\omega}^T$ and the first Pauli matrix ensures that the resulting corrections to $\check{g}$ do indeed describe the excitation of the phase mode. For instance, the normal part of $\check{g}$ will be proportional to $\hat{\tau}_0$, which will allow us to compute the change in the particle density due to the phase fluctuations of the order parameter in the transverse channel.  Indeed, 
as it is well known an excitation of the phase mode necessarily produces redistribution of electronic charge \cite{CarlsonGoldman73,Volkov1975,SchmidSchon1975,Volkov1979,SchmidSchon1979,Kulik1981,Kamenev2011}. This process is accounted for by the corresponding variation of the Coulomb potential which is described by a function 
\beg\label{dPhi}
\delta\Phi(\br,t)=\delta\Phi_{\bk\omega}e^{2i(\bk\br-\omega t)}.
\en
With these provisions, using quasiclassical equation \eqref{EilenMain} it can be shown that the second-order corrections for each element of the matrix $\check{g}$ can be found by solving the following equation
\beg\label{Eq4g2RAT}
\begin{aligned}
\left[\eps\hat{\tau}_3-\hat{\Delta}_\bn,\hat{g}_2\right]-2{v}_F(\bn.\bk)\hat{\tau}_0\hat{g}_2+\omega\left\{\hat{\tau}_3,\hat{g}_2\right\}&=\delta\hat{\Delta}_\bn^T\hat{g}_{\bn\eps-\omega}-\hat{g}_{\bn\eps+\omega}\delta\hat{\Delta}_\bn^T\\&+\delta\Phi_{\bk\omega}\hat{g}_{\bn\eps-\omega}-\hat{g}_{\bn\eps+\omega}\delta\Phi_{\bk\omega}.
\end{aligned}
\en
Variation of the order parameter is determined self-consistently via 
\beg\label{SelfT}
\delta\Delta_{\bk\omega}^T=\frac{g\nu_F}{4}\int\limits_0^{2\pi}\gamma(\theta_\bn)\frac{d\theta_\bn}{2\pi}\int\limits_{-\infty}^\infty{d\eps}\textrm{Tr}\left\{i\hat{\tau}_1\hat{g}_2^K(\bn\eps;\bk\omega)\right\},
\en
while the variation of the scalar potential is given by the solution of the Poisson equation:
\beg\label{Poisson}
\left(-\frac{k^{d-1}}{2^{d-1}\pi e^2}\right)\delta\Phi_{\bk \omega}=\nu_F\delta\Phi_{\bk\omega}+\frac{\pi\nu_F}{4}\int\limits_0^{2\pi}\frac{d\phi_\bn}{2\pi}\int\limits_{-\infty}^\infty{d\eps}\textrm{Tr}\left\{\hat{\tau}_0\hat{g}_2^K(\bn\eps;\bk\omega)\right\}
\en
and $d=2,3$ refers to the spatial dimensionality of the problem. 
Here, the first term on the right-hand side accounts for the finite polarizability of the electronic band \cite{Kamenev2009,Kamenev2011}.

\paragraph{$D=2$.} We start with the two-dimensional superconductor. After we determine the second order corrections to the propagator $\check{g}$ and insert the resulting expressions into the self-consistency condition \eqref{SelfT} and Poisson equation \eqref{Poisson}, we find the following system of equations (see Appendix \ref{AppendixB} for details):
\beg\label{TSystem}
\begin{aligned}
&\chi_{\textrm{AB}}^{-1}(\bq,\Omega)\cdot\delta\Delta_{\bq\Omega}^T+i\rho_{\bq\Omega}\cdot\delta\Phi_{\bq\Omega}=0, \quad
\chi_{\textrm{CG}}^{-1}(\bq,\Omega)\cdot\delta\Phi_{\bq\Omega}-i\rho_{\bq\Omega}\cdot\delta\Delta_{\bq\Omega}^T=0.
\end{aligned}
\en
Here $\chi_{\textrm{AB}}^{-1}(\bq,\Omega)$ is defined as
\beg\label{chiABdwave}
\begin{split}
&\chi_{\textrm{AB}}^{-1}(\bq,\Omega)=-\frac{2}{g\nu_F}+\int\limits_{0}^{2\pi}{\gamma}^2(\theta_\bn)\frac{d\theta_\bn}{2\pi}\int\limits_{-\omega_D}^{\omega_D}d\eps\\&\times\left\{
\frac{\left(\eta_{\bn\eps+\Omega}^{R}+\eta_{\bn\eps-\Omega}^{A}\right){A}_\bn^K(\eps_+,\eps_-)(t_{\eps+\Omega}-t_{\eps-\Omega})}{\left(\eta_{\bn\eps+\Omega}^{R}+\eta_{\bn\eps-\Omega}^{A}\right)^2-{v}_F^2(\bn.\bq)^2}\right.\\&\left.+\frac{\left(\eta_{\bn\eps+\Omega}^{R}+\eta_{\bn\eps-\Omega}^{R}\right){A}_\bn^R(\eps_+,\eps_-)t_{\eps-\Omega}}{\left(\eta_{\bn\eps+\Omega}^{R}+\eta_{\bn\eps-\Omega}^{R}\right)^2-{v}_F^2(\bn.\bq)^2}-
\frac{\left(\eta_{\bn\eps+\Omega}^{A}+\eta_{\bn\eps-\Omega}^{A}\right){A}_\bn^A(\eps_+,\eps_-)t_{\eps+\Omega}}{\left(\eta_{\bn\eps+\Omega}^{A}+\eta_{\bn\eps-\Omega}^{A}\right)^2-{v}_F^2(\bn.\bq)^2}\right\}.
\end{split}
\en
where
\beg\label{AKRAT}
\begin{aligned}
{A}_\bn^{R(A)}(\eps_{+},\eps_{-})=g_{\bn\eps+\Omega}^{R(A)}g_{\bn\eps-\Omega}^{R(A)}-f_{\bn\eps+\Omega}^{R(A)}f_{\bn\eps-\Omega}^{R(A)}+1, \\
{A}_\bn^{K}(\eps_{+},\eps_{-})=g_{\bn\eps+\Omega}^{R}g_{\bn\eps-\Omega}^{A}-f_{\bn\eps+\Omega}^{R}f_{\bn\eps-\Omega}^{A}+1.
\end{aligned}
\en
The definitions of the functions $\rho_{\bq\Omega}$ and $\chi_{\textrm{CG}}^{-1}(\bq,\Omega)$ can be found in Appendix \ref{AppendixB}.
Note that in a charge neutral superconductor  $\delta\Phi_{\bq\Omega}$ must be zero, so the function $\chi_{\textrm{AB}}^{-1}(\bq,\Omega)$, Eqs. (\ref{TSystem},\ref{chiABdwave}), not only defines the transverse pair susceptibility, but its frequency and momentum dependence determine the dispersion of the transverse collective mode in this case. 

For this reason it is instructive to consider the limit $\Omega\to 0$ of the function $\chi_{\textrm{AB}}^{-1}(\bq,\Omega)$ first. In this limit, the first term under the integral vanishes identically, while ${A}_\bn^{R(A)}(\eps,\eps)=2$ by virtue of the normalization condition. Given that the self-consistency condition \eqref{Coupling} holds, it is easy to see that equation $\chi_{\textrm{AB}}^{-1}(\bq,\Omega\to 0)\cdot\delta\Delta_{\bq\Omega}^T=0$ is satisfied identically for $\bq\to0$. 

%%%%%%%%%%%%% Fig -  Re[\chi_{AB}(q,w)] %%%%%%%%%%%%%%%%%
\begin{figure}
\includegraphics[width=0.450\linewidth]{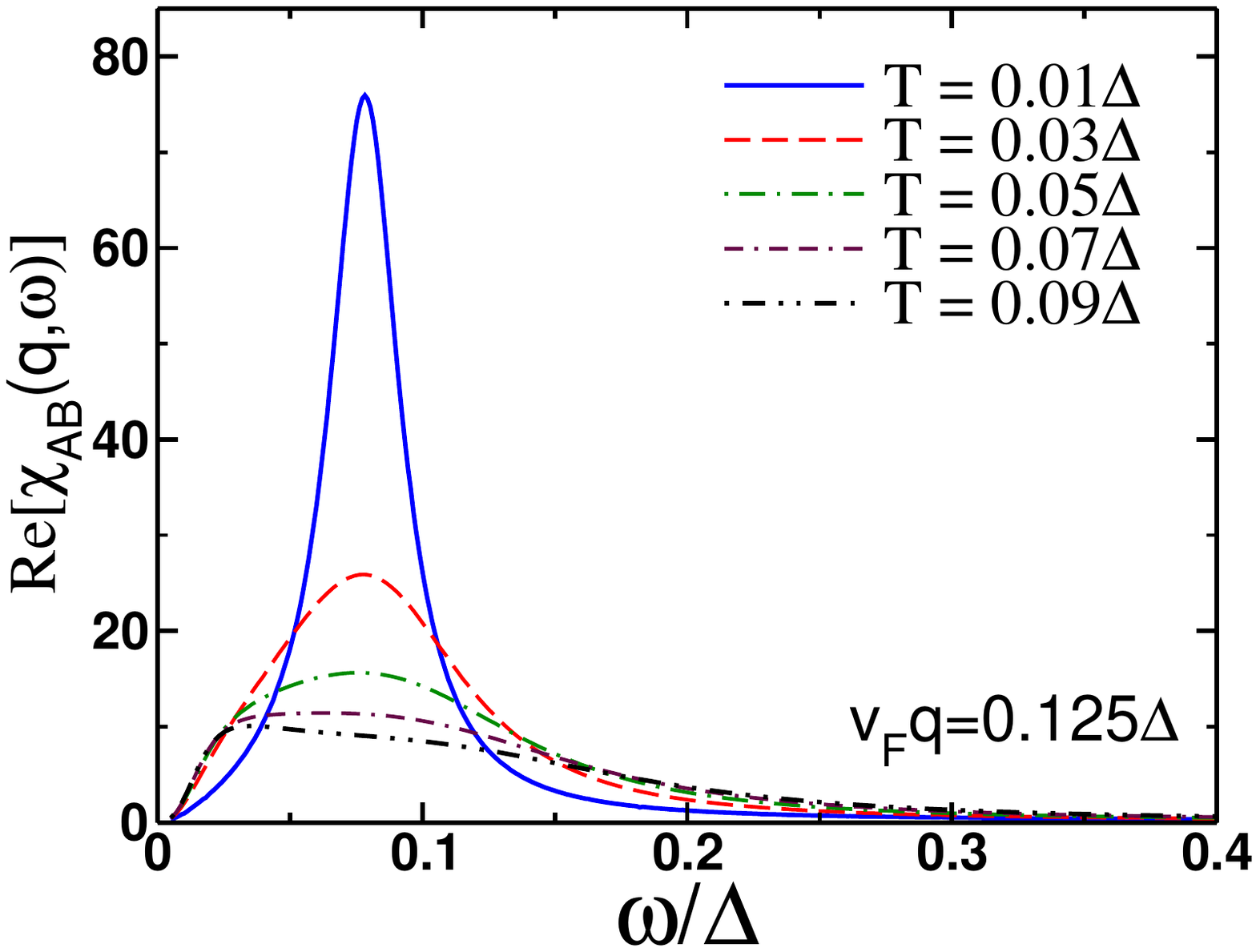} 
\includegraphics[width=0.45\linewidth]{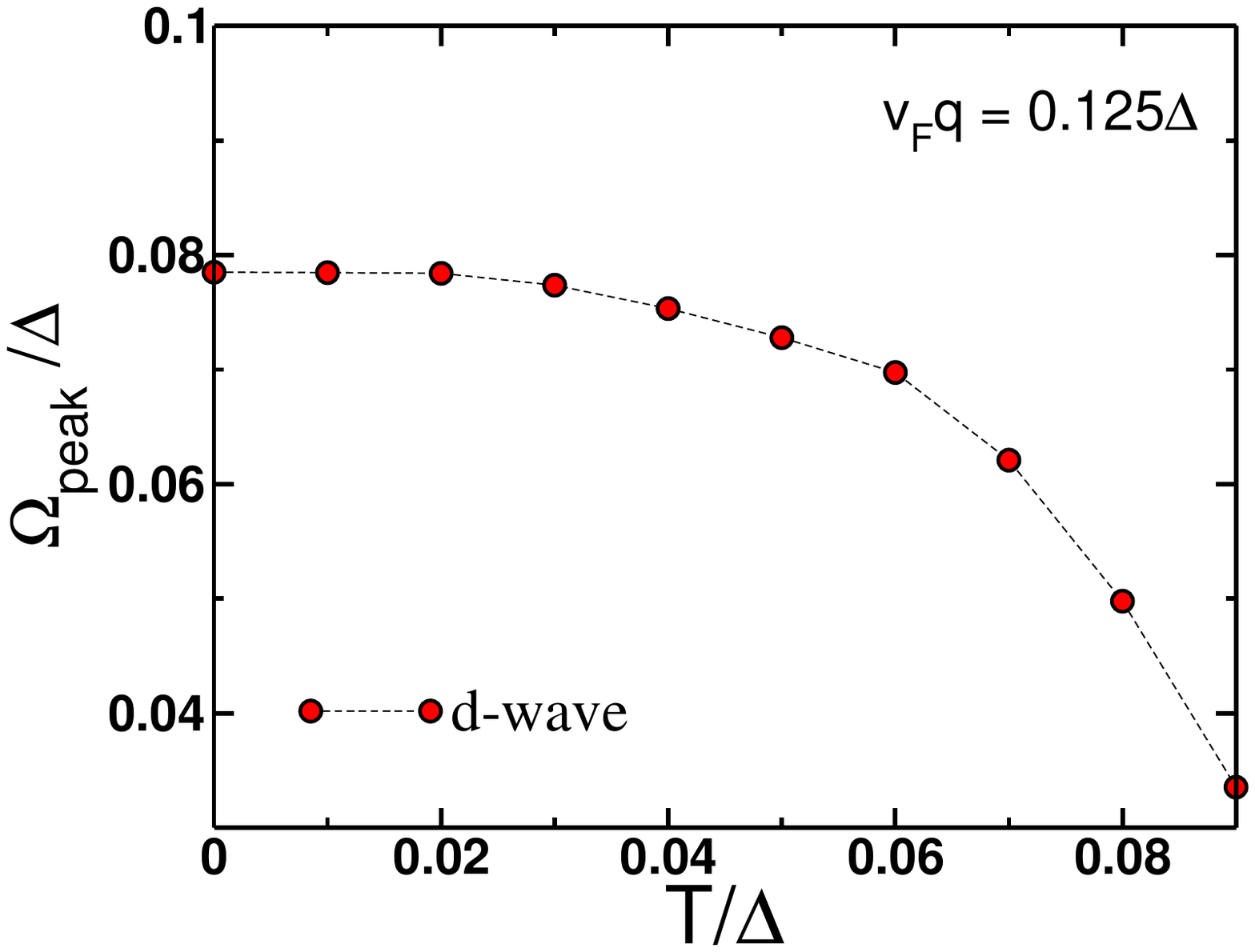} 
\caption{Left panel: real part of transverse pair susceptibility $\chi_{\textrm{AB}}(\bq,\omega)$ for $d$-wave superconductor plotted as a function of frequency and at a fixed value of momentum $v_Fq=0.125\Delta$ for different values of temperatures. All energies are given in the units of the pairing amplitude at zero temperature. Function $\chi_{\textrm{AB}}(\bq,\omega)$ exhibits a peak at $\omega=\Omega_{\mathrm{peak}}$ which corresponds to the energy of the collective transverse mode. Right panel: the dependence of $\Omega_{\mathrm{peak}}$ on temperature.} 
\label{Fig-Main-AB}
\end{figure}

 In order to determine the dispersion relation, we expand the function under the integral \eqref{chiABdwave} up to the second order in powers of $\bq$. Using the auxiliary expressions listed in Appendix \ref{AppendixC} and Appendix \ref{AppendixD}, we can cast an expression for the function 
$\chi_{\textrm{AB}}^{-1}(\bq,\Omega)$ into the following form
\beg\label{InvchiABapprox}
\chi_{\textrm{AB}}^{-1}(\bq,\Omega)\approx\Omega^2-\zeta(\Omega,T)v_F^2q^2.
\en
At small frequencies $\Omega\sim\Delta$ function $\zeta(\Omega,T)$ weakly varies with $\Omega$ and for this reason it is well approximated by 
%the following expression:
\beg\label{zetaw}
{\zeta}(T)\approx\int\limits_{0}^{2\pi}(\bn.\hat{\bq})^2\left\{1-\frac{|\Delta_\bn|}{2T}
\int\limits_{1}^{\infty}\frac{\cosh^{-2}(y|\Delta_\bn|/2T)dy}{y^3\sqrt{y^2-1}}\right\}\frac{d\theta_\bn}{2\pi},
\en
where $\hat{\bq}=\bq/q$. 
  At $T=0$, ${\zeta}(T) =1/2$. Substituting into (\ref{InvchiABapprox}), we find that 
  in a charge-neutral system 
   the collective mode is massless with the dispersion $\Omega =v_Fq/\sqrt{2}$ at the smallest $q$  and $\Omega$
  ($v_Fq, \Omega \ll \Delta$).
     This dispersion is the same as the one in an 
      $s$-
      wave superfluid. 

 This equivalence, however, does not extend to finite $T$.  In an $s-$ superfluid,  
the second term in \eqref{zetaw}  is exponentially small 
  at $T\ll \Delta$. 
   {This is not so for a $d$-wave gap function as there is always a range of $\theta_\bn$ in (\ref{zetaw}) where the argument of the hyperbolic cosine function is of order one}.  
   Limiting the angular integration to this region, we find
\beg\label{zetaTAsymptote}
{\zeta}(T\ll \Delta)\approx \frac{1}{2}-\left(\frac{3\log2}{
%AC pls check 16
%AC_l pls check
 8 \sqrt{2}}\right)\frac{T}{\Delta}-O(T^2)
\en
%\textcolor{red}{KRI: Do you want me to go over the prefactor calculation?} MD: no need
(see Appendix \ref{AppendixD} for details).
This result implies that at a finite $T$, a transverse collective mode in a $d-$wave superfluid propagates with a smaller velocity than in an $s-$wave superfluid due to thermal velocity reduction by nodal quasiparticles.

In Fig. \ref{Fig-Main-AB} (left panel), we present the dependence of the real part of the function $\chi_{\textrm{AB}}(\bq,\Omega)$ on frequency for a fixed value of momentum. Clearly, the peak in $\mathrm{Re}[\chi_{\textrm{AB}}(\bq,\Omega)]$ accounts for the excitation of the transverse mode in a charge-neutral $d$-wave superfluid. The frequency at which $\mathrm{Re}[\chi_{\textrm{AB}}(\bq,\Omega)]$ matches the value predicted using the approximate expression \eqref{zetaw}. We also note that the peak is red shifted with increasing temperature, Fig.  \ref{Fig-Main-AB} (right panel),  in agreement with our discussion above. 

Let us now turn our attention to the case of charged superconductors. With the help of the auxiliary expressions listed in Appendix \ref{AppendixC} for this case our system of equations \eqref{TSystem} reads ($d=2$):
\beg\label{TransverseMode}
\begin{aligned}
&\left[\Omega^2-v_F^2q^2\zeta(T)\right]\cdot \delta \Delta_{\bq\Omega}^T+\left({2i\Omega\Delta}\right)\cdot\delta\Phi_{\bq\Omega}=0, \\
&\left({2i\Omega}\Delta\right)\cdot\delta\Delta_{\bq\Omega}^T-\Delta^2\left\{4+\frac{2q}{\pi\nu_F e^2}\right\}\cdot\delta\Phi_{\bq \Omega}=0.
\end{aligned}
\en
%\textcolor{red}{KRI: in Eq. 39 second line, I fix the typo of $\Delta$ to $\delta \Delta$}
For a nontrivial solution to exist we require that the determinant of this system vanishes. This yields %\textcolor{red}{KRI: According to Eq. 39, you will get a factor of 2 $\pi$, instead of $4\pi$ inside the parenthesis of Eq. 40. I don't why it is written $4\pi$ which does not match with the Diagraamatic and Eq. 54. So I am changing it to $2\pi$. }
\beg\label{d2scplasmon}
\Omega^{(\mathrm{2D})}(\bq)=\sqrt{
%AC added the factor of 4
2
\pi\nu_Fe^2v_F^2\zeta(T)|\bq|},
\en
%AC_l pls check
%\textcolor{green}{KRI: Taken care of.}
The frequency of the mode scales as $\sqrt{q}$, like in an s-wave superconductor. However, the temperature variation of the effective velocity of the mode is again different: using   
  \eqref{zetaTAsymptote}, 
  We find that it decreases linearly with $T$ rather than exponentially.
  We emphasize that the $T$ variation of $\Omega_{\mathrm{p}}^{(2\mathrm{D})} (q)$ is described by the same function $\zeta(T)$ as in the charge neutral case.

\paragraph{$D=3$.} Let us now turn our attention to the case of the three-dimensional superconductor. For the charge neutral case, the dispersion of the transverse mode follows directly from \eqref{zetaw} where additional angular averaging must be performed. Upon averaging over the directions of $\bn$ at $T=0$ gives $\Omega^{({\mathrm{3D}})}=v_Fq/\sqrt{3}$. At finite temperatures, we find that the magnitude of $\Omega^{({\mathrm{3D}})}$ decreases with temperature just like in the 2D case. 

In the charged case, the system of equations is very similar to \eqref{TransverseMode} where we only need to change $2q\to q^2$ in the last term in the second equation in \eqref{TransverseMode}. Simple algebra gives the following formula for the dispersion of the transverse mode:
\beg\label{3dplasmak}
  \Omega_{\mathrm{p}}^{(3\mathrm{D})}(\bq)=\Omega_{\mathrm{p}}^{(3\mathrm{D})}(0)\sqrt{1+\frac{q^2}{4\pi\nu_Fe^2}},
\en   
where $\Omega_{\mathrm{p}}^{(3\mathrm{D})}(0)$ is the superconducting plasma frequency 
%\textcolor{red}{I double checked it. It should be $4\pi$ inside the parenthesis. Also, I change $\zeta(T)=\zeta(0)$ as we are calculating plasmon mode at zero temp.}:
\beg\label{VanishDet3}
\Omega_{\mathrm{p}}^{(3\mathrm{D})}(0)=\sqrt{
%AC 4
4
\pi\zeta(0) \nu_Fe^2v_F^2} = \sqrt{4\pi n e^2/m}
\en
%AC_l pls check
%\textcolor{green}{KRI: Taken care of.}
 where  $\zeta(0)=1/3$ in 3D, fermionic density $n = p^3_F/(3\pi^2)$ and $v_F=(3\pi^2n)^{1/3}\, m/\pi^2$.
At a finite $T$,
the  
 transverse mode in a $d-$wave superconductor is 
 again
located at a lower energy than in an $s-$wave superconductor. This result has a simple physical interpretation: at a finite $T$, nodal quasi-particles in a $d$-wave superconductor partially screen the Coulomb potential, leading to a smaller value of the superconducting plasma frequency.

\subsection{Diagrammatic approach}
We use Eq.~\eqref{susceptibility_equation} for the computation for the transverse pair-susceptibility, $\chi_T (q)$.  After analytical continuation to real frequencies, it becomes
\begin{align}
    \chi_T(\bq,\Omega)=\dfrac{\bar{\Pi}_T(\bq,\Omega)}{1-g\,\bar{\Pi}_T(\bq,\Omega)},
    \label{chi_T_expression}
\end{align}
where $\quad \bar{\Pi}_T(\bq,\Omega)=\Pi_T(\bq,\Omega)-8\,
\dfrac{\Pi^2_{GF}(\bq,\Omega)}{V^{-1}_\bq-2\,\Pi_0(\bq,\Omega)}$ 
  (see  Eq.~\eqref{Pi_T_expression}). We first present the results at $T=0$ and then at a finite $T$. 
\subsubsection{$T=0$}
For the polarization bubbles $\Pi_T,\Pi_{GF}$ and $\Pi_0$ at zero temperature, we obtain  
  after frequency integration and analytical continuation
 (see Appendix \ref{Bubble_calculation} for details) 
\begin{align}
   \Pi_T(\bq,\Omega)&=\dfrac{\nu_F}{2}\int_0^{2\pi}\gamma^2(\theta_\bk)\dfrac{d\theta_\bk}{2\pi}\int_{-\Lambda}^{\Lambda} d\xi_\bk \dfrac{E_++ E_-}{E_+\, E_-}\quad  \dfrac{E_+\, E_-+\xi_+\, \xi_-+\Delta^2_\bk}{(E_++E_-)^2-\Omega^2},\\
      \Pi_{GF}(\bq,\Omega)&=-\dfrac{\nu_F\, \Omega}{4}\int_0^{2\pi}\gamma (\theta_\bk)\dfrac{d\theta_\bk}{2\pi}\int_{-\Lambda}^{\Lambda} d\xi_\bk \dfrac{E_++ E_-}{E_+\, E_-}\quad  \dfrac{\Delta_\bk}{(E_++E_-)^2-\Omega^2},
      \label{delta_pi_m_expression}\\
      \Pi_0(\bq,\Omega)&= -\dfrac{\nu_F}{2}\int_0^{2\pi}\dfrac{d\theta_\bk}{2\pi}\int_{-\Lambda}^{\Lambda} d\xi_\bk \dfrac{E_++ E_-}{E_+\, E_-}\quad  \dfrac{E_+\, E_--\xi_+\, \xi_-+\Delta^2_\bk}{(E_++E_-)^2-\Omega^2},
      \label{delta_pi_0_expression}
\end{align}
where $\Omega$ should be understood as $\Omega + i \delta$,  $E_\pm=E_{\bk\pm \bq/2}$, $\xi_\pm=\xi_{\bk\pm \bq/2}$, 
 and, we remind, 
$\gamma(\theta_\bk)=\sqrt{2}\, \cos2\theta_\bk$ and the gap function $\Delta_\bk=\Delta\, \gamma(\theta_\bk)$.\\

We first consider a neutral superfluid and neglect 
the Coulomb interaction in Eq.~\eqref{chi_T_expression}. In this case, 
we expect that the pole of $\chi_{T}(\bq,\Omega)$ 
 yields a 
 Goldstone AB mode.
To find its dispersion, we split $\Pi_T(\bq,\Omega)$ into $\Pi_T(\bq,\Omega) = \Pi_T(0)+\delta\Pi_T(\bq,\Omega)$ and use the gap equation~\eqref{Gap equation s-d} in the form
 $1=g\,\Pi_T(0)$. After a straightforward algebra, we find 
\begin{align}
    \chi_T(\bq,\Omega)=-\dfrac{1}{g^2 \delta\Pi_T(\bq,\Omega)}-\dfrac{1}{g},
\end{align}
where 
\begin{align}
    \delta \Pi_T(\bq,\Omega)=-&\dfrac{\nu_F}{4}\int\limits_0^{2\pi}\gamma^2(\theta_\bk)\dfrac{d\theta_\bk}{2\pi}\int\limits_{-\infty}^\infty d\xi_\bk \left(\dfrac{E_++ E_-}{E_+\, E_-}\right) \dfrac{(\Omega+i\delta)^2-\left(\xi_+-\xi_-\right)^2}{(\Omega+i\delta)^2-\left(E_++ E_-\right)^2}.
     \label{delta piT equation s-d}
\end{align}
The corresponding expression for an $s-$wave superfluid has been obtained in Ref. \cite{Kulik1981}.
The dispersion of the AB mode is extracted from
 $\delta\Pi_T(\bq,\Omega)=0$. 

For small $\bq$ ($v_F|\bq|\ll\Delta$)  we approximate
$(\xi_+-\xi_-)^2\approx v^2_F|\bq|^2 \cos^2(\theta_\bk-\theta_\bq)$ and 
  $(E_++E_-)^2 \approx 4 \Delta_\bk^2+4 \xi^2_\bk+v^2_F|\bq|^2 \cos^2(\theta_\bk-\theta_\bq)$  in 
 the
  integrand of Eq.~\eqref{delta piT equation s-d}. The rest of the integrand is non-singular and 
   can be evaluated at $\bq=0$. Integrating 
    over $\xi_\bk$, and further expanding in small $ v_F |\bq|/\Delta$ and $\Omega/\Delta$ we obtain, in $2D$,  
\beg 
\begin{aligned}
    \delta\Pi_T(q)= \nu_F &\int_0^{2\pi} \dfrac{d\theta_\bk}{2\pi}\gamma^2(\theta_\bk) \dfrac{\Omega^2-v_F^2|\bq|^2 \cos^2(\theta_\bk-\theta_\bq)}{4 \Delta^2 \gamma^2(\theta_\bk)}=
    \frac{\nu_F}{4 \Delta^2}\, \left(\Omega^2-\frac{v_F^2|\bq|^2}{2}\right),
    \label{delta_Pi_T_expression}
\end{aligned}
\en
 where $\theta_\bq$ is the direction of the momentum $\bq$. The mode dispersion is 
   $\Omega=v_F |\bq|/\sqrt{2}$.  
  The same analysis for $D=3$ yields  $\Omega=v_F |\bq|/\sqrt{3}$.
    Both expressions are the same as we found quasi-classically and coincide with the corresponding expressions for 
      $s-$wave superfluids.   
  
  We next 
  include the Coulomb interaction into consideration and 
   use the full expression for the 
   transverse polarization bubble $\bar{\Pi}_T(\bq,\Omega)$, defined in Eq.~\eqref{chi_T_expression}. 
   We 
    use the fact that $\bar{\Pi}_T(0,0)=\Pi_T(0,0)$ and 
   express $\bar{\Pi}_T(\bq,\Omega)$ as 
   $\bar{\Pi}_T(\bq,\Omega)=\Pi_T(0,0)+\delta\bar{\Pi}_T(\bq,\Omega)$. Using the gap equation \eqref{Gap equation s-d}, we then obtain
 \begin{align}
     \chi_T(\bq,\Omega)=-\dfrac{1}{g^2 \delta\bar{\Pi}_T(\bq,\Omega)}-\dfrac{1}{g}. 
 \end{align}
 where \begin{align}
     \delta\bar{\Pi}_T(\bq,\Omega)=\delta\Pi_T(\bq,\Omega)-\dfrac{8\,\Pi_{GF}(\bq,\Omega)^2}{V^{-1}_\bq-2\, \Pi_0(\bq,\Omega)}.
     \label{delta_pi_bar_eq1}
 \end{align}
 The pole of the pair susceptibility $\chi_T(\bq,\Omega)$ is given by the condition $\delta\bar{\Pi}_T(\bq,\Omega)=0$. 
  At small $v_F|\bq|\ll \Delta $
    we expand $\delta\bar{\Pi}_T(\bq,\Omega)$ in small $\bq$ and ignore terms of the order $O(\Omega^2 |\bq|^2)$. 
    Approximating $\Pi_{GF}(\bq,\Omega)\approx -\nu_F\, \Omega/4\, \Delta$ and  $\Pi_0(\bq,\Omega)\approx -\nu_F$, using Eqs.~\eqref{delta_pi_m_expression}- \eqref{delta_pi_0_expression}, and further expanding Eq.~\eqref{delta_pi_bar_eq1} in $V^{-1}_\bq/2\nu_F$,  we find that 
 \begin{align}
    \delta\bar{\Pi}_T(\bq,\Omega)=\dfrac{|\bq|^2}{8 \, \Delta^2}\left( \dfrac{\Omega^2\, V^{-1}_\bq}{|\bq|^2}-\nu_F\, v^2_F\right).
\end{align}
The dispersion of the longitudinal mode is 
 \begin{align}
    \Omega(\bq)=\sqrt{\nu_F\, v^2_F\, |\bq|^2\, V_\bq}.
\label{jj_1}
\end{align}
The Coulomb potential in 2D is $V_\bq=2 \pi e^2/|\bq|$. 
 Substituting into (\ref{jj_1}), we find
 \begin{align}
    \Omega_{\mathrm{p}}^{(\mathrm{2D})}(\bq)=\sqrt{2\,\pi\,e^2\,\nu_F\, v^2_F|\bq|},
\end{align}

Expending this analysis to $D=3$, we find $\Omega_{\mathrm{p}}^{(\mathrm{3D})}=
\sqrt{(2/3) \nu_F v^2_F  q^2 V_q}$. Substituting $V_q = 4\pi e^2/q^2$ and using $p_F =(3\pi^2 n)^{1/3}$, 
we obtain $\Omega_{\mathrm{p}}^{(\mathrm{3D})}=\sqrt{4\,\pi\,e^2\, n/m}$. 

Comparing the results of the {quasiclassical} and diagrammatic approaches, we see that they are equivalent, as it indeed should be.
\subsubsection{finite $T$}

 We first discuss 
 charge neutral $d$-wave superfluid. The mode dispersion is determined from $\Pi_T (\bq, \Omega) = 1/g$, where we suppress the explicit temperature dependence of $\Pi_T$ for notational simplicity, and has to be obtained by summing over Matsubara frequencies instead of integrating over the frequency. At finite $T$, we find the following expression (see Appendix \ref{Bubble_calculation} for details), 
 \begin{align}
   \label{tranverse_bubble}
   \Pi_T(\bq,\Omega)&=\dfrac{1}{2} \int\dfrac{d^2\bk}{(2\pi)^2}\, \gamma^2_\bk \left[\left\{\left[1-n(E_+)-n(E_-)\right]  \left(\dfrac{E_++E_-}{E_+\, E_-}\right)\, \left(\dfrac{E_+ \, E_-+\xi_+\, \xi_-+\Delta_\bk^2(T)}{(E_++E_-)^2-\Omega^2}\right)\right\} \right. \nn & \left.-\left\{\left[n(E_+)-n(E_-)\right]\, \left(\dfrac{E_+-E_-}{E_+\, E_-}\right)\,\left( \dfrac{E_+\, E_--\xi_+\, \xi_--\Delta_\bk^2(T)}{(E_+-E_-)^2-\Omega^2}\right)  \right\}\right],
   \end{align}
where $\xi_\pm=\xi_{\bk\pm \bq/2}$, $E_\pm=E_{\bk\pm\bq/2}$, and $n(E_\pm)=1/\{\exp(\beta\, E_\pm)+1\}$ is the Fermi function. As before, we express $\Pi_T (\bq, \Omega)$ as $\Pi_T (0,0) + \delta \Pi_T (\bq, \Omega)$ and use the gap equation 
$\Pi_T (0,0) = 1/g$ such that the mode dispersion comes from $\delta \Pi_T  (\bq, \Omega) =0$. For small $\bq$, we approximate $\xi_\pm=\xi_\bk+\mathbf{v}_F.\bq/2$ and $E_{\pm}=E_\bk\pm \delta E_{\bk}$ where $\delta E_{\bk}= \xi_\bk\mathbf{v}_F.\bq/2\, E_\bk $, and express 
 $\delta\Pi_T(\bq,\Omega)$ as $\delta\Pi_T^{\mathrm{(a)}}(\bq,\Omega)+\delta\Pi_T^{\mathrm{(b)}}(\bq,T)$, where
 \begin{align}
   \label{deltaPiTa} 
   \delta\Pi_T^{(\mathrm{a})}(\bq,\Omega)&=\dfrac{\nu_F}{2}\int_{-\infty}^\infty   d\xi_\bk\int\limits_0^{2\pi}\gamma^2(\theta_\bk)\frac{d\theta_\bk}{2\pi}
\left[1-2 n(E_\bk)\right] \\&\times\left[ \left(\dfrac{E_++E_-}{E_+\, E_-}\right) \,  \dfrac{E_+ \, E_-+\xi_+\, \xi_-+\Delta_\bk^2(T)}{(E_++E_-)^2-\Omega^2} -\dfrac{1}{E_\bk}\right]
\end{align}
 
and
 \begin{align}
    \label{deltaPiTb}
    \delta\Pi_T^{(\mathrm{b})}(\bq,\Omega)&=-\nu_F\int_{-\infty}^\infty   d\xi_\bk\int\limits_0^{2\pi}\gamma^2(\theta_\bk)\dfrac{d\theta_\bk}{2\pi} \, \left[\dfrac{\partial n(E_\bk)}{\partial E_\bk}\right]  \delta E_{\bk}\\&\times \left(\dfrac{E_+-E_-}{E_+\, E_-}\right) \,  \dfrac{E_+ \, E_--\xi_+\, \xi_--\Delta_\bk^2(T)}{(E_+-E_-)^2-\Omega^2}.
\end{align} 
We expand the integrand of $\delta \Pi^\mathrm{(a,b)}_T$  to  second order in $\bq$ and treat $T$ dependence as perturbation ($T<< \Delta,  n(E_\bk)\approx 0$). This gives for   
 $\delta\Pi_T^{(\mathrm{a})}(\bq,\Omega)$ an the overall factor $\Omega^2-(\mathbf{v}_F.\bq)^2$, and equal to 
 \begin{align}
     \delta\Pi^\mathrm{a}_T(\bq,\Omega)=\dfrac{\nu_F}{4} \int_0^{2\pi}\dfrac{d\theta_\bk}{2\pi} \gamma^2(\theta_\bk) \left[\Omega^2-(\mathbf{v}_F.\bq)^2\right].
 \end{align}In the second term, $\delta\Pi_T^{(\mathrm{b})}$, we obtain
\beg\label{deltaPiTbapp}
\begin{aligned}
\delta\Pi_T^{(\mathrm{b})}(\bq,T)\approx -\dfrac{\,\nu_F\Delta^2}{4}\int\limits_0^{2\pi}\frac{d\theta_\bk}{2\pi}({\mathbf v}_F.\bq)^2\gamma^4(\theta_\bk)\int\limits_{-\infty}^{\infty}\frac{d\xi_\bk}{E_\bk^4}\left[\dfrac{\partial n(E_\bk)}{\partial E_\bk}\right],
\end{aligned}
\en
where we approximate the gap amplitude $\Delta(T)\approx \Delta(0)=\Delta$. Combining \eqref{deltaPiTa} and \eqref{deltaPiTb} and introducing new integration variables $\xi_\bk=x|\gamma(\theta_\bk)|\Delta$ and  $y=\sqrt{x^2+1}$, we obtain the dispersion relation at a finite $T$ in the form $\Omega = v_F q \sqrt{\zeta (T)}$, where ($\hat{\mathbf{n}}\equiv\bk/|\bk|$ and $\Delta_\mathbf{n}\equiv\Delta(\theta_\bk)$) 
\beg\label{FiniteTAB}
\zeta (T) = \int\limits_0^{2\pi}(\hat{\bn}.\hat{\bq})^2\frac{d\theta_\bk}{2\pi}
\left\{1-\left(\frac{|\Delta_\bn|}{2T}\right)\int\limits_{1}^{\infty}\frac{\cosh^{-2}\left(\frac{|\Delta_\bn|}{2T}y\right)dy}{y^3\sqrt{y^2-1}}\right\},
\en
 is the same as  we found in the quasiclassical formalism,
Eq. (\ref{zetaw}). 
{For the charged $ d$-wave superconductor, we find that the temperature corrections to the bubbles, $\Pi_{GF}$ and $\Pi_0$, are subleading and can be ignored}. As a result, we find that the diagrammatic result for the mode dispersion in a charged $d-$wave superconductor contains the same $\zeta (T)$, again in agreement with the result obtained using the quasiclassical formalism, Eq. \eqref{d2scplasmon}.  
\section{
 Longitudinal  susceptibility and dispersion of 
 the longitudinal mode}\label{SHQA}
\subsection{Quasiclassical approach}
We now turn our discussion to the derivation of the expression for the pair susceptibility of the longitudinal (Schmid-Higgs)  mode. Calculation of the 
 longitudinal susceptibility is essentially identical to that for the transverse mode. The only difference lies in the choice of the matrix structure for $\delta\hat{\Delta}(\br,t)$. Namely, in the case of longitudinal fluctuations must be given by:
\beg\label{dDLTL}
\delta\hat{\Delta}_\bn(\br,t)=\left(-i\hat{\tau}_2\right)\gamma(\theta_\bn)\delta\Delta_{\bk\omega}^Le^{2i(\bk\br-\omega t)}.
\en
The expression for the longitudinal
 susceptibility can be derived from the self-consistency equation 
\beg\label{Self}
\delta\Delta_{\bk\omega}^L=\frac{g\nu_F}{4}\int\limits_0^{2\pi}\frac{d\theta_\bn}{2\pi}\gamma(\theta_\bn)\int\limits_{-\infty}^\infty{d\eps}\textrm{Tr}\left\{-i\hat{\tau}_2\hat{g}_2^K(\bn\eps;\bk\omega)\right\},
\en
where $\hat{g}_2^K(\bn\eps;\bk\omega)$ is formally still given by \eqref{g2KansatzT} along with Eqs. (\ref{g2RAT},\ref{dg2RAKFinalTPoisson}), where we have to set $\delta\Phi_{\bk\omega}$ to zero and replace $\delta\hat{\Delta}_{\bk\omega}^T\to(-i\hat{\tau}_2)\delta{\Delta}_{\bk\omega}^L$. Inserting the resulting expression for $\hat{g}_2^K$ into \eqref{Self} yields the following expression for the 
 longitudinal  susceptibility:
\beg\label{chiSHdwave}
\begin{aligned}
&\chi_{\textrm{L}}^{-1}(\bq,\Omega)=-\frac{4}{g\nu_F}+\int\limits_{-\Lambda}^{\Lambda}d\eps\int\limits_{0}^{2\pi}\frac{d\theta_\bn}{2\pi}{\gamma}^2(\theta_\bn)\\&\times\left\{
\frac{\left(\eta_{\bn\eps+\Omega/2}^{R}+\eta_{\bn\eps-\Omega/2}^{A}\right){\cal A}_\bn^K(\eps_+,\eps_-)(t_{\eps+\Omega/2}-t_{\eps-\Omega/2})}{\left(\eta_{\bn\eps+\Omega/2}^{R}+\eta_{\bn\eps-\Omega/2}^{A}\right)^2-{v}_F^2(\bn.\bq)^2}\right.\\&\left.+\frac{\left(\eta_{\bn\eps+\Omega/2}^{R}+\eta_{\bn\eps-\Omega/2}^{R}\right){\cal A}_\bn^R(\eps_+,\eps_-)t_{\eps-\Omega/2}}{\left(\eta_{\bn\eps+\Omega/2}^{R}+\eta_{\bn\eps-\Omega/2}^{R}\right)^2-{v}_F^2(\bn.\bq)^2}-
\frac{\left(\eta_{\bn\eps+\Omega/2}^{A}+\eta_{\bn\eps-\Omega/2}^{A}\right){\cal A}_\bn^A(\eps_+,\eps_-)t_{\eps+\Omega/2}}{\left(\eta_{\bn\eps+\Omega/2}^{A}+\eta_{\bn\eps-\Omega/2}^{A}\right)^2-{v}_F^2(\bn.\bq)^2}\right\}.
\end{aligned}
\en
Here we introduced auxiliary functions
\beg\label{AKRAL}
\begin{aligned}
&{\cal A}_\bn^{R(A)}(\eps,\eps')=g_{\bn\eps}^{R(A)}g_{\bn\eps'}^{R(A)}+f_{\bn\eps}^{R(A)}f_{\bn\eps'}^{R(A)}+1, \quad
{\cal A}_\bn^{K}(\eps,\eps')=g_{\bn\eps}^{R}g_{\bn\eps'}^{A}+f_{\bn\eps}^{R}f_{\bn\eps'}^{A}+1
\end{aligned}
\en
and $\eps_{\pm}=\eps\pm\Omega/2$. The term which contains the coupling constant can be expressed in terms of the ground state corrections functions using the self-consistency condition \ref{Coupling}. 
It is worth noting that in the limit $\bq=0$ and taking ${\gamma}(\theta_\bn)=1$ we recover the previously derived  expression for the 
 longitudinal 
 susceptibility of a 
  $s$-wave superconductor. Note also that while the integrals in Eqs. \eqref{chiSHdwave} and \eqref{Coupling} need to be
cut off at some ultraviolet energy scale $\Lambda$, being taken together they yield
a UV convergent integral. Thus expression for the inverse
susceptibility, $\chi_{\textrm{L}}^{-1}(\bq,\Omega)$, is in fact cutoff independent. 

In Fig. \ref{Fig-chiSH}, we show the frequency dependence of 
  the longitudinal susceptibility in the $d$-wave case for different values of momentum and compare it with the corresponding results for the $s$-wave case. The fact that $\chi_{\mathrm{L}}(\Omega,\bq)$ has negligible dependence on $\bq/q$ can be understood from the analysis of \eqref{chiSHdwave} for small values of momentum. Expanding the expression under the integral, it follows that due to the square of the form factor under the integral, the angular average for $\hat{\bq}=\bq/q$ along the nodal and anti-nodal directions yields essentially the same answer. As it is well known in the $s$-wave case (and in the absence of disorder), the Higgs resonance is at $\omega_{\mathrm{peak}}=2\Delta$ for $\bq=0$ and it broadens and shifts to higher frequencies as the value of $\bq$ increases \cite{Phan2023,Pasha2025}. It is worth noting here that in the case of a diffusive $s$-wave superconductor, the Higgs resonance shifts to lower frequencies with increasing values of momentum \cite{Pasha2025,Kamenev2025}. In the $d$-wave case the Higgs resonance is at frequency $\omega\approx 2\Delta_{\mathrm{max}}$ with $\Delta_{\mathrm{max}}$ given by the value of the pairing amplitude in the antinodal direction, $\Delta_{\mathrm{max}}=\sqrt{2}\Delta$.  

\subsection{Diagrammatic approach}

We restrict our discussion here to the case of $T=0$ as the differences between $d-$wave and $s-$wave cases are already manifested at zero temperature. 

\begin{figure}
\includegraphics[width=0.4750\linewidth]{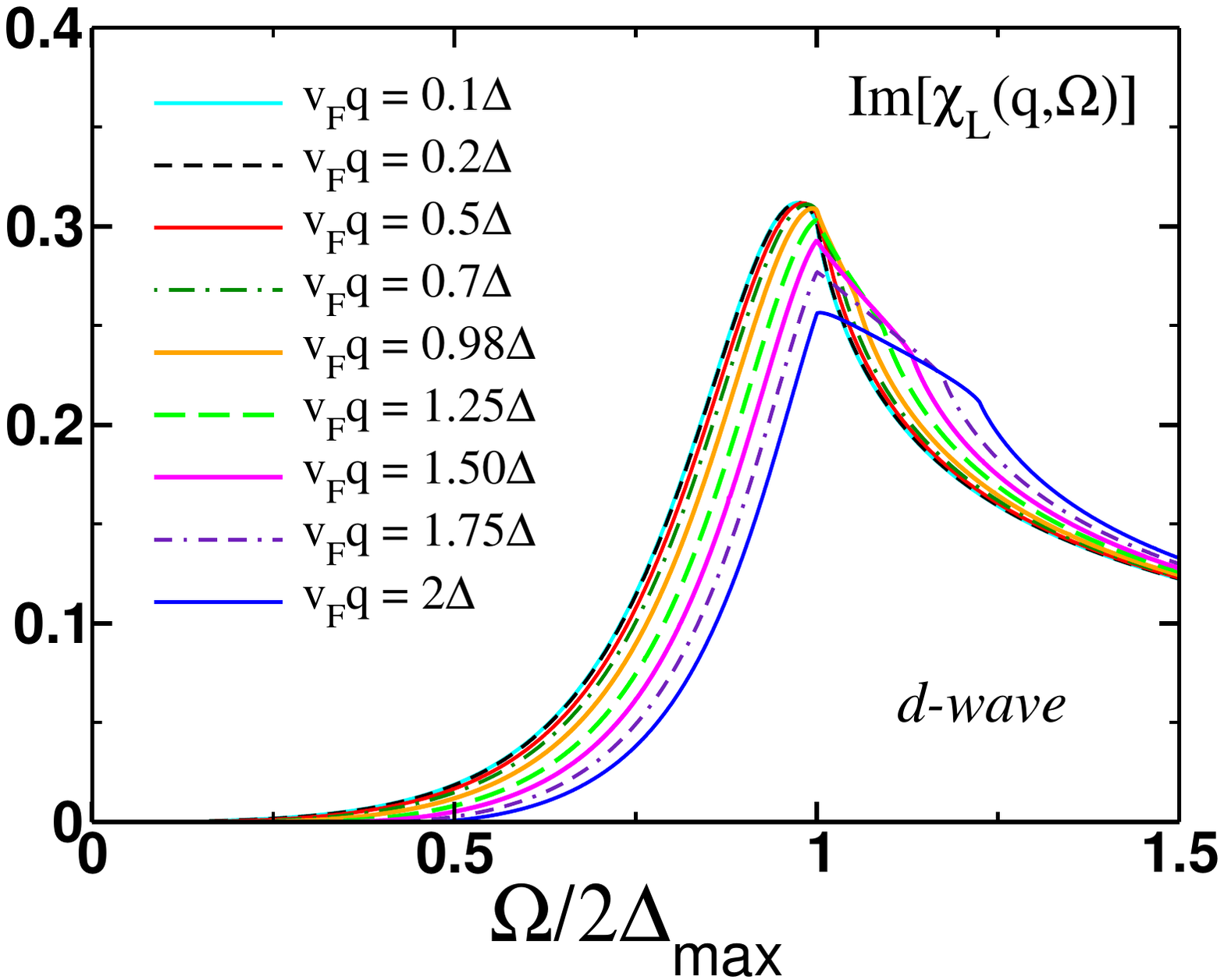} \quad 
\includegraphics[width=0.4750\linewidth]{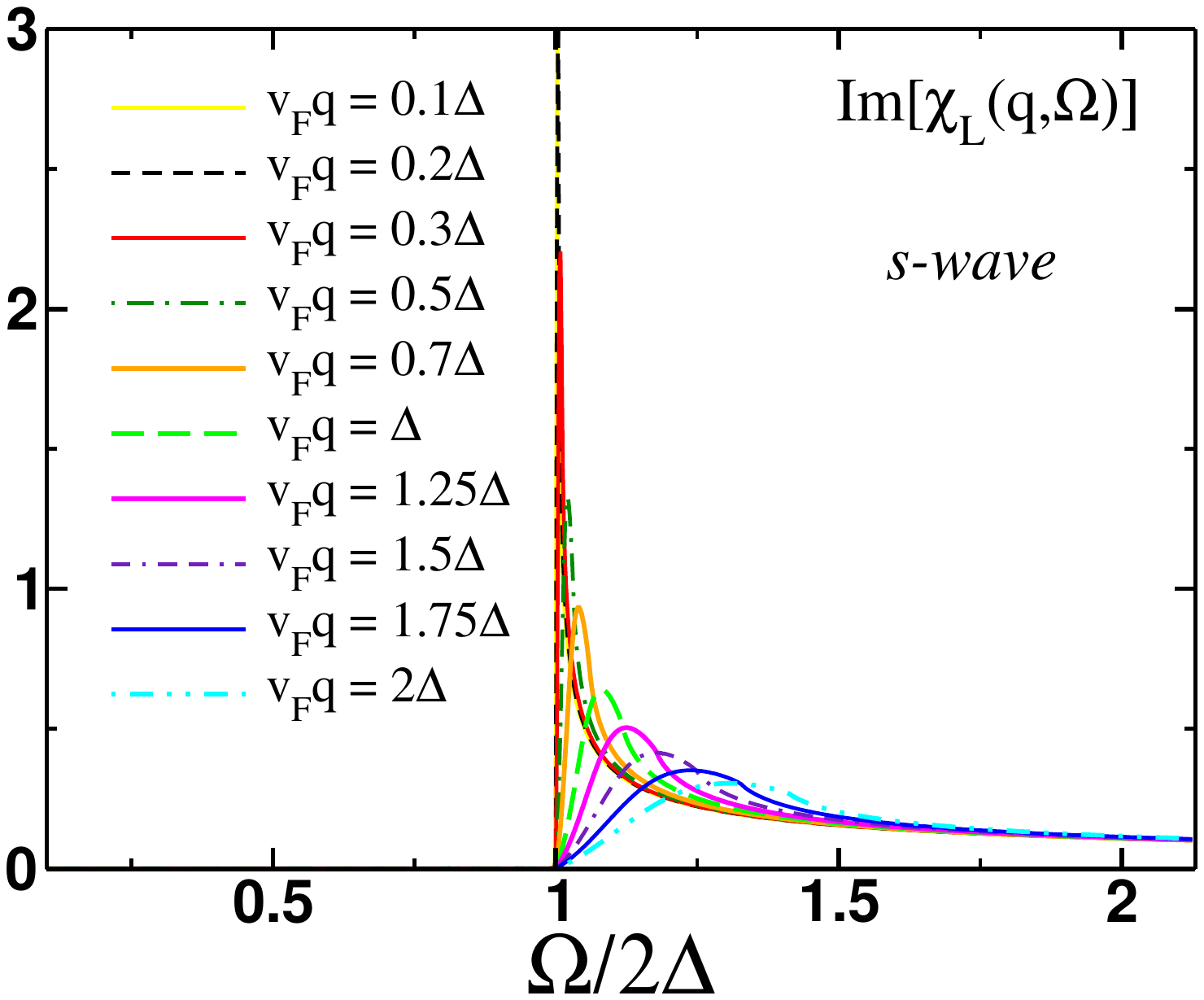}
\includegraphics[width=0.4750\linewidth]{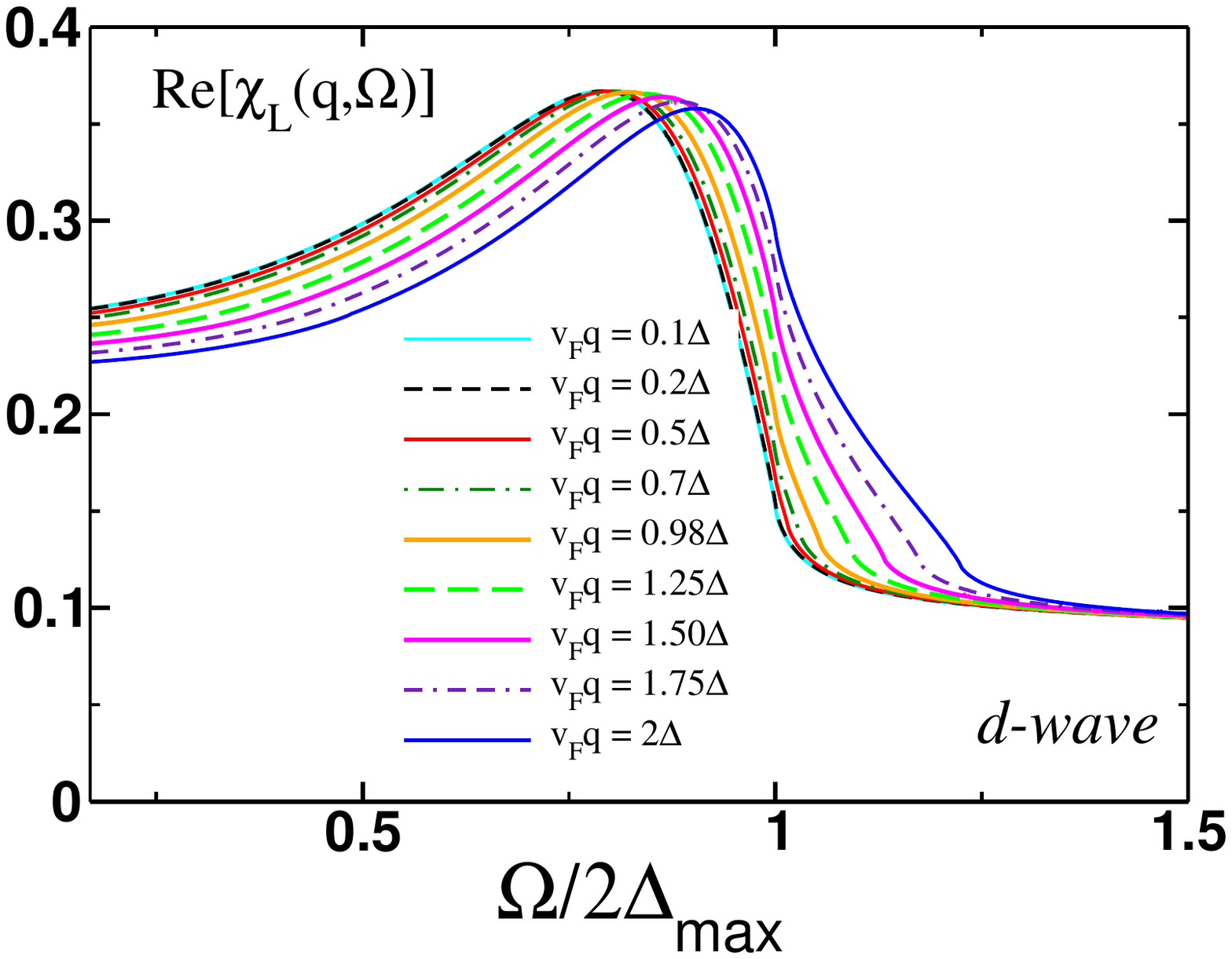} \quad 
\includegraphics[width=0.4750\linewidth]{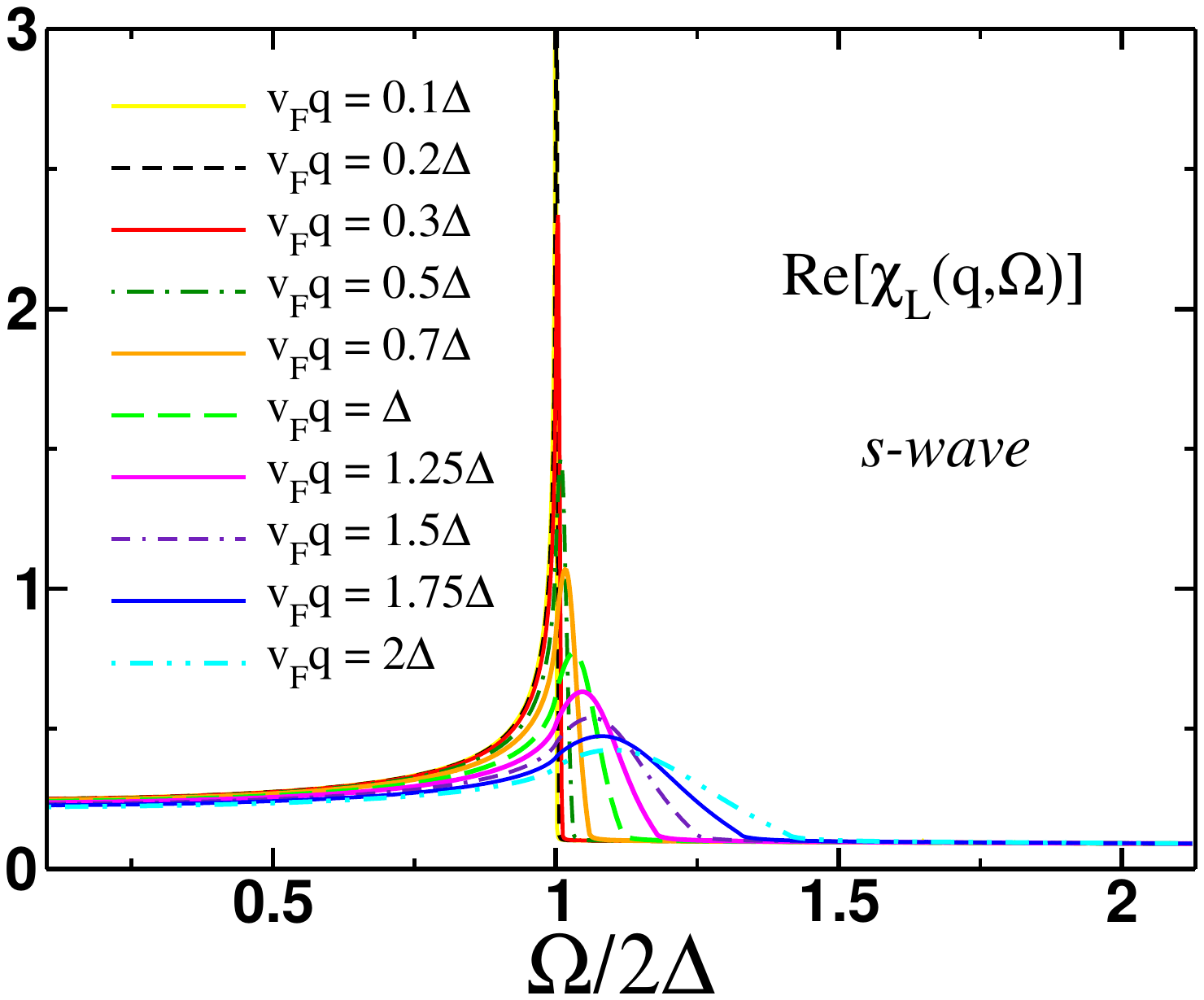} \\
\caption{Frequency dependence of the imaginary and real parts of the longitudinal susceptibility, evaluated at different values of momentum $\bq$ for both $s$-wave (right panel) and $d$-wave superconductors (left panel). The results for the $d$-wave case were obtained for $\bq$ pointing along the anti-nodal direction $\bq=q(1,0)$. We remind the reader that $\Delta_{\mathrm{max}}=\sqrt{2}\Delta$.} 
\label{Fig-chiSH}
\end{figure}

\begin{figure}
\includegraphics[width=0.450\linewidth]{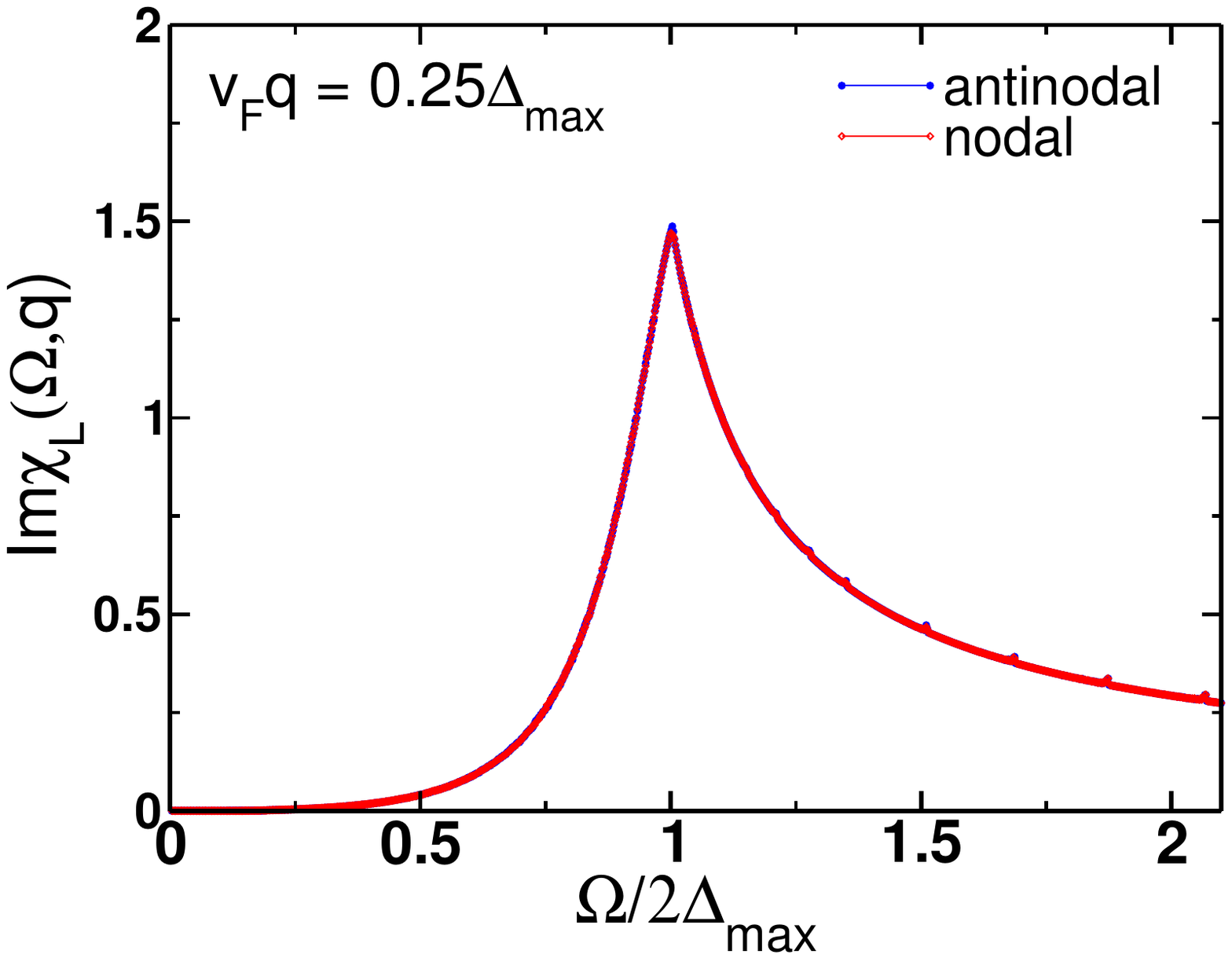} 
\includegraphics[width=0.50\linewidth]{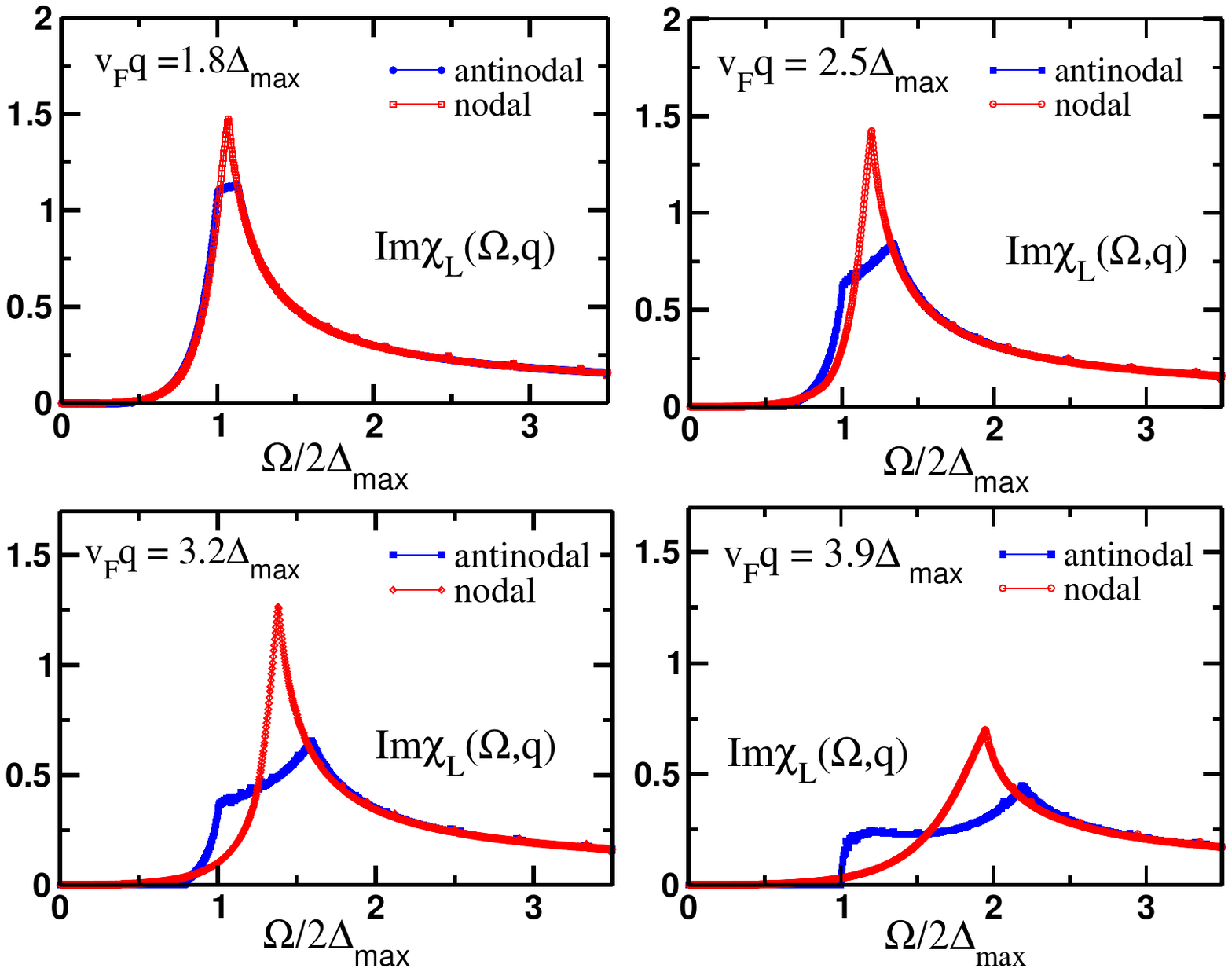} 
\caption{Frequency dependence of the imaginary part of the $d$-wave 
 longitudinal 
susceptibility evaluated for five different values of momentum. At small values of momentum $v_Fq<\Delta$, the difference between these two directions is negligible, however it becomes pronounced for $v_Fq>\Delta$. } 
\label{Fig-NodalvsAntiNodal} 
\end{figure}

\begin{figure}
\includegraphics[width=0.45\linewidth]{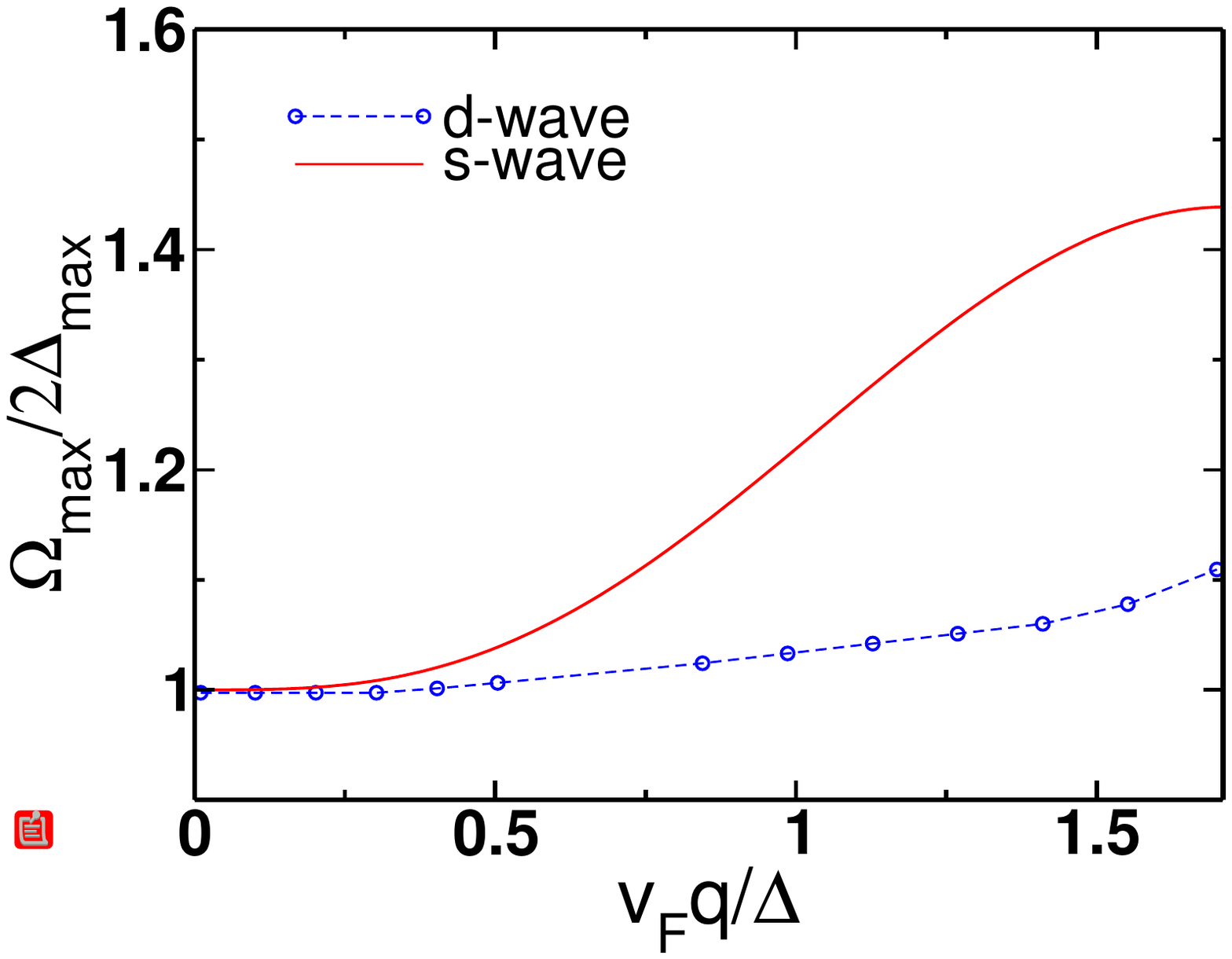} 
\includegraphics[width=0.45\linewidth]{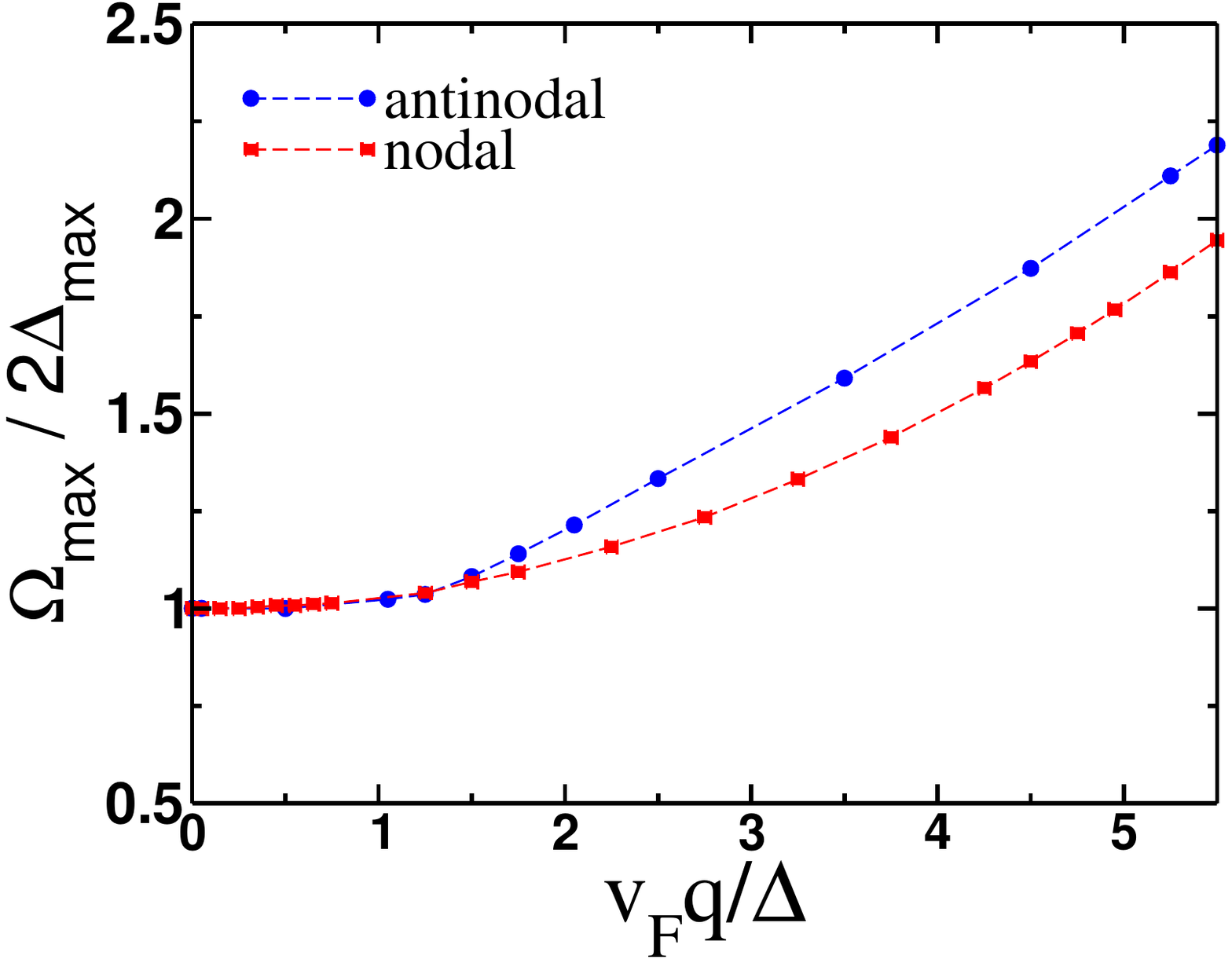} 
\caption{Left panel: dependence of the position of the maximum in $\mathrm{Im}\chi_{\mathrm{L}}(\omega,\bq)$ for $s$-wave and $d$-wave case. For the $s$-wave case $\Delta_{\mathrm{max}}=\Delta$ while in the $d$-wave case $\Delta_{\mathrm{max}}=\sqrt{2}\Delta$. Right panel: dependence of the position of the maximum in  
$\mathrm{Im}\chi_{\mathrm{L}}(\omega,\bq)$ 
 for two different directions of $\bq$:  antinodal $\bq=q(1,0)$ and nodal $\bq=q(1,1)/\sqrt{2}$. 
 The resonance frequency in the nodal direction is lower than that in the antinodal direction. This agrees with our analytical reasoning. }
\label{Fig-wmax}
\end{figure}

The longitudinal part of the pair susceptibility $\chi_L $ is obtained by analytical continuation of 
  Eq.~(\ref{susceptibility_equation}) from  Matsubara to real axis ($i\Omega_m \to \Omega + i \delta$) 
  We have 
\beg\label{CHISHDIA}
    \chi_{\textrm{L}}(q)=\dfrac{\Pi_L(q)}{1-g\,\Pi_L(q)},
\en
where $q=(\bq,\Omega)$. As before, we split the polarization bubble into two parts: $\Pi_L(q)=\Pi_T(0)+\delta \Pi_L(q)$. Using the gap equation  $1=g \,\Pi_T(0)$, we 
 find 
\begin{align}
     \chi_{\textrm{L}}(q)=-\dfrac{1}{g^2\delta\Pi_L(q)}-\frac{1}{g},
     \label{pi and chi relation}
\end{align}
where 
\begin{align}
    \delta \Pi_L(q)=-&\dfrac{\nu_F}{4}\int\limits_0^{2\pi}\gamma^2(\theta_\bk)\dfrac{d\theta_\bk}{2\pi}\int\limits_{-\infty}^\infty d\xi_\bk \left(\dfrac{E_++ E_-}{E_+\, E_-}\right) \dfrac{(\Omega+i\delta)^2-4 \Delta_\bk^2-\left(\xi_+-\xi_-\right)^2}{(\Omega+i\delta)^2-\left(E_++ E_-\right)^2}.
     \label{delta piL equation s-d}
\end{align}
For small $\bq$ ($v_Fq\ll\Delta $)  we again approximate  in the integrand of Eq.~\eqref{delta piL equation s-d} $(\xi_+-\xi_-)^2$ by 
$v^2_F|\bq|^2 \cos^2(\theta_\bk-\theta_\bq)$ and 
$(E_++E_-)^2$ by  
$4 \Delta_\bk^2+4 \xi^2_\bk+v^2_F|\bq|^2 \cos^2(\theta_\bk-\theta_\bq)$,  
  and set $\bq=0$ in 
  the rest of the integrand.
   Integrating over $\xi_\bk$, we 
    obtain
\beg\label{delta piL for d}
\begin{aligned}
    \delta\Pi_L(q)=-\nu_F&\int_0^{2\pi}\gamma^2(\theta_\bk)\dfrac{\sqrt{4 \Delta_\bk^2+v^2_F|\bq|^2 \cos^2(\theta_\bk-\theta_\bq)-(\Omega+i\delta)^2}}{\sqrt{(\Omega+i\delta)^2-v^2_F|\bq|^2 \cos^2(\theta_\bk-\theta_\bq)}}\\&\times \sec^{-1}\left(\dfrac{2 \Delta_\bk}{\sqrt{4 \Delta_\bk^2+v^2_F|\bq|^2 \cos^2(\theta_\bk-\theta_\bq)-(\Omega+i\delta)^2}}\right)\dfrac{d\theta_\bk}{2\pi},
\end{aligned}
\en
 where $\theta_\bq$ is the direction of the momentum $\bq$. 
   Inserting this expression into \eqref{CHISHDIA} and evaluating the remaining integrals numerically, we
      obtain a complex $\chi_{\textrm{L}}(q)$, which we 
        show in Fig. \ref{Fig-chiSH} 
        for $\bq$ in the antinodal direction.
        For comparison, in the right panel of this figure we show a complex $\chi_{\textrm{L}}(q)$ for an 
        $s-$wave superconductor.  It is customary to associate the position of the peak in $\mathrm{Im}[\chi_{\textrm{L}}(\bq. \Omega)]$ with the frequency of the longitudinal collective mode and the width of the peak at half-maximum with the damping rate of the mode.   We see from  Fig. \ref{Fig-chiSH} that in a $d-$wave superconductor the mode is at $\Omega =2\Delta_{\mathrm{max}}$ at $q=0$ (we recall that in our notations, $\Delta_{\mathrm{max}} = \sqrt{2} \Delta$).
         As $q$ increases, the peak becomes anisotropic, with more weight at larger $\Omega$. Yet, the position of the peak and its width do not change as long as $v_F q <  2\Delta_{max}$.
        At larger $q$, the situation changes, see Fig. \ref{Fig-NodalvsAntiNodal}. Now $\mathrm{Im}[\chi_{\textrm{L}}(\bq, \Omega)]$ 
         for $\bq$ in the antinodal direction (blue line in Fig. \ref{Fig-NodalvsAntiNodal}) has
       a cusp at a smaller frequency and a peak at a larger  $\Omega_{max}$.  The cusp frequency 
        remains at $2 \Delta_{\mathrm{max}}$,  
         and the peak position at $\Omega_{max}$ disperses with $v_F q$ towards larger values.  In this range of $v_F q$, there is also a clear variation of Im $\chi_{\textrm{L}}(\bq, \Omega)$ with the direction of $\bq$. We see from Fig. \ref{Fig-NodalvsAntiNodal} that the cusp/peak structure holds for $\bq$ in the antinodal direction, while for $\bq$ in the nodal direction, there is a single peak at a frequency, which also increases with $v_F q$, but remains smaller than the peak frequency for the antinodal direction of $\bq$.  We illustrate this in Fig. \ref{Fig-wmax}, where we plot momentum variation of $\Omega_{max}$ along the two directions of $\bq$.
       
  Some of the results are shown in Figs. \ref{Fig-chiSH} - \ref{Fig-wmax} can be understood analytically by analyzing Eq.       
    \eqref{delta piL for d}. We consider the cases $\bq =0$ and finite $\bq$ separately.  
   
   %%%%%%%%%%%%% Fig. -  \chiSH(q,w) Summary %%%%%%%%%%%%%%%%%
%%%%%%%%%%%%%%%%%%%%%%%%%%%%%%%%%%%%%%%%%%%%%%%

 \paragraph{$\bq=0$.}
 \label{q is zero case}
At zero momentum, 
\begin{align}
  \delta\Pi_L(0,\Omega)=-\nu_F\, \int\limits_0^{2\pi} \gamma^2(\theta_\bk)\dfrac{d\theta_\bk}{2\pi} & \dfrac{\sqrt{4 \Delta_\bk^2-(\Omega+i\delta)^2}}{\Omega}\sec^{-1}\left(\dfrac{2 \Delta_\bk}{\sqrt{4 \Delta_\bk^2-(\Omega+i\delta)^2}}\right), 
    \label{delta-piL-for-d}
\end{align} 
A simple experiment shows that at small frequencies, relevant angles $\theta_k$ are around nodal points 
$\theta=(2n+1)\pi/4, n=0-3$.  Expanding near a given nodal point (say, $\theta = \pi/4$) as $\theta_\bk =\pi/4 + \delta\theta_\bk$, approximating $|\cos 2\theta_\bk| \approx 2 \delta\theta_\bk$ and  $\sqrt{4 \Delta_\bk^2-\Omega^2}\approx -i \sqrt{2\Omega} \sqrt{\Omega-4 \Delta\delta\theta_\bk}$ and evaluating the integral for  $|\delta\theta_\bk|< \Omega/4\Delta$, for which Im $\delta\Pi_L(\Omega)$ is non-zero, we obtain
\begin{align}
    \text{Im}\delta\Pi_L(0,\Omega)\propto \dfrac{1}{\sqrt{\Omega}} \int\limits_0^{\Omega/4\Delta} (\delta\theta_\bk)^2d(\delta\theta_\bk)\sqrt{\Omega-4\sqrt{2}\Delta\delta\theta_\bk}\propto \Omega^3. 
    \label{deltaPiLIm at low omega}
\end{align}
The real part of $\delta\Pi_L(0,\Omega)$ comes from all angles on the Fermi surface and is obtained by just expanding the integrand of Eq.~\eqref{delta piL for d} in small $\Omega$ up to order $\Omega^2$.  Evaluating the integrals, we obtain
 \begin{align}
    \text{Re}\delta\Pi_L(0,\Omega)\propto \int\limits_0^{2\pi} \dfrac{d\theta_\bk}{2\pi}  \left(\cos^22\theta_\bk-\dfrac{\Omega^2}{24\Delta^2}\right)= \dfrac{1}{2}-\dfrac{\Omega^2}{24 \Delta^2}.
    \label{deltaPiLRe at low omega}
\end{align}
Combining Eq.~\eqref{deltaPiLIm at low omega} and \eqref{deltaPiLRe at low omega}, we find that $\mathrm{Im}\left[\chi_{\mathrm{L}}(0,\Omega)\right] \propto \Omega^3$ at low frequencies. 
 A similar result has been obtained for Raman susceptibility~\cite{Devereaux2007}.
 We recall that in an $s-$wave superconductor the imaginary part of the 
 longitudinal susceptibility is zero for $\Omega\leq 2\Delta$. 

Consider next $\Omega \sim 2\Delta_{\mathrm{max}} = 2\sqrt{2} \Delta$.  We introduce 
 $\Omega=2\Delta_{\mathrm{max}} +\nu$.  A simple experiment shows that in this case the largest contribution to $\delta \Pi_L$ comes from antinodal regions $\theta_{\mathrm h}= n \pi/2, n=0-3$. Expanding the  gap function near an antinodal point at $n=0$ as 
 $\Delta_\bk=\Delta_{\mathrm{max}}(1-2 \theta^2_\bk)$
   and expanding $\sqrt{(2\Delta_{\mathrm{max}} )^2\cos^22\theta_\bk-\Omega^2}\approx \sqrt{2}\Delta_{\mathrm{max}} \sqrt{-2\nu/\Delta_{\mathrm{max}} -4\theta^2_\bk}$,
    we obtain for positive $\nu$
 \begin{align}
\delta\Pi_L(0,\nu>0) \propto -i\dfrac{\pi}{2} \left[{\cal W}^2+\dfrac{1}{8}\left(1+4 \log\, 2+\log \dfrac{{\cal W}^2}{\overline{\nu}}\right)\overline{\nu}\right]+\overline{\nu}{\cal W}+\dfrac{4 {\cal W}^3}{3},
\label{delta piL for d at above 2}
\end{align}
where $\overline{\nu}=2{\nu}/{\Delta_{\mathrm{max}}}$ and ${\cal W}$ is an ultraviolet cutoff. For $\nu<0$, we have, similarly
\begin{align}
\delta\Pi_L(0,\nu<0) &\propto\dfrac{\pi^2\, |\overline{\nu}|}{16}-i\dfrac{\pi}{2} \left[{\cal W}^2-\dfrac{1}{8}\left(1+4 \log\, 2+\log \dfrac{{\cal W}^2}{|\overline{\nu}|}\right){|\overline{\nu}|}\right]-{|\overline{\nu}|}{\cal W}+\dfrac{4 {\cal W}^3}{3}.
\label{delta piL for d at below 2}
\end{align}
Combining these expressions and substituting into (\ref{CHISHDIA}), we 
we find that $\mathrm{Im}\chi_{\mathrm{L}}(0,\Omega)$ is a non-analytic function of $\nu$: 
\begin{align}
    \textrm{Im}\chi_{\mathrm{L}}(0,\nu)= A_1+A_2 \,\nu \,  \log |\nu|+\Theta(-\nu) A_3 |\nu|, 
\end{align}
 where $A_j>0$ are positive and $\Theta(x)$ is the Heaviside step-function \cite{Irene1964}. 
   This gives rise to a cusp in Im $\chi_L (0, \Omega)$ at $\Omega =2 \Delta_{\mathrm{max}}$.
  We checked numerically that  $\partial\mathrm{Im}\left[\chi_{\mathrm{L}}(0,\Omega)\right]/\partial \Omega$ 
   diverges logarithmically
     at $\nu =0$, i.e., at $\Omega = 2 \Delta_{\mathrm{max}}$. 
{This agrees with the results obtained using quasiclassical theory}. 

 \paragraph{$\bq\not=0$.} 
  We now analyze Eq.~\eqref{delta piL for d} at a finite $\bq$. 
 Since  $\delta\Pi_L$ is $C_4$ symmetric  with respect to  directions of $\bq$, we keep $0\leq\theta_\bq\leq \pi/4$. 
 First, we verified that the integrand of Eq.~\eqref{delta piL for d} does not have an imaginary part if $\Omega < \textrm{min}\left\{A(\theta_\bk,\theta_\bq)\right\}$ where $ A(\theta_\bk,\theta_\bq)= \sqrt{4\Delta_\bk^2+v^2_F |\bq|^2 \cos^2(\theta_\bk-\theta_
     \bq)}$. As a result, $\mathrm{Im}\chi_{\mathrm{L}}(\bq,\Omega)$ is zero below this threshold.
%%%%%%%%%%%%% Fig. -  Peaks: Nodal & Antinodal %%%%%%%%%%%%%%%%%
%%%%%%%%%%%%%%%%%%%%%%%%%%%%%%%%%%%%%%%%%%%%%%%
It is instructive to compare the cases when $\bq$ points along the antinodal and nodal directions ($\theta_\bq=0$ and $\theta_\bq=\pi/4$, respectively).
  A simple analysis shows that $\theta_\bq=0$, the frequency, below which 
Im $\chi_{\mathrm{L}}(\bq,\Omega)$ vanishes, becomes finite at $\bq \neq 0$ and reaches 
 $\Omega =2\Omega_{max}$ at $v_F |\bq| =  4\Delta_{max}$.
  However, when $\bq$ is along the
    nodal  direction, $\theta_\bq=\pi/4$, 
  Im $\chi_{\mathrm{L}}(\bq,\Omega)$  is non-zero for all $\Omega$, 
 irrespective of value of $|\bq|$. For an intermediate direction of $\bq$, the threshold is at 
  some finite $\Omega$ below the value of the threshold in the antinodal direction.
  
 Next, we find that the cusp and the peak in $\mathrm{Im}\left[\chi_{\mathrm{L}}(\bq,\Omega)\right]$  for $\bq$ along an antinodal direction come from non-equal contributions from antinodal regions near 
  $\theta=0$ (or $\pi$) and near  $\theta_{\bk}=\pi/2$ (or $3\pi/2$).
   Integration near of these two sets of points gives rise to 
    non-analytic $\epsilon \log\epsilon$ behavior, but for the cusp $\epsilon = \Omega -2 \Omega_{max}$ 
     while for the peak,
      $\epsilon = \Omega -(2 \Omega_{\mathrm{max}} + v^2_F |\bq|^2/(2 \Delta_{\mathrm{max}}))$.  
     For a generic  direction of $\bq$, i.e., generic $\theta_\bq$ within $0 <\theta_\bq < \pi/4$,  the cusp and the peak are located at  $\Omega_1  =2 \Omega_{\mathrm{max}} + v^2_F  \sin^2{\theta_\bq} |\bq|^2/(2 \Delta_{\mathrm{max}})$ and 
     $\Omega_2  =2 \Omega_{\mathrm{max}} + v^2_F  \cos^2{\theta_\bq} |\bq|^2/(2 \Delta_{\mathrm{max}})$, respectively.  For $\theta_\bq = \pi/4$, $\Omega_1 = \Omega_2$, and  $\mathrm{Im}\left[\chi_{\mathrm{L}}(\bq,\Omega)\right]$ contains a single peak at $\Omega  =2 \Omega_{\mathrm{max}} + v^2_F |\bq|^2/(2 \sqrt{2} \Delta_{\mathrm{max}})$. This agrees with the plots in Fig. \ref{Fig-NodalvsAntiNodal}.

%%%%%%%%%%%%%%%%%%%%%%%%%%%%%%%%%%%%%%%%%%%%%%%

\subsection{Spatially resolved dynamics of the 
%Schmid-Higgs
 longitudinal  mode}

It is also of interest to investigate the dynamics of the 
 longitudinal mode at finite momentum. 
  The longitudinal 
   susceptibility of an 
  $s$-wave  superconductor 
    decays at long times  as 
     $1/\sqrt{t}$ at $\bq=0$ and as 
     $1/t^2$ at  finite $\bq$ 
     \cite{Pasha2025,Kamenev2025}.
    For a $d-$wave superconductor, earlier  study \cite{Awelewa2025} found that at $\bq =0$, 
 the
longitudinal susceptibility decays as $1/t^2$. {We have computed the decay rate at non-zero $\bq$ and found that it remains independent on the direction of momentum}.

 The time-dependent longitudinal susceptibility $\chi_{\mathrm{L}}(t,\bq)$ at $t >0$ is the Fourier transform of $\chi_{\mathrm{L}}(\omega+i0,\bq)$:
\beg\label{chiSHt}
\chi_{\mathrm{L}}(t,\bq)=\int\limits_{-\infty}^\infty\frac{d\omega}{2\pi}\chi_{\mathrm{L}}(\omega+i0,\bq)e^{-i\omega t},
\en
 Because $\chi_{\textrm{L}}(\omega+i0,\bq)$ is analytic in the upper half-plane of frequency,  we re-express 
 $\chi_{\mathrm{L}}(t,\bq)$ as 
\beg\label{chiSHtRe}
\chi_{\textrm{L}}(t,\bq)=\frac{1}{\pi}\int\limits_{0}^\infty\mathrm{Im}[\chi_{\textrm{L}}(\omega+i0,\bq)]\sin(\omega t){d\omega}.
\en

%%%%%%%%%%%%% Fig. -  \chiSH(t,q)  %%%%%%%%%%%%%%%%%
\begin{figure}
\includegraphics[width=0.4750\linewidth]{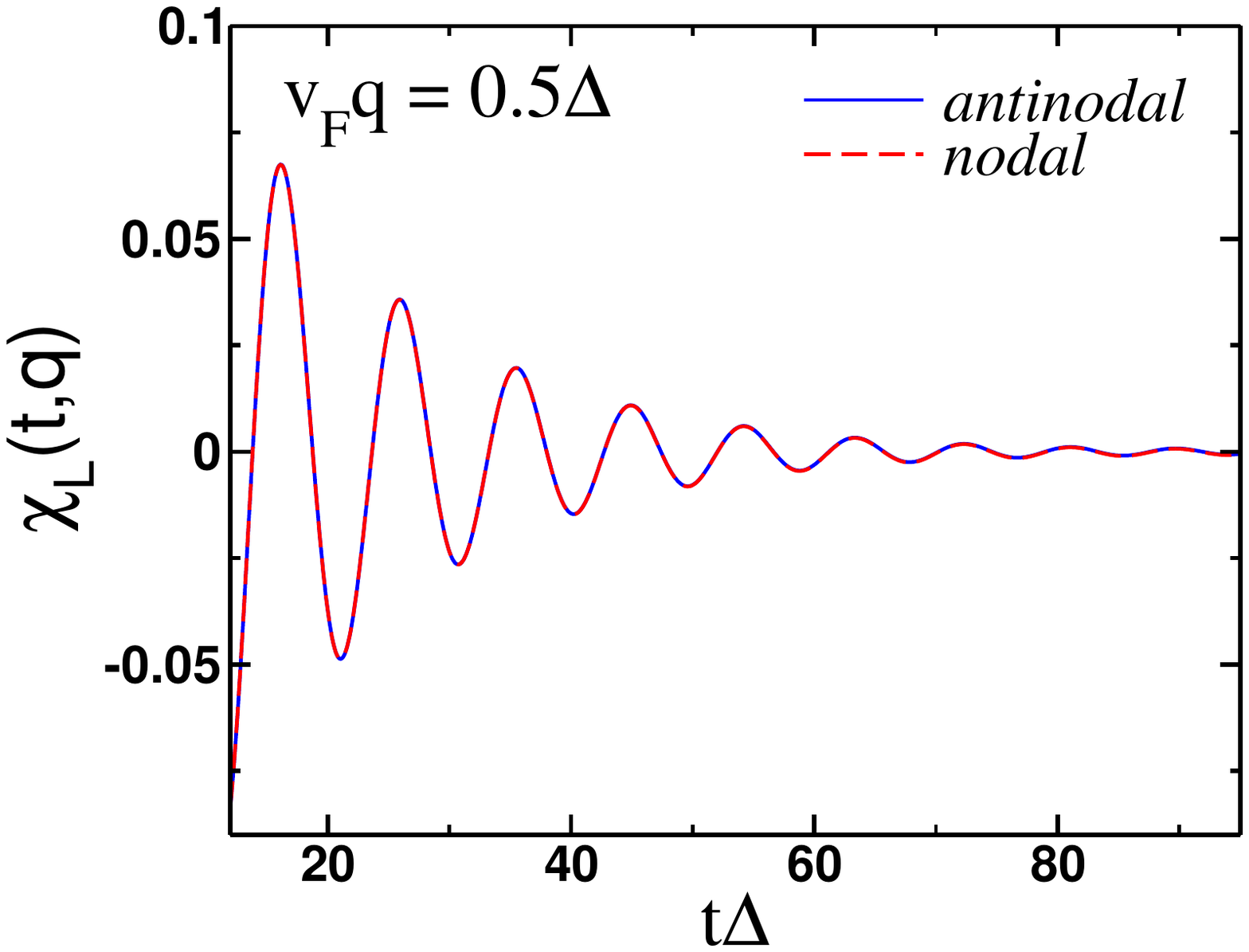} \quad 
\includegraphics[width=0.4750\linewidth]{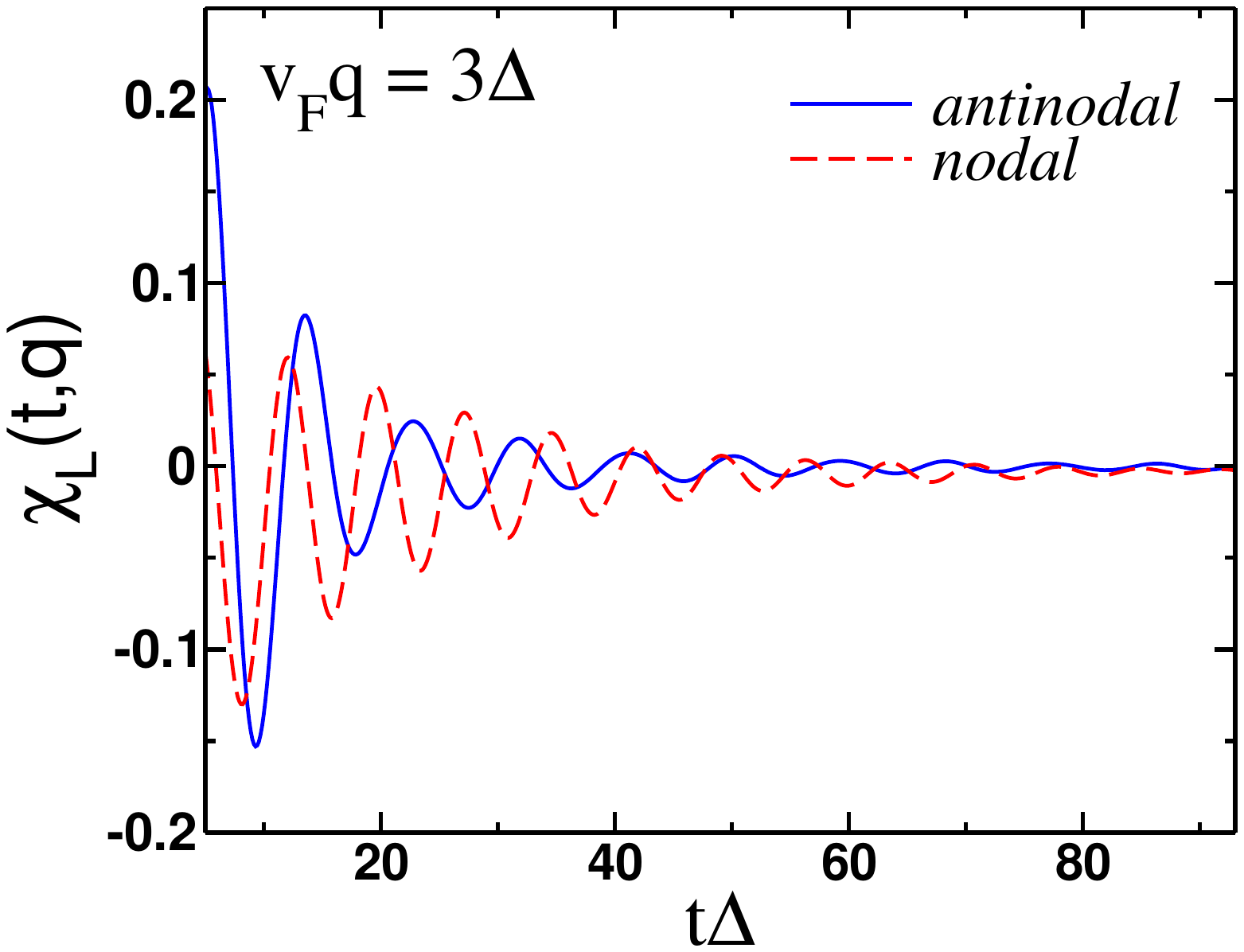} 
\caption{Time dependence of the
 longitudinal 
 susceptibility plotted for different values of momentum and its direction. Note that the frequency of the oscillations is different for $q=3\Delta/v_F$ when the momentum lies in the nodal and antinodal direction. This result is in qualitative agreement with those shown in Fig. \ref{Fig-NodalvsAntiNodal}.} 
\label{Fig-Dynamics}
\end{figure}
%%%%%%%%%%%%%%%%%%%%%%%%%%%%%%%%%%%%%%%%%%%%%%%

In Fig. \ref{Fig-Dynamics}, we show the results of numerical computation of  $\chi_{\textrm{L}}(t,\bq)$ for two different values of momentum. The 
frequency of oscillations of $\chi_{\textrm{L}}(t,\bq)$ depends on  
the direction of momentum, yet the 
 decay rate of the amplitude of the 
  oscillations remains $1/t^2$ - the same  as at $q=0$.

\section{Discussion}\label{Discussion}
 The longitudinal mode has been recently probed experimentally 
  in Raman and optical studies of 2D  $d$-wave 
   cuprate 
  superconductors Bi$_2$Sr$_2$CaCu$_2$O$_{8+x}$ ($T_c=65$ K) \cite{KotaHighTc1,KotaHighTc2}, 
 La$_{1.84}$Sr$_{0.16}$CuO$_4$ ($T_c=45$ K), DyBa$_2$Cu$_3$O$_{7-x}$ ($T_c=90$ K) and YBa$_2$Cu$_3$O$_{7-x}$ ($T_c=88$K) \cite{Cuprates2020}. In  optical experiments, 
  an oscillatory time dependence of optical reflectivity has been detected below $T_c$ and 
  associated with the excitation of the Higgs mode.
     The decay of the oscillations is faster than in 
      $s-$wave superconductors \cite{Shimano2020} which
      qualitatively agrees with our findings.
   To get a more detailed comparison with  the experiments,  one would need to know the 
   relation between the oscillation frequency and the value of the pairing gap -- it  has not been yet 
     extracted from the experiments.

Unconventional $d$-wave superconductivity usually emerges in materials with strong electron-electron correlations, such as high-$T_c$ 
  cuprates and heavy-fermion Ce-based 
   materials \cite{Petrovic2001,Movshovich2001,Miyake2007,Morr2014,Sarrao2007}. In our
   treatment of the 
   longitudinal mode, we 
      neglected the effects associated with electronic correlations. These correlations may 
     change the exponent in the 
     power law decay of the amplitude mode compared to that we report here for free electrons. This opens up an  
      exciting opportunity to explore a possible connection between the value of the exponent 
        and the strength of the electron-electron correlations.

Potential disorder also leads to a fast suppression of 
$d-$wave 
 superconductivity and as such, may also significantly affect the dynamics of the 
   longitudinal mode similar to 
   the effect of paramagnetic impurities in an $s-$wave 
   superconductor.
   In the latter case, it has been found~\cite{Kamenev2025}
   that the damping rate of the longitudinal mode  at $\bq=0$ in the presence of  pair-breaking paramagnetic impurities 
   is the same as that of the mode at a certain  finite $\bq$ in the clean case, 
      determined from the relation
       $(\xi q)^6\longleftrightarrow 1/(\tau_s\veps_{\mathrm g})$, where $\tau_s$ is the 
      scattering time from paramagnetic impurities, $\varepsilon_{\mathrm g}$ is the single-particle threshold
       and $\xi=v_F/\Delta_0$ is the coherence length. 
        Given that 
        we  found that 
        the decay rate in 
        a $d$-wave superconductor does not change with momentum, it would be interesting to analyze how  
         pair-breaking effects due to disorder affect 
           the dynamics of the longitudinal mode in the $d-$wave case.
            Finally, it is worth exploring the possibility that  
             the frequency of the 
             longitudinal mode in a $d-$wave superconductor may decrease with momentum due to a subtle interplay between the pair breaking and the dispersing amplitude mode, as was found for an $s-$wave superconductor~\cite{Pasha2025}.  
            We leave these 
             items for 
             future study.

\section{Conclusions}\label{Conclusions}
In this communication, we presented the results of our study of collective excitations in 
  in a clean $d$-wave superconductor in the presence of long-range Coulomb interaction,  neglecting other interaction effects.
   We used diagrammatic and quasiclassical approaches and obtained the dispersions of both 
    transverse (phase)  and longitudinal (amplitude) excitations. 
      For the transverse mode, we showed explicitly how a Goldstone phase mode in the absence of Coulomb interaction transforms into a plasma mode once Coulomb interaction is included.  We found that at $T=0$  the mode dispersion is the same as in an $s-$wave superconductor, but at a finite $T$ the dispersion in the $d-$wave case is softer and with a higher decay rate due to partial screening of the Coulomb potential by nodal quasiparticles.
        These $d-$wave-specific features  should be detectable by measuring the 
      Carlson-Goldman mode at $T \leq T_c$. 

         For the longitudinal mode,  we found that it is a resonance within the continuum and the mode frequency at $q=0$ is $2\Delta_{\text{max}}$, where $\Delta{\text{max}}$ is the value of the pairing gap in the anti-nodal direction.
          This is similar to an $s-$wave case, but in distinction to it,  the imaginary part of the longitudinal susceptibility of a d-wave superconductor is non-zero for all $\Omega$.  
           We analyzed the dispersion of the amplitude mode in the $d-$wave case and found that it is quite different from that in an $s-$wave superconductor. 
         Specifically,
          over some range of momentum $\bq$,  the frequency
            of the amplitude mode and the shape of the longitudinal susceptibility strongly depend on the direction of $\bq$,
             e.g.,  
              the resonance in  the longitudinal susceptibility is at 
              different frequency for momenta along nodal and antinodal directions.
               Momentum-dependent  resonance frequencies give rise to momentum-dependent periodicity of  oscillations of
               time dependent  $\chi_{\textrm{L}}(\omega,\bq)$. 
                At the same time, the decay rate of the amplitude of the oscillations 
                is $1/t^2$ independent of the 
                magnitude of momentum and its 
                 direction.

\section{Acknowledgments}
M.D. is indebted to Peter Gordon for numerous stimulating discussions. The work of SA and MD was financially supported by the National Science Foundation Grant No. DMR-2400484. One of us (MD) has performed the main part of this work at Aspen Center for Physics, which is supported by the National Science Foundation Grant No. PHY-2210452. This work of K.R.I and A.V.C was supported by U.S. Department
of Energy, Office of Science, Basic Energy Sciences,
under Award No. DE-SC0014402. K.R.I. acknowledges support
from the Doctoral Dissertation Fellowship by the University of
Minnesota, which provided funding during the course of this
research.

\section{Data availability}
The data that support the findings of this paper are not
publicly available upon publication because it is not technically feasible. The data are available from the authors
upon reasonable request.

\begin{appendix}
\section{Computation of the polarization bubbles}\label{AppendixA}
\label{Bubble_calculation}
In this section, we list the expression for a set of polarization bubbles used in the main text. For the convenience of the readers, we redefine them below 
using Eqs.~\eqref{Pi_G}-\eqref{Pi_m},\eqref{pi_0_def}
\begin{align}
&\Pi_T(q)= \int_k \,\gamma^2_\bk \left[G(k+\dfrac{q}{2})\,G(-k+\dfrac{q}{2}) + F(k+\dfrac{q}{2})\,F(-k+\dfrac{q}{2})\right], \\
&\Pi_L(q)= \int_k\, \gamma^2_\bk \left[G(k+\dfrac{q}{2})\,G(-k+\dfrac{q}{2}) - F(k+\dfrac{q}{2})\,F(-k+\dfrac{q}{2})\right], \\
&\Pi_{GF}(q) = \dfrac{1}{2}\int_k\,\gamma_\bk\, \left[ G(k+\dfrac{q}{2})\, F(-k+\dfrac{q}{2})+G(-k+\dfrac{q}{2})\, F(k+\dfrac{q}{2})\right],\\
& \Pi_0(q) = \int_k\,\left[G(k+\dfrac{q}{2})\, G(k-\dfrac{q}{2})-F(k+\dfrac{q}{2})\, F(k-\dfrac{q}{2})\right],\\
& \Pi_{GG}(q) = \dfrac{1}{2} \left[\Pi_T(q)+\Pi_L(q)\right],\\
& \Pi_{FF}(q) = \dfrac{1}{2} \left[\Pi_T(q)-\Pi_L(q)\right],\\
\end{align}
where $\Pi_T$, $\Pi_L$, and $\Pi_0$ are called transverse, longitudinal, and density polarization bubble, respectively, in the main text (we have not given any specific name to $\Pi_{GF},\Pi_{GG}$ and $\Pi_{FF}$, although). We use the combined notation for the frequency-momentum variable: $k=(\bk,\omega_m)$ and $q=(\bq,\Omega_m)$, where $\bk,\bq$ are the 2-dimensional momentum vector, $\omega_m= (2\, m+1)\pi \,T$ is the Fermionic Matsubara frequency and $\Omega_m= 2\, m\,\pi \,T$ is the Bosonic Matsubara frequency at temperature $T$. $\gamma_\bk=\sqrt{2}\, \cos2\theta_\bk$ is the d-wave pairing form factor. The integration sign  stands for $\int_k= T \sum_{\omega_m} \int \dfrac{d^2\bk}{(2\pi)^2}$.  The normal($G$) and anomalous ($F$) Green's functions are defined in Eq.~\eqref{Green's function s-d}, and have the following expression, 
 \beg
G(\bk,\omega_m)=\dfrac{i\, \omega_m+\xi_\bk}{(i\, \omega_m)^2-E^2_{\bk}},\quad  F(\bk,\omega_m)=\dfrac{\Delta_\bk}{(i\, \omega_m)^2-E^2_{\bk}}.
\en
We compute the frequency sum over $\omega_m$ in a straight-forward manner and perform the analytic continuation $i\Omega_m\rightarrow \Omega+i\delta$ to find the following expressions for the polarization bubbles (\cite{Kulik1981})
\begin{align}
   \label{tranverse_bubble_Appendix}
   \Pi_T(\bq,\Omega)&=\dfrac{1}{4} \int\dfrac{d^2\bk}{(2\pi)^2}\, \gamma^2_\bk \left(\left\{\left[1-n(E_+)-n(E_-)\right] \left[\dfrac{1}{E_++E_--\Omega-i\delta}+\dfrac{1}{E_++E_-+\Omega+i\delta}\right]\right. \right. \nn & \left. \left. \times \left[1+\dfrac{\xi_+\, \xi_-+\Delta_\bk^2(T)}{E_+\, E_-}\right]\right\}-\left\{\left[n(E_+)-n(E_-)\right]\left[1-\dfrac{\xi_+\, \xi_-+\Delta_\bk^2(T)}{E_+\, E_-}\right] \right. \right. \nn & \left.\left.  \times \left[\dfrac{1}{E_+-E_--\Omega-i\delta}+\dfrac{1}{E_+-E_-+\Omega+i\delta}\right]  \right\}\right)
   \end{align}
   \begin{align}
   \label{longitudinal_bubble_Appendix}
  \Pi_L(\bq,\Omega)&=\dfrac{1}{4} \int\dfrac{d^2\bk}{(2\pi)^2}\, \gamma^2_\bk \left(\left\{\left[1-n(E_+)-n(E_-)\right]\left[\dfrac{1}{E_++E_--\Omega-i\delta}+\dfrac{1}{E_++E_-+\Omega+i\delta}\right]\right. \right. \nn & \left. \left. \times \left[1+\dfrac{\xi_+\, \xi_--\Delta_\bk^2(T)}{E_+\, E_-}\right]\right\}-\left\{\left[n(E_+)-n(E_-)\right]\left[1-\dfrac{\xi_+\, \xi_--\Delta_\bk^2(T)}{E_+\, E_-}\right] \right. \right. \nn & \left.\left.  \times \left[\dfrac{1}{E_+-E_--\Omega-i\delta}+\dfrac{1}{E_+-E_-+\Omega+i\delta}\right]  \right\}\right)\\   
\end{align}
\begin{align}
 \label{PiGF_bubble_Appendix}
 \Pi_{GF}(\bq,\Omega)&=-\dfrac{1}{8} \int\dfrac{d^2\bk}{(2\pi)^2}\, \gamma_\bk \left(\left\{\left[1-n(E_+)-n(E_-)\right]\,\Delta_\bk(T) \,\left[\dfrac{1}{E_+}+\dfrac{1}{E_-}\right]\right. \right.\nn & \left.\left. \times \left[\dfrac{1}{E_++E_--\Omega-i\delta}-\dfrac{1}{E_++E_-+\Omega+i\delta}\right]\right\}-\left\{\left[n(E_+)-n(E_-)\right] \,\Delta_\bk(T) \right.\right. \nn & \left. \left.  \times\left[\dfrac{1}{E_+}-\dfrac{1}{E_-}\right]\times \left[\dfrac{1}{E_+-E_--\Omega-i\delta}-\dfrac{1}{E_+-E_-+\Omega+i\delta}\right]  \right\}\right)\\
\end{align}

\begin{align}
 \label{Pi0_bubble_Appendix} 
   \Pi_0(\bq,\Omega)&=-\dfrac{1}{4} \int\dfrac{d^2\bk}{(2\pi)^2}\, \gamma^2_\bk \left(\left\{\left[1-n(E_+)-n(E_-)\right]\, \left[\dfrac{1}{E_++E_--\Omega-i\delta}+\dfrac{1}{E_++E_-+\Omega+i\delta}\right]\right. \right. \nn & \left. \left. \times \left[1-\dfrac{\xi_+\, \xi_--\Delta_\bk^2(T)}{E_+\, E_-}\right]\right\}-\left\{\left[n(E_+)-n(E_-)\right] \,\left[1+\dfrac{\xi_+\, \xi_--\Delta_\bk^2(T)}{E_+\, E_-}\right] \right. \right. \nn & \left.\left.  \times \left[\dfrac{1}{E_+-E_--\Omega-i\delta}+\dfrac{1}{E_+-E_-+\Omega+i\delta}\right]  \right\}\right),\\ 
\end{align}
where $\xi_{\pm}=\xi_{\bk\pm \bq/2}$, $E_\pm=E_{\bk\pm \bq/2}$ and $n(E_\pm)= \dfrac{1}{e^{\beta\, E_\pm}+1}$ is the Fermi function. At $T=0$, $n(E_\pm)=0$ since the quasi-particle excitation energy $E_\pm>0$.

%AC_l  I cut the commented part here. Pls double check

\section{Transverse mode: derivation of the main equations}\label{AppendixB} 
Solution of the Eilenberger equation \eqref{Eq4g2RAT} for the retarded and advanced components of $\check{g}_2$ reads
\beg\label{g2RAT}
\begin{aligned}
\hat{g}_2^{R(A)}(\bn\eps;\bk\omega)=&\frac{\left(\eta_{\bn\eps+\omega}^{R(A)}+\eta_{\bn\eps-\omega}^{R(A)}\right)\left(\hat{g}_{\bn\eps+\omega}^{R(A)}\delta\hat{\Delta}_\bn^T\hat{g}_{\bn\eps-\omega}^{R(A)}-\delta\hat{\Delta}_\bn^T\right)}{\left(\eta_{\bn\eps+\omega}^{R(A)}+\eta_{\bn\eps-\omega}^{R(A)}\right)^2-4{v}_F^2(\bn\bk)^2}\\&+\frac{\left(\eta_{\bn\eps+\omega}^{R(A)}+\eta_{\bn\eps-\omega}^{R(A)}\right)\left(\hat{g}_{\bn\eps+\omega}^{R(A)}\hat{g}_{\bn\eps-\omega}^{R(A)}-\hat{\tau}_0\right)}{\left(\eta_{\bn\eps+\omega}^{R(A)}+\eta_{\bn\eps-\omega}^{R(A)}\right)^2-4{v}_F^2(\bn\bk)^2}\delta\Phi_{\bk\omega}.
\end{aligned}
\en
At the same time, the correction to the Keldysh component does not have the same form as \eqref{g2RAT} due to the different form of the normalization condition that $\hat{g}_2^K$ must satisfy. 
Thus it is customary to look for $\hat{g}_2^K$ in the following form \cite{Moore2017,Eremin2024}:
\beg\label{g2KansatzT}
{\hat{g}_2^K(\bn\eps;\bk\omega)=\hat{g}_2^R(\bn\eps;\bk\omega)t_{\eps-\omega}-t_{\eps+\omega}\hat{g}_2^A(\bn\eps;\bk\omega)+\delta \hat{g}_2^K(\bn\eps;\bk\omega)},
\en
where we introduced the shorthand notation $t_\eps=\tanh(\eps/2T)$ for brevity.
After a somewhat involved but otherwise straightforward calculation we find 
\beg\label{dg2RAKFinalTPoisson}
\begin{aligned}
\delta\hat{g}_2^K(\bn\eps;\bk\omega)=&\frac{\left(\eta_{\bn\eps+\omega}^{R}+\eta_{\bn\eps-\omega}^{A}\right)\left(\delta\hat{\Delta}_\bn^T-\hat{g}_{\bn\eps+\omega}^R\delta\hat{\Delta}_\bn^T\hat{g}_{\bn\eps-\omega}^A\right)(t_{\eps-\omega}-t_{\eps+\omega})}{\left(\eta_{\bn\eps+\omega}^{R}+\eta_{\bn\eps-\omega}^{A}\right)^2-4{v}_F^2(\bn\bk)^2}\\&+
\frac{\left(\eta_{\bn\eps+\omega}^{R}+\eta_{\bn\eps-\omega}^{A}\right)\left(\hat{\tau}_0-\hat{g}_{\bn\eps+\omega}^R\hat{g}_{\bn\eps-\omega}^A\right)(t_{\eps-\omega}-t_{\eps+\omega})}{\left(\eta_{\bn\eps+\omega}^{R}+\eta_{\bn\eps-\omega}^{A}\right)^2-4{v}_F^2(\bn\bk)^2}\delta\Phi_{\bk\omega}.
\end{aligned}
\en
We can now insert these expressions into the self-consistency equations \eqref{SelfT} and \eqref{Poisson}. As we have pointed out in the main text, in order to derive the equations which determine the dispersion of the trasverse mode we need to insert the expression for $\hat{g}_2^K(\bn\eps;\bk\omega)$ given by (\ref{g2KansatzT}) into the self-consistency equations (\ref{SelfT},\ref{Poisson}). 

The computation of traces yields
\beg\label{FirstTrace}
\begin{aligned}
&\int\limits_0^{2\pi}\frac{d\theta_\bn}{2\pi}{\gamma}(\theta_\bn)\int\limits_{-\infty}^\infty{d\eps}\textrm{Tr}\left\{i\hat{\tau}_1\hat{g}_2^K(\bn\eps;\bq\omega)\right\}=2\left\{\frac{1}{\lambda}+\chi_{\textrm{AB}}^{-1}(\bq,\omega)\right\}\delta\Delta_{\bq\omega}^T+2i\rho_{\bq\omega}\cdot\delta\Phi_{\bq\omega}, \\
&\int\limits_0^{2\pi}\frac{d\theta_\bn}{2\pi}\int\limits_{-\infty}^\infty{d\eps}\textrm{Tr}\left\{\hat{\tau}_0\hat{g}_2^K(\bn\eps;\bq\omega)\right\}=-2i\rho_{\bq\omega}\cdot\delta\Delta_{\bq\omega}^T+2\chi_{\textrm{CG}}^{-1}(\bq,\omega)\delta\Phi_{\bq\omega}.
\end{aligned}
\en
We thus recover the system of equations \eqref{TSystem} in the main text. 

Function $\chi_{\textrm{AB}}^{-1}(\bq,\omega)$ has been defined in the main text. The remaining functions are defined according to
\beg\label{zetaetc}
\begin{aligned}
&\rho_{\bq\omega}=\int\limits_{0}^{2\pi}\frac{d\theta_\bn}{2\pi}{\gamma}(\theta_\bn)\int\limits_{-\omega_D}^{\omega_D}d\eps\left\{
\frac{\left(\eta_{\bn\eps+\Omega/2}^{R}+\eta_{\bn\eps-\Omega/2}^{A}\right){B}_\bn^K(\eps_+,\eps_-)(t_{\eps+\Omega/2}-t_{\eps-\Omega/2})}{\left(\eta_{\bn\eps+\Omega/2}^{R}+\eta_{\bn\eps-\Omega/2}^{A}\right)^2-{v}_F^2(\bn.\bq)^2}\right.\\&\left.+\frac{\left(\eta_{\bn\eps+\Omega/2}^{R}+\eta_{\bn\eps-\Omega/2}^{R}\right){B}_\bn^R(\eps_+,\eps_-)t_{\eps-\Omega/2}}{\left(\eta_{\bn\eps+\Omega/2}^{R}+\eta_{\bn\eps-\Omega/2}^{R}\right)^2-{v}_F^2(\bn.\bq)^2}-
\frac{\left(\eta_{\bn\eps+\Omega/2}^{A}+\eta_{\bn\eps-\Omega/2}^{A}\right){B}_\bn^A(\eps_+,\eps_-)t_{\eps+\Omega/2}}{\left(\eta_{\bn\eps+\Omega/2}^{A}+\eta_{\bn\eps-\Omega/2}^{A}\right)^2-{v}_F^2(\bn.\bq)^2}\right\}, \\ 
\end{aligned}
\en
and
\beg\label{chiCC}
\begin{aligned}
&\chi_{\textrm{CG}}^{-1}(\bq,\omega)=\int\limits_{0}^{2\pi}\frac{d\theta_\bn}{2\pi}\int\limits_{-\infty}^{\infty}d\eps\left\{
\frac{\left(\eta_{\bn\eps+\Omega/2}^{R}+\eta_{\bn\eps-\Omega/2}^{A}\right)\widetilde{A}_\bn^K(\eps_+,\eps_-)(t_{\eps+\Omega/2}-t_{\eps-\Omega/2})}{\left(\eta_{\bn\eps+\Omega/2}^{R}+\eta_{\bn\eps-\Omega/2}^{A}\right)^2-{v}_F^2(\bn.\bq)^2}\right.\\&\left.+\frac{\left(\eta_{\bn\eps+\Omega/2}^{R}+\eta_{\bn\eps-\Omega/2}^{R}\right)\widetilde{A}_\bn^R(\eps_+,\eps_-)t_{\eps-\Omega/2}}{\left(\eta_{\bn\eps+\Omega/2}^{R}+\eta_{\bn\eps-\Omega/2}^{R}\right)^2-{v}_F^2(\bn.\bq)^2}-
\frac{\left(\eta_{\bn\eps+\Omega/2}^{A}+\eta_{\bn\eps-\Omega/2}^{A}\right)\widetilde{A}_\bn^A(\eps_+,\eps_-)t_{\eps+\Omega/2}}{\left(\eta_{\bn\eps+\Omega/2}^{A}+\eta_{\bn\eps-\Omega/2}^{A}\right)^2-{v}_F^2(\bn.\bq)^2}\right\}.
\end{aligned}
\en
Here
\beg\label{epspm}
\eps_\pm=\eps\pm\frac{\Omega}{2}
\en
and we introduced functions
\beg\label{tilAKRA}
\begin{aligned}
&\widetilde{A}_\bn^{R(A)}(\eps,\eps')=g_{\bn\eps}^{R(A)}g_{\bn\eps'}^{R(A)}-f_{\bn\eps}^{R(A)}f_{\bn\eps'}^{R(A)}-1, \quad
\widetilde{A}_\bn^{K}(\eps,\eps')=g_{\bn\eps}^{R}g_{\bn\eps'}^{A}-f_{\bn\eps}^{R}f_{\bn\eps'}^{A}-1, \\
&{B}_\bn^K(\eps,\eps')=g_\eps^Rf_{\eps'}^A-f_\eps^Rg_{\eps'}^A, \quad {B}_\bn^{R(A)}(\eps,\eps')=g_\eps^{R(A)}f_{\eps'}^{R(A)}-f_\eps^{R(A)}g_{\eps'}^{R(A)}.
\end{aligned}
\en
Note that function $\rho_{\bq\omega}$ vanishes identically in the limit $\omega\to 0$ and also must exhibit very weak momentum dependence because the kernel of the integral over $\phi_\bn$ is proportional to \textcolor{blue}{$\gamma_{\bn}$} while the expression under the integral involves even powers of $\bn.\bq$. This is confirmed by the numerical analysis of $\rho_{\bq\omega}$.

\section{Auxiliary expressions}\label{AppendixC}
For our subsequent discussion, it will be useful to simplify the expressions under the integrals defined above. Since we are mainly interested in studying the behavior of the susceptibilities at small momenta, we first consider the corresponding expressions evaluated at $\bq=0$.

Let us start with $\chi_{\mathrm{AB}}^{-1}(\omega,\bq=0)$. We introduce the following notations:
\beg\label{KamenevNotes}
u_\eps=g_\eps^R, \quad u_\eps^*=-g_\eps^A, \quad v_\eps=f_\eps^R, \quad v_{\eps}^*=-f_\eps^A.
\en
For the first term under the integral in \eqref{chiABdwave} we have
\beg\label{LeshaKeldysh}
\frac{A^K(\eps,\eps')}{\eta_\eps^R+\eta_{\eps'}^A}=\frac{(\eta_\eps^R-\eta_{\eps'}^A)\left(g_\eps^Rg_{\eps'}^A-f_\eps^Rf_{\eps'}^A+1\right)}{(\eps-\eps')(\eps+\eps')}=\frac{u_\eps-u_{\eps'}^*}{\eps+\eps'}
\en
and on the last step we used the normalization condition $g_\eps^2-f_\eps^2=1$. Using the same trick for the remaining terms in \eqref{chiABdwave} we have
\beg\label{LeshaRA}
\frac{A^R(\eps,\eps')}{\eta_\eps^R+\eta_{\eps'}^R}=\frac{u_\eps+u_{\eps'}}{\eps+\eps'}, \quad \frac{A^A(\eps,\eps')}{\eta_\eps^A+\eta_{\eps'}^A}=-\frac{u_\eps^*+u_{\eps'}^*}{\eps+\eps'}.
\en
Combining these results, for the expression under the integral \eqref{chiABdwave} we find
\beg\label{UnderAB}
\frac{A^K(\eps,\eps')(t_\eps-t_{\eps'})}{\eta_\eps^R+\eta_{\eps'}^A}+\frac{A^R(\eps,\eps')t_{\eps'}}{\eta_\eps^R+\eta_{\eps'}^R}
-\frac{A^A(\eps,\eps')t_{\eps}}{\eta_\eps^A+\eta_{\eps'}^A}=\frac{(u_\eps+u_{\eps}^*)t_\eps+(u_{\eps'}+u_{\eps'}^*)t_{\eps'}}{\eps+\eps'}.
\en
This expression matches the corresponding expressions listed in \cite{Kamenev2011}. 

We continue with function $\rho_{\bq\omega}$. At $\bq=0$ it obtains:
\beg\label{BKKeldysh}
\frac{B^K(\eps,\eps')}{\eta_\eps^R+\eta_{\eps'}^A}=\frac{(\eta_\eps^R-\eta_{\eps'}^A)\left(g_\eps^Rf_{\eps'}^A-f_\eps^Rg_{\eps'}^A\right)}{(\eps-\eps')(\eps+\eps')}=\frac{\eps f_{\eps'}^A-\Delta_\bn g_{\eps'}^A-\Delta_\bn g_\eps^R+\eps'f_\eps^R}{(\eps-\eps')(\eps+\eps')}=\frac{f_{\eps'}^A-f_{\eps}^R}{\eps+\eps'}.
\en
Here on the last step we used $\eps f_{\bn\eps}=\Delta_\bn g_{\bn\eps}$. Finally, for the remaining two terms, it obtains:
\beg\label{BRAzeta}
\begin{aligned}
\frac{B^R(\eps,\eps')}{\eta_\eps^R+\eta_{\eps'}^R}=\frac{f_{\eps'}^R-f_{\eps}^R}{\eps+\eps'}, \quad \frac{B^A(\eps,\eps')}{\eta_\eps^A+\eta_{\eps'}^A}=\frac{f_{\eps'}^A-f_{\eps}^A}{\eps+\eps'}.
\end{aligned}
\en
Collecting these terms we have
\beg\label{UnderRho}
\frac{B^K(\eps,\eps')(t_\eps-t_{\eps'})}{\eta_\eps^R+\eta_{\eps'}^A}+\frac{B^R(\eps,\eps')t_{\eps'}}{\eta_\eps^R+\eta_{\eps'}^R}
-\frac{B^A(\eps,\eps')t_{\eps}}{\eta_\eps^A+\eta_{\eps'}^A}=\frac{(v_{\eps'}+v_{\eps'}^*)t_{\eps'}-(v_\eps+v_{\eps}^*)t_\eps}{\eps+\eps'}.
\en
In passing, we note that these expressions agree with the corresponding expressions listed in \cite{Kamenev2011}.

 Let us now repeat the same steps to simplify function $\chi_{\mathrm{CG}}^{-1}(\omega,\bq=0)$. We start with the Keldysh part
\beg\label{LeshaKeldysh2}
\begin{aligned}
&\frac{\tilde{A}^K(\eps,\eps')}{\eta_\eps^R+\eta_{\eps'}^A}=\frac{(\eta_\eps^R-\eta_{\eps'}^A)\left(g_\eps^Rg_{\eps'}^A-f_\eps^Rf_{\eps'}^A-1\right)}{(\eps-\eps')(\eps+\eps')}=\frac{g_{\eps'}^A-g_\eps^R}{\eps-\eps'}+2\frac{\eps g_\eps^R-\eps' g_{\eps'}^A+\eta_{\eps'}^A-\eta_\eps^R}{(\eps-\eps')(\eps+\eps')}\\&=\frac{g_{\eps'}^A-g_\eps^R}{\eps-\eps'}+2\frac{\eta_\eps^R (g_\eps^R)^2-\eta_{\eps'}^A (g_{\eps'}^A)^2+\eta_{\eps'}^A-\eta_\eps^R}{(\eps-\eps')(\eps+\eps')}=-\frac{g_{\bn\eps}^R-g_{\bn\eps'}^A}{\eps-\eps'}+
\frac{2\Delta_{\bn}(f_{\bn\eps}^R-f_{\bn\eps'}^A)}{(\eps-\eps')(\eps+\eps')}.
\end{aligned}
\en
Similar expressions can be found for the remaining two terms. As a result we find
\beg\label{Fin4chiCG}
\begin{aligned}
&\chi_{\textrm{CG}}^{-1}(\omega,\bq=0)=\int\limits_{0}^{2\pi}\frac{d\theta_\bn}{2\pi}\int\limits_{-\infty}^{\infty}d\eps\left(\frac{\tilde{A}^K(\eps,\eps')(t_\eps-t_{\eps'})}{\eta_\eps^R+\eta_{\eps'}^A}+\frac{\tilde{A}^R(\eps,\eps')t_{\eps'}}{\eta_\eps^R+\eta_{\eps'}^R}
-\frac{\tilde{A}^A(\eps,\eps')t_{\eps}}{\eta_\eps^A+\eta_{\eps'}^A}\right)\\&=\int\limits_{0}^{2\pi}\frac{d\theta_\bn}{2\pi}\int\limits_{-\infty}^{\infty}d\eps\left[\frac{(u_{\eps'}+u_{\eps'}^*)t_{\eps'}}{\eps-\eps'}+\frac{2\Delta_\bn(v_\eps+v_{\eps}^*)t_\eps}{(\eps-\eps')(\eps+\eps')}+(\eps\leftrightarrow\eps')\right].
\end{aligned}
\en

Now, in order to derive the system of equations \eqref{TransverseMode} we will use the expressions (\ref{UnderAB},\ref{UnderRho},\ref{Fin4chiCG})  listed above. 
%\textcolor{red}{KRI: MD, please take care of it. } 
%AC_l  Pls re-phrase.  The sentence above sounds strange.   
We rewrite the first equation in \eqref{FirstTrace} as follows:
\beg\label{FirstEq}
\begin{aligned}
&\int\limits_{0}^{2\pi}{\gamma}^2(\theta_\bn)\frac{d\theta_\bn}{2\pi}\int\limits_{-\infty}^\infty d\eps\left[\frac{(g_{\bn\eps}^R-g_{\bn\eps}^A)t_\eps}{\eps+\eps'}+\frac{(g_{\bn\eps'}^R-g_{\bn\eps'}^A)t_{\eps'}}{\eps+\eps'}-(f_{\bn\eps}^R-f_{\bn\eps}^A)\frac{t_\eps}{\Delta}\right]\left(i\Delta_{\bq\omega}^T\right)\\&-\int\limits_{0}^{2\pi}{\gamma}(\theta_\bn)\frac{d\theta_\bn}{2\pi}\int\limits_{-\infty}^\infty d\eps\left[\frac{(f_{\bn\eps}^R-f_{\bn\eps}^A)t_\eps}{\eps+\eps'}-\frac{(f_{\bn\eps'}^R-f_{\bn\eps'}^A)t_{\eps'}}{\eps+\eps'}\right]\delta\Phi_{\bq\omega}\\&=\left(\frac{v_Fq}{\Delta}\right)^2\zeta(\omega,T)\left(i\Delta_{\bq\omega}^T\right).
\end{aligned}
\en
Here we changed the integration variable $\eps+\omega/2\to \eps$, so that  $\eps-\omega/2\to\eps'=\eps-\omega$. 
The dimensionless function ${\zeta}(\omega)$ has been defined according to
\beg\label{calBw}
\begin{aligned}
{\zeta}(\omega,T)&=\int\limits_{0}^{2\pi}|\Delta_\bn|^2\cos^2(\theta_\bn-\vartheta_\bq)\frac{d\theta_\bn}{2\pi}\int\limits_{-\infty}^{\infty}d\eps\left\{
\frac{{A}_\bn^K(\eps_+,\eps_-)(t_{\eps+\omega/2}-t_{\eps-\omega/2})}{\left(\eta_{\bn\eps+\omega/2}^{R}+\eta_{\bn\eps-\omega/2}^{A}\right)^3}\right. \\& \left.+\frac{{A}_\bn^R(\eps_+,\eps_-)t_{\eps-\omega/2}}{\left(\eta_{\bn\eps+\omega/2}^{R}+\eta_{\bn\eps-\omega/2}^{R}\right)^3}
-\frac{{A}_\bn^A(\eps_+,\eps_-)t_{\eps+\omega/2}}{\left(\eta_{\bn\eps+\omega/2}^{A}+\eta_{\bn\eps-\omega/2}^{A}\right)^3}\right\}
\end{aligned}
\en
with $\eps_{\pm}=\eps\pm\omega/2$.  Here we took into account that in function $\chi_{\mathrm{AB}}^{-1}(\omega,\bq)$ we can safely ignore its dependence on $\hat{\bq}=\bq/q$. We note that the difference in the values of this function computed for $s$- and $d$-wave pairings essentially determines the difference in the values of the gap in the dispersion of the transverse mode between $s$- and $d$-wave superconductors.

The integral in the first equation \eqref{FirstEq} can be very well  approximated by
\beg\label{aatwom}
\begin{aligned}
\int\limits_{0}^{2\pi}{\gamma}^2(\theta_\bn)\frac{d\theta_\bn}{2\pi}\int\limits_{-\infty}^\infty d\eps\left[\frac{(g_{\bn\eps}^R-g_{\bn\eps}^A)t_\eps}{\eps+\eps'}+\frac{(g_{\bn\eps'}^R-g_{\bn\eps'}^A)t_{\eps'}}{\eps+\eps'}-(f_{\bn\eps}^R-f_{\bn\eps}^A)\frac{t_\eps}{\Delta}\right]\simeq\frac{\omega^2}{\Delta^2}.
\end{aligned}
\en
In addition, for the last integral in the left-hand side of \eqref{FirstEq} we approximately find
\beg\label{INtRHS}
\int\limits_{0}^{2\pi}{\gamma}(\theta_\bn)\frac{d\theta_\bn}{2\pi}\int\limits_{-\infty}^\infty d\eps\left[\frac{(f_{\bn\eps}^R-f_{\bn\eps}^A)t_\eps}{\eps+\eps'}-\frac{(f_{\bn\eps'}^R-f_{\bn\eps'}^A)t_{\eps'}}{\eps+\eps'}\right]\simeq\frac{2\omega}{\Delta}.
\en
In view of eqs. \eqref{aatwom} and \eqref{INtRHS},  equation \eqref{FirstEq} acquires the following simple form:
\beg\label{FirstEqApproximate}
\begin{aligned}
&\left[\omega^2-(v_Fq)^2\zeta(\omega,T)\right]\Delta_{\bq\omega}^T+\left(2i\omega\Delta\right)\delta\Phi_{\bq\omega}=0.
\end{aligned}
\en

We now proceed with the discussion of the 
the remaining Poisson equation. In its general form, it reads
\beg\label{Poisson2}
\begin{aligned}
&\left(-\frac{q^{d-1}}{2^{d-1}\pi \nu_Fe^2}\right)\delta\Phi_{\bq \omega}=\left(
1-\frac{1}{4}\int\limits_{0}^{2\pi}\frac{d\theta_\bn}{2\pi}\int\limits_{-\infty}^\infty d\eps\left[\frac{(g_{\bn\eps}^R-g_{\bn\eps}^A)t_\eps}{\eps-\eps'}-\frac{(g_{\bn\eps'}^R-g_{\bn\eps'}^A)t_{\eps'}}{\eps-\eps'}\right]\right)\delta\Phi_{\bq\omega}
\\&-\frac{\omega}{2\Delta}\left(i\Delta_{\bq\omega}^T\right)+
\frac{\Delta}{2}\int\limits_{0}^{2\pi}{\gamma}(\theta_\bn)\frac{d\theta_\bn}{2\pi}\int\limits_{-\infty}^\infty d\eps\left[\frac{(f_{\bn\eps}^R-f_{\bn\eps}^A)t_\eps}{(\eps-\eps')(\eps+\eps')}-\frac{(f_{\bn\eps'}^R-f_{\bn\eps'}^A)t_{\eps'}}{(\eps-\eps')(\eps+\eps')}\right]\delta\Phi_{\bq\omega}.
\end{aligned}
\en
Here we have ignored the $q$-dependence in the expressions under the integrals, for they are small compared to the term on the left-hand side, and took into account \eqref{aatwom}. One of the reasons as to why we are using these expressions instead of the ones listed above is that it has now become clear that the first term on the right-hand side is zero due to the fact that
\beg\label{CuriousInt}
\frac{1}{4}\int\limits_{0}^{2\pi}\frac{d\theta_\bn}{2\pi}\int\limits_{-\infty}^\infty d\eps\left[\frac{(g_{\bn\eps}^R-g_{\bn\eps}^A)t_\eps}{\eps-\eps'}-\frac{(g_{\bn\eps'}^R-g_{\bn\eps'}^A)t_{\eps'}}{\eps-\eps'}\right]=1
\en
Note that this is, of course, independent on the type of pairing, i.e. $s$-wave vs. $d$-wave. 

Thus, with the help of the definition \eqref{Fin4chiCG} equation \eqref{Poisson2} becomes
\beg\label{Poisson3}
\begin{aligned}
&\left\{\frac{q^{d-1}}{2^{d-3}\pi\nu_F e^2}+4-\left(\chi_{\textrm{CG}}^{-1}(0,\omega)-\chi_{\textrm{CG}}^{-1}(\bq,\omega)\right)\right\}\delta\Phi_{\bq \omega}=\frac{2\omega}{\Delta}\left(i\Delta_{\bq\omega}^T\right).
\end{aligned}
\en

To summarize, the system of equations that determines the energy and dispersion of the transverse mode reads
\beg\label{TransverseModeLong}
\begin{aligned}
&\left[\omega^2-v_F^2q^2\zeta(\omega,T)\right]\cdot \Delta_{\bq\omega}^T+\left({2i\omega\Delta}\right)\cdot\delta\Phi_{\bq\omega}=0, \\
&\left({2i\omega}\Delta\right) \delta\Delta_{\bq\omega}^T+\Delta^2\left\{\chi_{\textrm{CG}}^{-1}(0,\omega)-\chi_{\textrm{CG}}^{-1}(\bq,\omega)-4-\frac{q^{d-1}}{2^{d-3}\pi\nu_F e^2}\right\}\delta\Phi_{\bq \omega}=0.
\end{aligned}
\en
To find the dispersion, we demand that the determinant of this system goes to zero. As it turns out, one can also ignore the frequency dependence of the function $\zeta{(\omega, T)}$ (see discussion in \ref{AppendixD} below). We also take into account that function $\chi_{\textrm{CG}}^{-1}(0,\omega)-\chi_{\textrm{CG}}^{-1}(\bq,\omega)$ is weakly dependent on frequency for small values of momentum and therefore it can also be neglected in the limit $\bq\to 0$. This yields the linear system of equations \eqref{TransverseMode} in the main text.

\section{Transverse mode: function $\zeta(\omega,T)$}\label{AppendixD}
As it directly follows from the equations \eqref{TransverseMode} function $\zeta{(\omega,T)}$, Eq. \eqref{calBw}, plays a crucial role in determining the energy of the superconducting plasmon frequency. In this section we will provide a more detailed analysis of this function. Specifically, we are going to analyze the following integral
\beg\label{calJw}
\begin{aligned}
J(\omega)&=\int\limits_{-\infty}^{\infty}d\eps\left\{
\frac{{A}_\bn^K(\eps,\eps')(t_{\eps}-t_{\eps'})}{\left(\eta_{\bn\eps}^{R}+\eta_{\bn\eps'}^{A}\right)^3}+\frac{{A}_\bn^R(\eps,\eps')t_{\eps'}}{\left(\eta_{\bn\eps}^{R}+\eta_{\bn\eps'}^{R}\right)^3}
-\frac{{A}_\bn^A(\eps,\eps')t_{\eps'}}{\left(\eta_{\bn\eps}^{A}+\eta_{\bn\eps'}^{A}\right)^3}\right\}.
\end{aligned}
\en
Here $\eps'=\eps-\omega$.
As we have already mentioned above, this function displaces fairly weak frequency dependence across the wide range of frequencies. For this reason, in order to make progress with obtaining an analytical expressions we will analyze this integral in the limit $\omega\to 0$. Since the function under the integral is an even function of $\eps$, we can limit the integration to positive values of $\eps$. Given the definition of functions $\eta_{\bn\eps}^{R(A)}$ \eqref{etaRA} we can re-write \eqref{calJw} as follows
\beg\label{calJw2}
\begin{aligned}
J(\omega)&=2\int\limits_{|\Delta_\bn|}^{\infty}d\eps\left\{
\frac{{A}_\bn^K(\eps,\eps')(t_{\eps}-t_{\eps'})}{\left(\eta_{\bn\eps}^{R}+\eta_{\bn\eps'}^{A}\right)^3}+\frac{{A}_\bn^R(\eps,\eps')t_{\eps'}}{\left(\eta_{\bn\eps}^{R}+\eta_{\bn\eps'}^{R}\right)^3}
-\frac{{A}_\bn^A(\eps,\eps')t_{\eps'}}{\left(\eta_{\bn\eps}^{A}+\eta_{\bn\eps'}^{A}\right)^3}\right\},
\end{aligned}
\en
where we took into account that the integral for $\eps\in\left[0,|\Delta_\bn|\right]$ vanishes in the limit $\omega\to 0$ since $\eta_{\bn\eps}^R=\eta_{\bn\eps}^A$ for these values of $\eps$. Let us first consider the first term under the integral. 
We have $t_{\eps}-t_{\eps-\omega}\approx\omega t_\eps'$.
Furthermore using \eqref{LeshaKeldysh} we have
\beg\label{Simplify}
\begin{aligned}
&\frac{{A}_\bn^K(\eps,\eps')}{\left(\eta_{\bn\eps}^{R}+\eta_{\bn\eps'}^{A}\right)^3}=
\frac{g_\eps^R+g_{\eps'}^A}{(\eps+\eps')\left(\eta_{\bn\eps}^{R}+\eta_{\bn\eps'}^{A}\right)^2}=\frac{(g_\eps^R+g_{\eps'}^A)\left(\eta_{\bn\eps}^{R}-\eta_{\bn\eps'}^{A}\right)^2}{(\eps+\eps')^3(\eps-\eps')^2}
\end{aligned}
\en
It is now important to keep in mind that for $\eps\geq|\Delta_\bn|$ the following relation holds: $\eta_{\bn\eps}^R=-\eta_{\bn\eps}^A\equiv\eta_{\bn\eps}$.
Then in the limit $\omega\ll\Delta$ we find
\beg\label{Simplify2}
\begin{aligned}
&\frac{{A}_\bn^K(\eps_{+},\eps_{-})(t_{\eps+\omega/2}-t_{\eps-\omega/2})}{\left(\eta_{\bn\eps_{+}}^{R}+\eta_{\bn\eps_{-}}^{A}\right)^3}
\approx(g_{\eps+\omega/2}^R+g_{\eps-\omega/2}^A)\frac{\eta_{\bn\eps}^2 t_\eps'}{2\omega\eps^3}.
\end{aligned}
\en
For the sum of the retarded and advanced propagators, we have
\beg\label{Sum}
\begin{aligned}
&\lim\limits_{\omega\to 0}\frac{1}{\omega}\left(g_{\eps+\omega/2}^R+g_{\eps-\omega/2}^A\right)=\lim\limits_{\omega\to 0}\frac{1}{\omega}\left(\frac{\eps+{\omega}/{2}}{\eta_{\bn\eps+\omega/2}}-\frac{\eps-{\omega}/{2}}{\eta_{\bn\eps-\omega/2}}\right)=-\frac{|\Delta_\bn|^2}{\eta_{\bn\eps}^3},
\end{aligned}
\en
where on the last step we used the definition $\eta_{\bn\eps}=\sqrt{\eps^2-|\Delta_\bn|^2}$. Then we find
\beg\label{Simplify3}
\begin{aligned}
\lim\limits_{\omega\to 0}\frac{{A}_\bn^K(\eps_{+},\eps_{-})(t_{\eps+\omega/2}-t_{\eps-\omega/2})}{\left(\eta_{\bn\eps_{+}}^{R}+\eta_{\bn\eps_{-}}^{A}\right)^3}
=-\frac{|\Delta_\bn|^2t_\eps'}{2\eps^3\eta_{\bn\eps}}.
\end{aligned}
\en

Let us now discuss the remaining two contributions to $J(\omega)$, i.e. second and third terms under the integral in \eqref{calJw}. For $\eps\geq|\Delta_\bn|$ we have
\beg\label{ARARCancel}
\begin{aligned}
\lim\limits_{\omega\to 0}\left\{\frac{{A}_\bn^R(\eps_+,\eps_-)t_{\eps-\omega/2}}{\left(\eta_{\bn\eps+\omega/2}^{R}+\eta_{\bn\eps-\omega/2}^{R}\right)^3}
-\frac{{A}_\bn^A(\eps_+,\eps_-)t_{\eps+\omega/2}}{\left(\eta_{\bn\eps+\omega/2}^{A}+\eta_{\bn\eps-\omega/2}^{A}\right)^3}\right\}=
\frac{t_\eps}{2\eta_{\bn\eps}^3},
\end{aligned}
\en
where we used the normalization condition $\left(g_{\bn\eps}^{R(A)}\right)^2-\left(f_{\bn\eps}^{R(A)}\right)^2=1$. For the case when $\eps\leq|\Delta_\bn|$ the combination of functions in
\eqref{ARARCancel} vanishes identically, since in that case $g_{\bn\eps}^R=g_{\bn\eps}^A$.
Therefore, the integral $J(\omega)$ in the limit $\omega\ll\Delta$ reduces to the following expression:
\beg\label{Jw20}
J(\omega\ll\Delta)=\frac{1}{2}\int\limits_{|\Delta_\bn|}^\infty\frac{t_\eps d\eps}{\left(\eps^2-|\Delta_\bn|^2\right)^{3/2}}+\frac{1}{2}\int\limits_{-\infty}^{-|\Delta_\bn|}\frac{t_{|\eps|} d\eps}{\left(\eps^2-|\Delta_\bn|^2\right)^{3/2}}-\int\limits_{|\Delta_\bn|}^{\infty}\frac{|\Delta_\bn|^2t_\eps'd\eps}{\eps^3\sqrt{\eps^2-|\Delta_\bn|^2}}.
\en
In the limit of low temperatures $T\ll \Delta$, we can approximate \textcolor{blue}{$t_{|\eps|}\approx 1$}. We also note that the first two integrals are convergent at $\eps\gg\Delta$. Then for this reason it will be convenient to shift the integration into the upper half plane of complex complex variable $\eps$ and change the integration to the integral along the imaginary axis. 
Making the change of variables $\eps=y|\Delta_\bn|$, taking into account that 
\beg
t_\eps'=\frac{1}{2T}\cosh^{-2}\left(\frac{\eps}{2T}\right) 
\en
equation \eqref{Jw20} acquires the following form 
\beg\label{Reduce}
\lim\limits_{\omega\to 0}J(\omega)=\frac{1}{|\Delta_\bn|^2}\left\{1-\frac{|\Delta_\bn|}{2T}
\int\limits_{1}^{\infty}\frac{\cosh^{-2}(y|\Delta_\bn|/2T)dy}{y^3\sqrt{y^2-1}}\right\}.
\en
Inserting this expression into \eqref{calBw} yields the formula \eqref{zetaw} in the main text.

\paragraph{Low temperature asymptotic of function $\zeta(T)$.} Let us now compute the low temperature asymptote of the integral in \eqref{zetaw}. Performing the change of the variables $y=\sqrt{x^2+1}$ we define the following function
\beg\label{DefineIntegra}
\begin{aligned}
F(x,\theta)&=\frac{|\cos(2\theta)|}{(x^2+1)^2}\cosh^{-2}\left(\frac{|\cos(2\theta)|}{2t}\sqrt{x^2+1}\right)
\end{aligned}
\en
and we also introduced the dimensionless parameter $\tau=T/\Delta$. Our goal is to evaluate the following integral 
\beg\label{IntegralAgain}
K(\tau,\Delta)=\int\limits_0^{2\pi}\cos^2(\theta-\vartheta_\bq)d\theta\int\limits_0^\infty F(x,\theta)dx
\en
in the limit when the dimensionless parameter $t$ is assumed to be small.  

We start by making the following observations. By symmetry, the integral 
\eqref{IntegralAgain} reduces to 
\beg\label{ReduceAgain}
K(\tau,\Delta)=4\int\limits_0^{\pi/2}\left[\cos^2(\vartheta_\bq)\cos^2(\theta)+
\sin^2(\vartheta_\bq)\sin^2(\theta)\right]d\theta\int\limits_0^\infty F(x,\theta)dx
\en
Splitting the interval of the angular integration into two - one for $\theta\in[0,\pi/4]$ and one for $\theta\in[\pi/4,\pi/2]$ - it follows 
 that we will only need to perform the angular integration over the interval [$0,\pi/4$] of the following function: 
\beg\label{IntegralAgain2}
K(\tau,\Delta)=4\int\limits_0^{\pi/4}d\theta\int\limits_0^\infty F(x,\theta)dx
\en
We remind the reader that this integral needs to be evaluated assuming $t\ll 1$. 

The integral \eqref{IntegralAgain2} over $\theta$ can be split into two parts: one with the integration over the interval  $0\leq\theta\leq\pi/4-\tau\varphi(\tau)$ and the other over the interval $\pi/4-\tau\varphi(\tau)\leq\theta\leq\pi/4$, where $\varphi(\tau)$ is some positive-valued unknown function of $\tau$. As it will become clear below, the choice of function $\varphi(\tau)$ is basically irrelevant to the lowest order correction to \eqref{IntegralAgain}. The only assumption we make about $\varphi(\tau)$ is that 
\beg\label{varphiT}
\lim\limits_{\tau\to0}\varphi(\tau)\to\infty.
\en
It then follows that when $|\cos2\theta|\geq \tau\varphi(\tau)$ the integral
over the interval $[0,\pi/4-\tau\varphi(\tau)]$ is exponentially suppressed: 
\beg\label{tI}
\int\limits_0^{\pi/4-\tau\varphi(\tau)}d\theta\int\limits_0^\infty F(x,\theta)dx\ll I(\tau), \quad I(\tau)=\int\limits_{\pi/4-\tau\varphi(\tau)}^{\pi/4}d\theta\int\limits_0^\infty F(x,\theta)dx.
\en
Lastly, making the change of variables $\theta=\pi/4-s$ and approximating $
|\cos(2\theta)|\approx 2s+O(s^3)$ we have
\beg\label{IntegralAgainApp}
I(\tau)=\int\limits_0^{\tau\varphi(\tau)}2sds\int\limits_0^\infty \frac{dx}{(1+x^2)^2\cosh^2\left(\frac{2s}{\tau}\sqrt{x^2+1}\right)}.
\en 

In expression \eqref{IntegralAgainApp}, we integrate over $s$ first. Keeping in mind condition \eqref{varphiT} we have
\beg\label{IntOvers}
\begin{aligned}
&\int\limits_0^{\tau\varphi(\tau)}\frac{2sds}{\cosh^2\left(\frac{2s}{T}\sqrt{x^2+1}\right)}=\frac{\tau^2}{2}\left\{\frac{2\varphi(\tau)}{\sqrt{1+x^2}}\tanh\left[2\varphi(\tau)\sqrt{1+x^2}\right]\right.\\&\left.-\frac{\log\left(\cosh\left[2\varphi(\tau)\sqrt{1+x^2}\right]\right)}{1+x^2}\right\}\approx\frac{\tau^2\log2}{2(1+x^2)}.
\end{aligned}
\en
Performing the remaining integration over $x$ yields
\beg\label{ResJ}
K(\tau,\Delta)\approx\left(\frac{3\pi}{8}\right)\tau^2\log(2).
\en
Inserting this result into \eqref{zetaw} and keeping in mind the extra pre-factor $1/2\pi\tau$ yields \eqref{zetaTAsymptote} in the main text.

\end{appendix}

\bibliography{dquench}

\begin{thebibliography}{106}%
\makeatletter
\providecommand \@ifxundefined [1]{%
 \@ifx{#1\undefined}
}%
\providecommand \@ifnum [1]{%
 \ifnum #1\expandafter \@firstoftwo
 \else \expandafter \@secondoftwo
 \fi
}%
\providecommand \@ifx [1]{%
 \ifx #1\expandafter \@firstoftwo
 \else \expandafter \@secondoftwo
 \fi
}%
\providecommand \natexlab [1]{#1}%
\providecommand \enquote  [1]{``#1''}%
\providecommand \bibnamefont  [1]{#1}%
\providecommand \bibfnamefont [1]{#1}%
\providecommand \citenamefont [1]{#1}%
\providecommand \href@noop [0]{\@secondoftwo}%
\providecommand \href [0]{\begingroup \@sanitize@url \@href}%
\providecommand \@href[1]{\@@startlink{#1}\@@href}%
\providecommand \@@href[1]{\endgroup#1\@@endlink}%
\providecommand \@sanitize@url [0]{\catcode `\\12\catcode `\$12\catcode
  `\&12\catcode `\#12\catcode `\^12\catcode `\_12\catcode `\%12\relax}%
\providecommand \@@startlink[1]{}%
\providecommand \@@endlink[0]{}%
\providecommand \url  [0]{\begingroup\@sanitize@url \@url }%
\providecommand \@url [1]{\endgroup\@href {#1}{\urlprefix }}%
\providecommand \urlprefix  [0]{URL }%
\providecommand \Eprint [0]{\href }%
\providecommand \doibase [0]{https://doi.org/}%
\providecommand \selectlanguage [0]{\@gobble}%
\providecommand \bibinfo  [0]{\@secondoftwo}%
\providecommand \bibfield  [0]{\@secondoftwo}%
\providecommand \translation [1]{[#1]}%
\providecommand \BibitemOpen [0]{}%
\providecommand \bibitemStop [0]{}%
\providecommand \bibitemNoStop [0]{.\EOS\space}%
\providecommand \EOS [0]{\spacefactor3000\relax}%
\providecommand \BibitemShut  [1]{\csname bibitem#1\endcsname}%
\let\auto@bib@innerbib\@empty
%</preamble>
\bibitem [{\citenamefont {Ginzburg}\ and\ \citenamefont {Landau}(2009)}]{GL}%
  \BibitemOpen
  \bibfield  {author} {\bibinfo {author} {\bibfnamefont {V.~L.}\ \bibnamefont
  {Ginzburg}}\ and\ \bibinfo {author} {\bibfnamefont {L.~D.}\ \bibnamefont
  {Landau}},\ }\bibinfo {title} {On the theory of superconductivity},\ in\
  \href {https://doi.org/10.1007/978-3-540-68008-6_4} {\emph {\bibinfo
  {booktitle} {On Superconductivity and Superfluidity: A Scientific
  Autobiography}}}\ (\bibinfo  {publisher} {Springer Berlin Heidelberg},\
  \bibinfo {address} {Berlin, Heidelberg},\ \bibinfo {year} {2009})\ pp.\
  \bibinfo {pages} {113--137}\BibitemShut {NoStop}%
\bibitem [{\citenamefont {Bardeen}\ \emph {et~al.}(1957)\citenamefont
  {Bardeen}, \citenamefont {Cooper},\ and\ \citenamefont
  {Schrieffer}}]{BCS1957}%
  \BibitemOpen
  \bibfield  {author} {\bibinfo {author} {\bibfnamefont {J.}~\bibnamefont
  {Bardeen}}, \bibinfo {author} {\bibfnamefont {L.~N.}\ \bibnamefont
  {Cooper}},\ and\ \bibinfo {author} {\bibfnamefont {J.~R.}\ \bibnamefont
  {Schrieffer}},\ }\bibfield  {title} {\bibinfo {title} {Theory of
  superconductivity},\ }\href@noop {} {\bibfield  {journal} {\bibinfo
  {journal} {Phys. Rev.}\ }\textbf {\bibinfo {volume} {108}},\ \bibinfo {pages}
  {1175} (\bibinfo {year} {1957})}\BibitemShut {NoStop}%
\bibitem [{\citenamefont {Anderson}(1958{\natexlab{a}})}]{Anderson1958}%
  \BibitemOpen
  \bibfield  {author} {\bibinfo {author} {\bibfnamefont {P.~W.}\ \bibnamefont
  {Anderson}},\ }\bibfield  {title} {\bibinfo {title} {Random-phase
  approximation in the theory of superconductivity},\ }\href@noop {} {\bibfield
   {journal} {\bibinfo  {journal} {Phys. Rev.}\ }\textbf {\bibinfo {volume}
  {112}},\ \bibinfo {pages} {1900} (\bibinfo {year}
  {1958}{\natexlab{a}})}\BibitemShut {NoStop}%
\bibitem [{\citenamefont {Anderson}(1958{\natexlab{b}})}]{Anderson1958b}%
  \BibitemOpen
  \bibfield  {author} {\bibinfo {author} {\bibfnamefont {P.~W.}\ \bibnamefont
  {Anderson}},\ }\bibfield  {title} {\bibinfo {title} {New method in the theory
  of superconductivity},\ }\href {https://doi.org/10.1103/PhysRev.110.985}
  {\bibfield  {journal} {\bibinfo  {journal} {Phys. Rev.}\ }\textbf {\bibinfo
  {volume} {110}},\ \bibinfo {pages} {985} (\bibinfo {year}
  {1958}{\natexlab{b}})}\BibitemShut {NoStop}%
\bibitem [{\citenamefont {Bogoliubov}(1958)}]{NNB1958}%
  \BibitemOpen
  \bibfield  {author} {\bibinfo {author} {\bibfnamefont {N.~N.}\ \bibnamefont
  {Bogoliubov}},\ }\bibfield  {title} {\bibinfo {title} {A new method in the
  theory of superconductivity.},\ }\href@noop {} {\bibfield  {journal}
  {\bibinfo  {journal} {Sov. Phys. JETP}\ }\textbf {\bibinfo {volume} {34}}
  (\bibinfo {year} {1958})}\BibitemShut {NoStop}%
\bibitem [{\citenamefont {Anderson}(1963)}]{AndersonGauge}%
  \BibitemOpen
  \bibfield  {author} {\bibinfo {author} {\bibfnamefont {P.~W.}\ \bibnamefont
  {Anderson}},\ }\bibfield  {title} {\bibinfo {title} {Plasmons, gauge
  invariance, and mass},\ }\href {https://doi.org/10.1103/PhysRev.130.439}
  {\bibfield  {journal} {\bibinfo  {journal} {Phys. Rev.}\ }\textbf {\bibinfo
  {volume} {130}},\ \bibinfo {pages} {439} (\bibinfo {year}
  {1963})}\BibitemShut {NoStop}%
\bibitem [{\citenamefont {Carlson}\ and\ \citenamefont
  {Goldman}(1973)}]{CarlsonGoldman73}%
  \BibitemOpen
  \bibfield  {author} {\bibinfo {author} {\bibfnamefont {R.~V.}\ \bibnamefont
  {Carlson}}\ and\ \bibinfo {author} {\bibfnamefont {A.~M.}\ \bibnamefont
  {Goldman}},\ }\bibfield  {title} {\bibinfo {title} {Superconducting
  order-parameter fluctuations below ${T}_{c}$},\ }\href
  {https://doi.org/10.1103/PhysRevLett.31.880} {\bibfield  {journal} {\bibinfo
  {journal} {Phys. Rev. Lett.}\ }\textbf {\bibinfo {volume} {31}},\ \bibinfo
  {pages} {880} (\bibinfo {year} {1973})}\BibitemShut {NoStop}%
\bibitem [{\citenamefont {Artemenko}\ and\ \citenamefont
  {Volkov}(1975)}]{Volkov1975}%
  \BibitemOpen
  \bibfield  {author} {\bibinfo {author} {\bibfnamefont {S.~N.}\ \bibnamefont
  {Artemenko}}\ and\ \bibinfo {author} {\bibfnamefont {A.~F.}\ \bibnamefont
  {Volkov}},\ }\bibfield  {title} {\bibinfo {title} {Collective excitations
  with a sound spectrum in superconductors},\ }\href@noop {} {\bibfield
  {journal} {\bibinfo  {journal} {Sov. Phys. JETP}\ }\textbf {\bibinfo {volume}
  {42}},\ \bibinfo {pages} {896} (\bibinfo {year} {1975})}\BibitemShut
  {NoStop}%
\bibitem [{\citenamefont {Schmid}\ and\ \citenamefont
  {Sch\"on}(1975)}]{SchmidSchon1975}%
  \BibitemOpen
  \bibfield  {author} {\bibinfo {author} {\bibfnamefont {A.}~\bibnamefont
  {Schmid}}\ and\ \bibinfo {author} {\bibfnamefont {G.}~\bibnamefont
  {Sch\"on}},\ }\bibfield  {title} {\bibinfo {title} {Collective oscillations
  in a dirty superconductor},\ }\href
  {https://doi.org/10.1103/PhysRevLett.34.941} {\bibfield  {journal} {\bibinfo
  {journal} {Phys. Rev. Lett.}\ }\textbf {\bibinfo {volume} {34}},\ \bibinfo
  {pages} {941} (\bibinfo {year} {1975})}\BibitemShut {NoStop}%
\bibitem [{\citenamefont {Artemenko}\ and\ \citenamefont
  {Volkov}(1979)}]{Volkov1979}%
  \BibitemOpen
  \bibfield  {author} {\bibinfo {author} {\bibfnamefont {S.~N.}\ \bibnamefont
  {Artemenko}}\ and\ \bibinfo {author} {\bibfnamefont {A.~F.}\ \bibnamefont
  {Volkov}},\ }\bibfield  {title} {\bibinfo {title} {Electric fields and
  collective oscillations in superconductors},\ }\href@noop {} {\bibfield
  {journal} {\bibinfo  {journal} {Sov. Phys. Usp.}\ }\textbf {\bibinfo {volume}
  {22}},\ \bibinfo {pages} {295} (\bibinfo {year} {1979})}\BibitemShut
  {NoStop}%
\bibitem [{\citenamefont {Schmid}\ and\ \citenamefont
  {Sch\"on}(1979)}]{SchmidSchon1979}%
  \BibitemOpen
  \bibfield  {author} {\bibinfo {author} {\bibfnamefont {A.}~\bibnamefont
  {Schmid}}\ and\ \bibinfo {author} {\bibfnamefont {G.}~\bibnamefont
  {Sch\"on}},\ }\bibfield  {title} {\bibinfo {title} {Linearized kinetic
  equations and relaxation processes of a superconductor near $t_c$,},\
  }\href@noop {} {\bibfield  {journal} {\bibinfo  {journal} {J. Low Temp.
  Phys.}\ }\textbf {\bibinfo {volume} {20}},\ \bibinfo {pages} {1747} (\bibinfo
  {year} {1979})}\BibitemShut {NoStop}%
\bibitem [{\citenamefont {Kulik}\ \emph {et~al.}(1981)\citenamefont {Kulik},
  \citenamefont {Entin-Wohlman},\ and\ \citenamefont {Orbach}}]{Kulik1981}%
  \BibitemOpen
  \bibfield  {author} {\bibinfo {author} {\bibfnamefont {I.~O.}\ \bibnamefont
  {Kulik}}, \bibinfo {author} {\bibfnamefont {O.}~\bibnamefont
  {Entin-Wohlman}},\ and\ \bibinfo {author} {\bibfnamefont {R.}~\bibnamefont
  {Orbach}},\ }\bibfield  {title} {\bibinfo {title} {Pair susceptibility and
  mode propagation in superconductors: A microscopic approach},\ }\href
  {https://doi.org/10.1007/BF00115617} {\bibfield  {journal} {\bibinfo
  {journal} {Journal of Low Temperature Physics}\ }\textbf {\bibinfo {volume}
  {43}},\ \bibinfo {pages} {591} (\bibinfo {year} {1981})}\BibitemShut
  {NoStop}%
\bibitem [{\citenamefont {Ohashi}\ and\ \citenamefont
  {Takada}(1997)}]{TakadaSwave1997}%
  \BibitemOpen
  \bibfield  {author} {\bibinfo {author} {\bibfnamefont {Y.}~\bibnamefont
  {Ohashi}}\ and\ \bibinfo {author} {\bibfnamefont {S.}~\bibnamefont
  {Takada}},\ }\bibfield  {title} {\bibinfo {title} {Goldstone mode in charged
  superconductivity: Theoretical studies of the carlson-goldman mode and
  effects of the landau damping in the superconducting state},\ }\href
  {https://doi.org/10.1143/JPSJ.66.2437} {\bibfield  {journal} {\bibinfo
  {journal} {Journal of the Physical Society of Japan}\ }\textbf {\bibinfo
  {volume} {66}},\ \bibinfo {pages} {2437} (\bibinfo {year} {1997})},\ \Eprint
  {https://arxiv.org/abs/https://doi.org/10.1143/JPSJ.66.2437}
  {https://doi.org/10.1143/JPSJ.66.2437} \BibitemShut {NoStop}%
\bibitem [{\citenamefont {Ohashi}\ and\ \citenamefont
  {Takada}(1998)}]{TakadaPlasma1998}%
  \BibitemOpen
  \bibfield  {author} {\bibinfo {author} {\bibfnamefont {Y.}~\bibnamefont
  {Ohashi}}\ and\ \bibinfo {author} {\bibfnamefont {S.}~\bibnamefont
  {Takada}},\ }\bibfield  {title} {\bibinfo {title} {On the plasma oscillation
  in superconductivity},\ }\href {https://doi.org/10.1143/JPSJ.67.551}
  {\bibfield  {journal} {\bibinfo  {journal} {Journal of the Physical Society
  of Japan}\ }\textbf {\bibinfo {volume} {67}},\ \bibinfo {pages} {551}
  (\bibinfo {year} {1998})},\ \Eprint
  {https://arxiv.org/abs/https://doi.org/10.1143/JPSJ.67.551}
  {https://doi.org/10.1143/JPSJ.67.551} \BibitemShut {NoStop}%
\bibitem [{\citenamefont {Kamenev}(2011)}]{Kamenev2011}%
  \BibitemOpen
  \bibfield  {author} {\bibinfo {author} {\bibfnamefont {A.}~\bibnamefont
  {Kamenev}},\ }\href@noop {} {\emph {\bibinfo {title} {Field Theory of
  Non-Equilibrium Systems}}}\ (\bibinfo  {publisher} {Cambridge University
  Press},\ \bibinfo {year} {2011})\BibitemShut {NoStop}%
\bibitem [{\citenamefont {Schmid}(1968)}]{ASchmid}%
  \BibitemOpen
  \bibfield  {author} {\bibinfo {author} {\bibfnamefont {A.}~\bibnamefont
  {Schmid}},\ }\bibfield  {title} {\bibinfo {title} {The approach to
  equilibrium in a pure superconductor the relaxation of the cooper pair
  density},\ }\href {https://doi.org/10.1007/BF02422735} {\bibfield  {journal}
  {\bibinfo  {journal} {Physik der kondensierten Materie}\ }\textbf {\bibinfo
  {volume} {8}},\ \bibinfo {pages} {129} (\bibinfo {year} {1968})}\BibitemShut
  {NoStop}%
\bibitem [{\citenamefont {Volkov}\ and\ \citenamefont
  {Kogan}(1974)}]{VolkovKogan1973}%
  \BibitemOpen
  \bibfield  {author} {\bibinfo {author} {\bibfnamefont {A.~F.}\ \bibnamefont
  {Volkov}}\ and\ \bibinfo {author} {\bibfnamefont {S.~M.}\ \bibnamefont
  {Kogan}},\ }\bibfield  {title} {\bibinfo {title} {Collisionless relaxation of
  the energy gap in superconductors},\ }\href@noop {} {\bibfield  {journal}
  {\bibinfo  {journal} {Zh. Eksp. Teor. Fiz}\ }\textbf {\bibinfo {volume}
  {65}},\ \bibinfo {pages} {2038} (\bibinfo {year} {1974})},\ \bibinfo {note}
  {{English} translation: Sov. Phys. JETP, {\bf 38}, 1018 (1974)}\BibitemShut
  {NoStop}%
\bibitem [{\citenamefont {Galaiko}(1972)}]{Galaiko1972}%
  \BibitemOpen
  \bibfield  {author} {\bibinfo {author} {\bibfnamefont {V.~P.}\ \bibnamefont
  {Galaiko}},\ }\bibfield  {title} {\bibinfo {title} {Kinetic equation for
  relaxation processes in superconductors},\ }\href@noop {} {\bibfield
  {journal} {\bibinfo  {journal} {Sov. Phys. JETP}\ }\textbf {\bibinfo {volume}
  {34}},\ \bibinfo {pages} {203} (\bibinfo {year} {1972})}\BibitemShut
  {NoStop}%
\bibitem [{\citenamefont {Galperin}\ \emph {et~al.}(1981)\citenamefont
  {Galperin}, \citenamefont {Kozub},\ and\ \citenamefont
  {Spivak}}]{Galperin1981}%
  \BibitemOpen
  \bibfield  {author} {\bibinfo {author} {\bibfnamefont {Y.~M.}\ \bibnamefont
  {Galperin}}, \bibinfo {author} {\bibfnamefont {V.~I.}\ \bibnamefont
  {Kozub}},\ and\ \bibinfo {author} {\bibfnamefont {B.~Z.}\ \bibnamefont
  {Spivak}},\ }\bibfield  {title} {\bibinfo {title} {Dissipationless bcs
  dynamics with large branch imbalance},\ }\href@noop {} {\bibfield  {journal}
  {\bibinfo  {journal} {Sov. Phys. JETP}\ }\textbf {\bibinfo {volume} {54}},\
  \bibinfo {pages} {1126} (\bibinfo {year} {1981})}\BibitemShut {NoStop}%
\bibitem [{\citenamefont {Shumeiko}(1990)}]{Shumeiko1990}%
  \BibitemOpen
  \bibfield  {author} {\bibinfo {author} {\bibfnamefont {V.~S.}\ \bibnamefont
  {Shumeiko}},\ }\href@noop {} {\emph {\bibinfo {title} {Dynamics of electronic
  system with off-diagonal order parameter and non-linear resonant phenomena in
  superconductors}}}\ (\bibinfo  {publisher} {Doctoral Thesis, Institute for
  Low Temperature Physics and Engineering},\ \bibinfo {address} {Kharkov,
  Ukraine},\ \bibinfo {year} {1990})\BibitemShut {NoStop}%
\bibitem [{\citenamefont {Barankov}\ \emph {et~al.}(2004)\citenamefont
  {Barankov}, \citenamefont {Levitov},\ and\ \citenamefont
  {Spivak}}]{Spivak2004}%
  \BibitemOpen
  \bibfield  {author} {\bibinfo {author} {\bibfnamefont {R.~A.}\ \bibnamefont
  {Barankov}}, \bibinfo {author} {\bibfnamefont {L.~S.}\ \bibnamefont
  {Levitov}},\ and\ \bibinfo {author} {\bibfnamefont {B.~Z.}\ \bibnamefont
  {Spivak}},\ }\bibfield  {title} {\bibinfo {title} {Solitons and rabi
  oscillations in a time-dependent bcs pairing problem},\ }\href@noop {}
  {\bibfield  {journal} {\bibinfo  {journal} {Phys. Rev. Lett.}\ }\textbf
  {\bibinfo {volume} {93}},\ \bibinfo {pages} {160401} (\bibinfo {year}
  {2004})}\BibitemShut {NoStop}%
\bibitem [{\citenamefont {Andreev}\ \emph {et~al.}(2004)\citenamefont
  {Andreev}, \citenamefont {Gurarie},\ and\ \citenamefont
  {Radzihovsky}}]{Andreev2004}%
  \BibitemOpen
  \bibfield  {author} {\bibinfo {author} {\bibfnamefont {A.~V.}\ \bibnamefont
  {Andreev}}, \bibinfo {author} {\bibfnamefont {V.}~\bibnamefont {Gurarie}},\
  and\ \bibinfo {author} {\bibfnamefont {L.}~\bibnamefont {Radzihovsky}},\
  }\bibfield  {title} {\bibinfo {title} {Nonequilibrium dynamics and
  thermodynamics of a degenerate fermi gas across a feshbach resonance},\
  }\href {https://doi.org/10.1103/PhysRevLett.93.130402} {\bibfield  {journal}
  {\bibinfo  {journal} {Phys. Rev. Lett.}\ }\textbf {\bibinfo {volume} {93}},\
  \bibinfo {pages} {130402} (\bibinfo {year} {2004})}\BibitemShut {NoStop}%
\bibitem [{\citenamefont {Yuzbashyan}\ \emph {et~al.}(2005)\citenamefont
  {Yuzbashyan}, \citenamefont {Altshuler}, \citenamefont {Kuznetsov},\ and\
  \citenamefont {Enolskii}}]{Enolski2005a}%
  \BibitemOpen
  \bibfield  {author} {\bibinfo {author} {\bibfnamefont {E.~A.}\ \bibnamefont
  {Yuzbashyan}}, \bibinfo {author} {\bibfnamefont {B.~L.}\ \bibnamefont
  {Altshuler}}, \bibinfo {author} {\bibfnamefont {V.~B.}\ \bibnamefont
  {Kuznetsov}},\ and\ \bibinfo {author} {\bibfnamefont {V.~Z.}\ \bibnamefont
  {Enolskii}},\ }\bibfield  {title} {\bibinfo {title} {Nonequilibrium cooper
  pairing in the nonadiabatic regime},\ }\href@noop {} {\bibfield  {journal}
  {\bibinfo  {journal} {Phys. Rev. B}\ }\textbf {\bibinfo {volume} {72}},\
  \bibinfo {pages} {220503(R)} (\bibinfo {year} {2005})}\BibitemShut {NoStop}%
\bibitem [{\citenamefont {Barankov}\ and\ \citenamefont
  {Levitov}(2006)}]{Levitov2006}%
  \BibitemOpen
  \bibfield  {author} {\bibinfo {author} {\bibfnamefont {R.~A.}\ \bibnamefont
  {Barankov}}\ and\ \bibinfo {author} {\bibfnamefont {L.~S.}\ \bibnamefont
  {Levitov}},\ }\bibfield  {title} {\bibinfo {title} {Dynamical selection in
  developing fermionic pairing},\ }\href@noop {} {\bibfield  {journal}
  {\bibinfo  {journal} {Phys. Rev. A}\ }\textbf {\bibinfo {volume} {73}},\
  \bibinfo {pages} {033614} (\bibinfo {year} {2006})}\BibitemShut {NoStop}%
\bibitem [{\citenamefont {Yuzbashyan}\ and\ \citenamefont
  {Dzero}(2006)}]{Gapless2006}%
  \BibitemOpen
  \bibfield  {author} {\bibinfo {author} {\bibfnamefont {E.~A.}\ \bibnamefont
  {Yuzbashyan}}\ and\ \bibinfo {author} {\bibfnamefont {M.}~\bibnamefont
  {Dzero}},\ }\bibfield  {title} {\bibinfo {title} {Dynamical vanishing of the
  order parameter in a fermionic condensate},\ }\href
  {https://doi.org/10.1103/PhysRevLett.96.230404} {\bibfield  {journal}
  {\bibinfo  {journal} {Phys. Rev. Lett.}\ }\textbf {\bibinfo {volume} {96}},\
  \bibinfo {pages} {230404} (\bibinfo {year} {2006})}\BibitemShut {NoStop}%
\bibitem [{\citenamefont {Barankov}\ and\ \citenamefont
  {Levitov}(2007)}]{Levitov2007}%
  \BibitemOpen
  \bibfield  {author} {\bibinfo {author} {\bibfnamefont {R.~A.}\ \bibnamefont
  {Barankov}}\ and\ \bibinfo {author} {\bibfnamefont {L.~S.}\ \bibnamefont
  {Levitov}},\ }\bibfield  {title} {\bibinfo {title} {Excitation of the
  dissipationless higgs mode in a fermionic condensate},\ }\href@noop {}
  {\bibfield  {journal} {\bibinfo  {journal} {arXiv:0704.1292}\ } (\bibinfo
  {year} {2007})}\BibitemShut {NoStop}%
\bibitem [{\citenamefont {Yuzbashyan}\ \emph {et~al.}(2006)\citenamefont
  {Yuzbashyan}, \citenamefont {Tsyplyatyev},\ and\ \citenamefont
  {Altshuler}}]{Yuzbashyan2006}%
  \BibitemOpen
  \bibfield  {author} {\bibinfo {author} {\bibfnamefont {E.~A.}\ \bibnamefont
  {Yuzbashyan}}, \bibinfo {author} {\bibfnamefont {O.}~\bibnamefont
  {Tsyplyatyev}},\ and\ \bibinfo {author} {\bibfnamefont {B.~L.}\ \bibnamefont
  {Altshuler}},\ }\bibfield  {title} {\bibinfo {title} {Relaxation and
  persistent oscillations of the order parameter in the non-stationary bcs
  theory},\ }\href@noop {} {\bibfield  {journal} {\bibinfo  {journal} {Phys.
  Rev. Lett.}\ }\textbf {\bibinfo {volume} {96}},\ \bibinfo {pages} {097005}
  (\bibinfo {year} {2006})},\ \bibinfo {note} {erratum: Phys. Rev. Lett. {\bf
  96}, 179905 (2006)}\BibitemShut {NoStop}%
\bibitem [{\citenamefont {Yuzbashyan}(2008)}]{Yuzbashyan2008}%
  \BibitemOpen
  \bibfield  {author} {\bibinfo {author} {\bibfnamefont {E.~A.}\ \bibnamefont
  {Yuzbashyan}},\ }\bibfield  {title} {\bibinfo {title} {Normal and anomalous
  solitons in the theory of dynamical cooper pairing},\ }\href@noop {}
  {\bibfield  {journal} {\bibinfo  {journal} {Phys. Rev. B}\ }\textbf {\bibinfo
  {volume} {78}},\ \bibinfo {pages} {184507} (\bibinfo {year}
  {2008})}\BibitemShut {NoStop}%
\bibitem [{\citenamefont {Yuzbashyan}\ \emph {et~al.}(2015)\citenamefont
  {Yuzbashyan}, \citenamefont {Dzero}, \citenamefont {Gurarie},\ and\
  \citenamefont {Foster}}]{Yuzbashyan2015}%
  \BibitemOpen
  \bibfield  {author} {\bibinfo {author} {\bibfnamefont {E.~A.}\ \bibnamefont
  {Yuzbashyan}}, \bibinfo {author} {\bibfnamefont {M.}~\bibnamefont {Dzero}},
  \bibinfo {author} {\bibfnamefont {V.}~\bibnamefont {Gurarie}},\ and\ \bibinfo
  {author} {\bibfnamefont {M.~S.}\ \bibnamefont {Foster}},\ }\bibfield  {title}
  {\bibinfo {title} {Quantum quench phase diagrams of an $s$-wave bcs-bec
  condensate},\ }\href {https://doi.org/10.1103/PhysRevA.91.033628} {\bibfield
  {journal} {\bibinfo  {journal} {Phys. Rev. A}\ }\textbf {\bibinfo {volume}
  {91}},\ \bibinfo {pages} {033628} (\bibinfo {year} {2015})}\BibitemShut
  {NoStop}%
\bibitem [{\citenamefont {Littlewood}\ and\ \citenamefont
  {Varma}(1981)}]{VarmaLit1}%
  \BibitemOpen
  \bibfield  {author} {\bibinfo {author} {\bibfnamefont {P.~B.}\ \bibnamefont
  {Littlewood}}\ and\ \bibinfo {author} {\bibfnamefont {C.~M.}\ \bibnamefont
  {Varma}},\ }\bibfield  {title} {\bibinfo {title} {Gauge-invariant theory of
  the dynamical interaction of charge density waves and superconductivity},\
  }\href {https://doi.org/10.1103/PhysRevLett.47.811} {\bibfield  {journal}
  {\bibinfo  {journal} {Phys. Rev. Lett.}\ }\textbf {\bibinfo {volume} {47}},\
  \bibinfo {pages} {811} (\bibinfo {year} {1981})}\BibitemShut {NoStop}%
\bibitem [{\citenamefont {Littlewood}\ and\ \citenamefont
  {Varma}(1982)}]{VarmaLit2}%
  \BibitemOpen
  \bibfield  {author} {\bibinfo {author} {\bibfnamefont {P.~B.}\ \bibnamefont
  {Littlewood}}\ and\ \bibinfo {author} {\bibfnamefont {C.~M.}\ \bibnamefont
  {Varma}},\ }\bibfield  {title} {\bibinfo {title} {Amplitude collective modes
  in superconductors and their coupling to charge-density waves},\ }\href
  {https://doi.org/10.1103/PhysRevB.26.4883} {\bibfield  {journal} {\bibinfo
  {journal} {Phys. Rev. B}\ }\textbf {\bibinfo {volume} {26}},\ \bibinfo
  {pages} {4883} (\bibinfo {year} {1982})}\BibitemShut {NoStop}%
\bibitem [{\citenamefont {Anderson}(2015)}]{AndersonAllThat2015}%
  \BibitemOpen
  \bibfield  {author} {\bibinfo {author} {\bibfnamefont {P.~W.}\ \bibnamefont
  {Anderson}},\ }\bibfield  {title} {\bibinfo {title} {Higgs, anderson and all
  that},\ }\href {https://doi.org/10.1038/nphys3247} {\bibfield  {journal}
  {\bibinfo  {journal} {Nature Physics}\ }\textbf {\bibinfo {volume} {11}},\
  \bibinfo {pages} {93} (\bibinfo {year} {2015})}\BibitemShut {NoStop}%
\bibitem [{\citenamefont {Pekker}\ and\ \citenamefont
  {Varma}(2015)}]{Varma2014}%
  \BibitemOpen
  \bibfield  {author} {\bibinfo {author} {\bibfnamefont {D.}~\bibnamefont
  {Pekker}}\ and\ \bibinfo {author} {\bibfnamefont {C.}~\bibnamefont {Varma}},\
  }\bibfield  {title} {\bibinfo {title} {Amplitude/higgs modes in condensed
  matter physics},\ }\href
  {https://doi.org/10.1146/annurev-conmatphys-031214-014350} {\bibfield
  {journal} {\bibinfo  {journal} {Annual Review of Condensed Matter Physics}\
  }\textbf {\bibinfo {volume} {6}},\ \bibinfo {pages} {269} (\bibinfo {year}
  {2015})},\ \Eprint
  {https://arxiv.org/abs/https://doi.org/10.1146/annurev-conmatphys-031214-014350}
  {https://doi.org/10.1146/annurev-conmatphys-031214-014350} \BibitemShut
  {NoStop}%
\bibitem [{\citenamefont {Matsunaga}\ and\ \citenamefont
  {Shimano}(2012)}]{Pioneers2012}%
  \BibitemOpen
  \bibfield  {author} {\bibinfo {author} {\bibfnamefont {R.}~\bibnamefont
  {Matsunaga}}\ and\ \bibinfo {author} {\bibfnamefont {R.}~\bibnamefont
  {Shimano}},\ }\bibfield  {title} {\bibinfo {title} {Nonequilibrium bcs state
  dynamics induced by intense terahertz pulses in a superconducting nbn film},\
  }\href {https://doi.org/10.1103/PhysRevLett.109.187002} {\bibfield  {journal}
  {\bibinfo  {journal} {Phys. Rev. Lett.}\ }\textbf {\bibinfo {volume} {109}},\
  \bibinfo {pages} {187002} (\bibinfo {year} {2012})}\BibitemShut {NoStop}%
\bibitem [{\citenamefont {Matsunaga}\ \emph {et~al.}(2013)\citenamefont
  {Matsunaga}, \citenamefont {Hamada}, \citenamefont {Makise}, \citenamefont
  {Uzawa}, \citenamefont {Terai}, \citenamefont {Wang},\ and\ \citenamefont
  {Shimano}}]{Shimano2013}%
  \BibitemOpen
  \bibfield  {author} {\bibinfo {author} {\bibfnamefont {R.}~\bibnamefont
  {Matsunaga}}, \bibinfo {author} {\bibfnamefont {Y.~I.}\ \bibnamefont
  {Hamada}}, \bibinfo {author} {\bibfnamefont {K.}~\bibnamefont {Makise}},
  \bibinfo {author} {\bibfnamefont {Y.}~\bibnamefont {Uzawa}}, \bibinfo
  {author} {\bibfnamefont {H.}~\bibnamefont {Terai}}, \bibinfo {author}
  {\bibfnamefont {Z.}~\bibnamefont {Wang}},\ and\ \bibinfo {author}
  {\bibfnamefont {R.}~\bibnamefont {Shimano}},\ }\bibfield  {title} {\bibinfo
  {title} {Higgs amplitude mode in the bcs superconductors
  ${\mathrm{nb}}_{1\mathrm{\text{-}}x}{\mathrm{ti}}_{x}\mathbf{N}$ induced by
  terahertz pulse excitation},\ }\href@noop {} {\bibfield  {journal} {\bibinfo
  {journal} {Phys. Rev. Lett.}\ }\textbf {\bibinfo {volume} {111}},\ \bibinfo
  {pages} {057002} (\bibinfo {year} {2013})}\BibitemShut {NoStop}%
\bibitem [{\citenamefont {Matsunaga}\ \emph {et~al.}(2014)\citenamefont
  {Matsunaga}, \citenamefont {Tsuji}, \citenamefont {Fujita}, \citenamefont
  {Sugioka}, \citenamefont {Makise}, \citenamefont {Uzawa}, \citenamefont
  {Terai}, \citenamefont {Wang}, \citenamefont {Aoki},\ and\ \citenamefont
  {Shimano}}]{Shimano2014}%
  \BibitemOpen
  \bibfield  {author} {\bibinfo {author} {\bibfnamefont {R.}~\bibnamefont
  {Matsunaga}}, \bibinfo {author} {\bibfnamefont {N.}~\bibnamefont {Tsuji}},
  \bibinfo {author} {\bibfnamefont {H.}~\bibnamefont {Fujita}}, \bibinfo
  {author} {\bibfnamefont {A.}~\bibnamefont {Sugioka}}, \bibinfo {author}
  {\bibfnamefont {K.}~\bibnamefont {Makise}}, \bibinfo {author} {\bibfnamefont
  {Y.}~\bibnamefont {Uzawa}}, \bibinfo {author} {\bibfnamefont
  {H.}~\bibnamefont {Terai}}, \bibinfo {author} {\bibfnamefont
  {Z.}~\bibnamefont {Wang}}, \bibinfo {author} {\bibfnamefont {H.}~\bibnamefont
  {Aoki}},\ and\ \bibinfo {author} {\bibfnamefont {R.}~\bibnamefont
  {Shimano}},\ }\bibfield  {title} {\bibinfo {title} {Light-induced collective
  pseudospin precession resonating with higgs mode in a superconductor},\
  }\href {https://doi.org/10.1126/science.1254697} {\bibfield  {journal}
  {\bibinfo  {journal} {Science}\ }\textbf {\bibinfo {volume} {345}},\ \bibinfo
  {pages} {1145} (\bibinfo {year} {2014})},\ \Eprint
  {https://arxiv.org/abs/https://www.science.org/doi/pdf/10.1126/science.1254697}
  {https://www.science.org/doi/pdf/10.1126/science.1254697} \BibitemShut
  {NoStop}%
\bibitem [{\citenamefont {Beck}\ \emph {et~al.}(2013)\citenamefont {Beck},
  \citenamefont {Rousseau}, \citenamefont {Klammer}, \citenamefont {Leiderer},
  \citenamefont {Mittendorff}, \citenamefont {Winnerl}, \citenamefont {Helm},
  \citenamefont {Gol'tsman},\ and\ \citenamefont {Demsar}}]{THz3}%
  \BibitemOpen
  \bibfield  {author} {\bibinfo {author} {\bibfnamefont {M.}~\bibnamefont
  {Beck}}, \bibinfo {author} {\bibfnamefont {I.}~\bibnamefont {Rousseau}},
  \bibinfo {author} {\bibfnamefont {M.}~\bibnamefont {Klammer}}, \bibinfo
  {author} {\bibfnamefont {P.}~\bibnamefont {Leiderer}}, \bibinfo {author}
  {\bibfnamefont {M.}~\bibnamefont {Mittendorff}}, \bibinfo {author}
  {\bibfnamefont {S.}~\bibnamefont {Winnerl}}, \bibinfo {author} {\bibfnamefont
  {M.}~\bibnamefont {Helm}}, \bibinfo {author} {\bibfnamefont {G.~N.}\
  \bibnamefont {Gol'tsman}},\ and\ \bibinfo {author} {\bibfnamefont
  {J.}~\bibnamefont {Demsar}},\ }\bibfield  {title} {\bibinfo {title}
  {Transient increase of the energy gap of superconducting nbn thin films
  excited by resonant narrow-band terahertz pulses},\ }\href
  {https://doi.org/10.1103/PhysRevLett.110.267003} {\bibfield  {journal}
  {\bibinfo  {journal} {Phys. Rev. Lett.}\ }\textbf {\bibinfo {volume} {110}},\
  \bibinfo {pages} {267003} (\bibinfo {year} {2013})}\BibitemShut {NoStop}%
\bibitem [{\citenamefont {Sherman}\ \emph {et~al.}(2015)\citenamefont
  {Sherman}, \citenamefont {Pracht}, \citenamefont {Gorshunov}, \citenamefont
  {Poran}, \citenamefont {Jesudasan}, \citenamefont {Chand}, \citenamefont
  {Raychaudhuri}, \citenamefont {Swanson}, \citenamefont {Trivedi},
  \citenamefont {Auerbach}, \citenamefont {Scheffler}, \citenamefont
  {Frydman},\ and\ \citenamefont {Dressel}}]{Sherman2015-Disorder}%
  \BibitemOpen
  \bibfield  {author} {\bibinfo {author} {\bibfnamefont {D.}~\bibnamefont
  {Sherman}}, \bibinfo {author} {\bibfnamefont {U.~S.}\ \bibnamefont {Pracht}},
  \bibinfo {author} {\bibfnamefont {B.}~\bibnamefont {Gorshunov}}, \bibinfo
  {author} {\bibfnamefont {S.}~\bibnamefont {Poran}}, \bibinfo {author}
  {\bibfnamefont {J.}~\bibnamefont {Jesudasan}}, \bibinfo {author}
  {\bibfnamefont {M.}~\bibnamefont {Chand}}, \bibinfo {author} {\bibfnamefont
  {P.}~\bibnamefont {Raychaudhuri}}, \bibinfo {author} {\bibfnamefont
  {M.}~\bibnamefont {Swanson}}, \bibinfo {author} {\bibfnamefont
  {N.}~\bibnamefont {Trivedi}}, \bibinfo {author} {\bibfnamefont
  {A.}~\bibnamefont {Auerbach}}, \bibinfo {author} {\bibfnamefont
  {M.}~\bibnamefont {Scheffler}}, \bibinfo {author} {\bibfnamefont
  {A.}~\bibnamefont {Frydman}},\ and\ \bibinfo {author} {\bibfnamefont
  {M.}~\bibnamefont {Dressel}},\ }\bibfield  {title} {\bibinfo {title} {The
  higgs mode in disordered superconductors close to a quantum phase
  transition},\ }\href {https://doi.org/10.1038/nphys3227} {\bibfield
  {journal} {\bibinfo  {journal} {Nature Physics}\ }\textbf {\bibinfo {volume}
  {11}},\ \bibinfo {pages} {188} (\bibinfo {year} {2015})}\BibitemShut
  {NoStop}%
\bibitem [{\citenamefont {Katsumi}\ \emph {et~al.}(2024)\citenamefont
  {Katsumi}, \citenamefont {Fiore}, \citenamefont {Udina}, \citenamefont
  {Romero}, \citenamefont {Barbalas}, \citenamefont {Jesudasan}, \citenamefont
  {Raychaudhuri}, \citenamefont {Seibold}, \citenamefont {Benfatto},\ and\
  \citenamefont {Armitage}}]{KatoKatsumi2024}%
  \BibitemOpen
  \bibfield  {author} {\bibinfo {author} {\bibfnamefont {K.}~\bibnamefont
  {Katsumi}}, \bibinfo {author} {\bibfnamefont {J.}~\bibnamefont {Fiore}},
  \bibinfo {author} {\bibfnamefont {M.}~\bibnamefont {Udina}}, \bibinfo
  {author} {\bibfnamefont {R.}~\bibnamefont {Romero}}, \bibinfo {author}
  {\bibfnamefont {D.}~\bibnamefont {Barbalas}}, \bibinfo {author}
  {\bibfnamefont {J.}~\bibnamefont {Jesudasan}}, \bibinfo {author}
  {\bibfnamefont {P.}~\bibnamefont {Raychaudhuri}}, \bibinfo {author}
  {\bibfnamefont {G.}~\bibnamefont {Seibold}}, \bibinfo {author} {\bibfnamefont
  {L.}~\bibnamefont {Benfatto}},\ and\ \bibinfo {author} {\bibfnamefont
  {N.~P.}\ \bibnamefont {Armitage}},\ }\bibfield  {title} {\bibinfo {title}
  {Revealing novel aspects of light-matter coupling by terahertz
  two-dimensional coherent spectroscopy: The case of the amplitude mode in
  superconductors},\ }\href {https://doi.org/10.1103/PhysRevLett.132.256903}
  {\bibfield  {journal} {\bibinfo  {journal} {Phys. Rev. Lett.}\ }\textbf
  {\bibinfo {volume} {132}},\ \bibinfo {pages} {256903} (\bibinfo {year}
  {2024})}\BibitemShut {NoStop}%
\bibitem [{\citenamefont {M\'easson}\ \emph {et~al.}(2014)\citenamefont
  {M\'easson}, \citenamefont {Gallais}, \citenamefont {Cazayous}, \citenamefont
  {Clair}, \citenamefont {Rodi\`ere}, \citenamefont {Cario},\ and\
  \citenamefont {Sacuto}}]{Measson2014}%
  \BibitemOpen
  \bibfield  {author} {\bibinfo {author} {\bibfnamefont {M.-A.}\ \bibnamefont
  {M\'easson}}, \bibinfo {author} {\bibfnamefont {Y.}~\bibnamefont {Gallais}},
  \bibinfo {author} {\bibfnamefont {M.}~\bibnamefont {Cazayous}}, \bibinfo
  {author} {\bibfnamefont {B.}~\bibnamefont {Clair}}, \bibinfo {author}
  {\bibfnamefont {P.}~\bibnamefont {Rodi\`ere}}, \bibinfo {author}
  {\bibfnamefont {L.}~\bibnamefont {Cario}},\ and\ \bibinfo {author}
  {\bibfnamefont {A.}~\bibnamefont {Sacuto}},\ }\bibfield  {title} {\bibinfo
  {title} {Amplitude higgs mode in the $2h\ensuremath{-}{\text{nbse}}_{2}$
  superconductor},\ }\href {https://doi.org/10.1103/PhysRevB.89.060503}
  {\bibfield  {journal} {\bibinfo  {journal} {Phys. Rev. B}\ }\textbf {\bibinfo
  {volume} {89}},\ \bibinfo {pages} {060503} (\bibinfo {year}
  {2014})}\BibitemShut {NoStop}%
\bibitem [{\citenamefont {Behrle}\ \emph {et~al.}(2018)\citenamefont {Behrle},
  \citenamefont {Harrison}, \citenamefont {Kombe}, \citenamefont {Gao},
  \citenamefont {Link}, \citenamefont {Bernier}, \citenamefont {Kollath},\ and\
  \citenamefont {K{\"o}hl}}]{Behrle2018}%
  \BibitemOpen
  \bibfield  {author} {\bibinfo {author} {\bibfnamefont {A.}~\bibnamefont
  {Behrle}}, \bibinfo {author} {\bibfnamefont {T.}~\bibnamefont {Harrison}},
  \bibinfo {author} {\bibfnamefont {J.}~\bibnamefont {Kombe}}, \bibinfo
  {author} {\bibfnamefont {K.}~\bibnamefont {Gao}}, \bibinfo {author}
  {\bibfnamefont {M.}~\bibnamefont {Link}}, \bibinfo {author} {\bibfnamefont
  {J.~S.}\ \bibnamefont {Bernier}}, \bibinfo {author} {\bibfnamefont
  {C.}~\bibnamefont {Kollath}},\ and\ \bibinfo {author} {\bibfnamefont
  {M.}~\bibnamefont {K{\"o}hl}},\ }\bibfield  {title} {\bibinfo {title} {Higgs
  mode in a strongly interacting fermionic superfluid},\ }\href
  {https://doi.org/10.1038/s41567-018-0128-6} {\bibfield  {journal} {\bibinfo
  {journal} {Nature Physics}\ }\textbf {\bibinfo {volume} {14}},\ \bibinfo
  {pages} {781} (\bibinfo {year} {2018})}\BibitemShut {NoStop}%
\bibitem [{\citenamefont {Grasset}\ \emph {et~al.}(2019)\citenamefont
  {Grasset}, \citenamefont {Gallais}, \citenamefont {Sacuto}, \citenamefont
  {Cazayous}, \citenamefont {Ma\~nas Valero}, \citenamefont {Coronado},\ and\
  \citenamefont {M\'easson}}]{Measson2019}%
  \BibitemOpen
  \bibfield  {author} {\bibinfo {author} {\bibfnamefont {R.}~\bibnamefont
  {Grasset}}, \bibinfo {author} {\bibfnamefont {Y.}~\bibnamefont {Gallais}},
  \bibinfo {author} {\bibfnamefont {A.}~\bibnamefont {Sacuto}}, \bibinfo
  {author} {\bibfnamefont {M.}~\bibnamefont {Cazayous}}, \bibinfo {author}
  {\bibfnamefont {S.}~\bibnamefont {Ma\~nas Valero}}, \bibinfo {author}
  {\bibfnamefont {E.}~\bibnamefont {Coronado}},\ and\ \bibinfo {author}
  {\bibfnamefont {M.-A.}\ \bibnamefont {M\'easson}},\ }\bibfield  {title}
  {\bibinfo {title} {Pressure-induced collapse of the charge density wave and
  higgs mode visibility in $2h\text{\ensuremath{-}}{\mathrm{tas}}_{2}$},\
  }\href {https://doi.org/10.1103/PhysRevLett.122.127001} {\bibfield  {journal}
  {\bibinfo  {journal} {Phys. Rev. Lett.}\ }\textbf {\bibinfo {volume} {122}},\
  \bibinfo {pages} {127001} (\bibinfo {year} {2019})}\BibitemShut {NoStop}%
\bibitem [{\citenamefont {Shimano}\ and\ \citenamefont
  {Tsuji}(2020)}]{Shimano2020}%
  \BibitemOpen
  \bibfield  {author} {\bibinfo {author} {\bibfnamefont {R.}~\bibnamefont
  {Shimano}}\ and\ \bibinfo {author} {\bibfnamefont {N.}~\bibnamefont
  {Tsuji}},\ }\bibfield  {title} {\bibinfo {title} {Higgs mode in
  superconductors},\ }\href
  {https://doi.org/10.1146/annurev-conmatphys-031119-050813} {\bibfield
  {journal} {\bibinfo  {journal} {Annual Review of Condensed Matter Physics}\
  }\textbf {\bibinfo {volume} {11}},\ \bibinfo {pages} {103} (\bibinfo {year}
  {2020})},\ \Eprint
  {https://arxiv.org/abs/https://doi.org/10.1146/annurev-conmatphys-031119-050813}
  {https://doi.org/10.1146/annurev-conmatphys-031119-050813} \BibitemShut
  {NoStop}%
\bibitem [{\citenamefont {Nakamura}\ \emph {et~al.}(2020)\citenamefont
  {Nakamura}, \citenamefont {Katsumi}, \citenamefont {Terai},\ and\
  \citenamefont {Shimano}}]{NonReciprocal2020}%
  \BibitemOpen
  \bibfield  {author} {\bibinfo {author} {\bibfnamefont {S.}~\bibnamefont
  {Nakamura}}, \bibinfo {author} {\bibfnamefont {K.}~\bibnamefont {Katsumi}},
  \bibinfo {author} {\bibfnamefont {H.}~\bibnamefont {Terai}},\ and\ \bibinfo
  {author} {\bibfnamefont {R.}~\bibnamefont {Shimano}},\ }\bibfield  {title}
  {\bibinfo {title} {Nonreciprocal terahertz second-harmonic generation in
  superconducting nbn under supercurrent injection},\ }\href
  {https://doi.org/10.1103/PhysRevLett.125.097004} {\bibfield  {journal}
  {\bibinfo  {journal} {Phys. Rev. Lett.}\ }\textbf {\bibinfo {volume} {125}},\
  \bibinfo {pages} {097004} (\bibinfo {year} {2020})}\BibitemShut {NoStop}%
\bibitem [{\citenamefont {Katsumi}\ \emph
  {et~al.}(2018{\natexlab{a}})\citenamefont {Katsumi}, \citenamefont {Tsuji},
  \citenamefont {Hamada}, \citenamefont {Matsunaga}, \citenamefont
  {Schneeloch}, \citenamefont {Zhong}, \citenamefont {Gu}, \citenamefont
  {Aoki}, \citenamefont {Gallais},\ and\ \citenamefont
  {Shimano}}]{Katsumi2018}%
  \BibitemOpen
  \bibfield  {author} {\bibinfo {author} {\bibfnamefont {K.}~\bibnamefont
  {Katsumi}}, \bibinfo {author} {\bibfnamefont {N.}~\bibnamefont {Tsuji}},
  \bibinfo {author} {\bibfnamefont {Y.~I.}\ \bibnamefont {Hamada}}, \bibinfo
  {author} {\bibfnamefont {R.}~\bibnamefont {Matsunaga}}, \bibinfo {author}
  {\bibfnamefont {J.}~\bibnamefont {Schneeloch}}, \bibinfo {author}
  {\bibfnamefont {R.~D.}\ \bibnamefont {Zhong}}, \bibinfo {author}
  {\bibfnamefont {G.~D.}\ \bibnamefont {Gu}}, \bibinfo {author} {\bibfnamefont
  {H.}~\bibnamefont {Aoki}}, \bibinfo {author} {\bibfnamefont {Y.}~\bibnamefont
  {Gallais}},\ and\ \bibinfo {author} {\bibfnamefont {R.}~\bibnamefont
  {Shimano}},\ }\bibfield  {title} {\bibinfo {title} {Higgs mode in the
  $d$-wave superconductor
  ${\mathrm{bi}}_{2}{\mathrm{sr}}_{2}{\mathrm{cacu}}_{2}{\mathrm{o}}_{8+x}$
  driven by an intense terahertz pulse},\ }\href
  {https://doi.org/10.1103/PhysRevLett.120.117001} {\bibfield  {journal}
  {\bibinfo  {journal} {Phys. Rev. Lett.}\ }\textbf {\bibinfo {volume} {120}},\
  \bibinfo {pages} {117001} (\bibinfo {year} {2018}{\natexlab{a}})}\BibitemShut
  {NoStop}%
\bibitem [{\citenamefont {Katsumi}\ \emph
  {et~al.}(2020{\natexlab{a}})\citenamefont {Katsumi}, \citenamefont {Li},
  \citenamefont {Raffy}, \citenamefont {Gallais},\ and\ \citenamefont
  {Shimano}}]{Katsumi2020}%
  \BibitemOpen
  \bibfield  {author} {\bibinfo {author} {\bibfnamefont {K.}~\bibnamefont
  {Katsumi}}, \bibinfo {author} {\bibfnamefont {Z.~Z.}\ \bibnamefont {Li}},
  \bibinfo {author} {\bibfnamefont {H.}~\bibnamefont {Raffy}}, \bibinfo
  {author} {\bibfnamefont {Y.}~\bibnamefont {Gallais}},\ and\ \bibinfo {author}
  {\bibfnamefont {R.}~\bibnamefont {Shimano}},\ }\bibfield  {title} {\bibinfo
  {title} {Superconducting fluctuations probed by the higgs mode in
  ${\mathrm{bi}}_{2}{\mathrm{sr}}_{2}\mathrm{Ca}{\mathrm{cu}}_{2}{\mathrm{o}}_{8+x}$
  thin films},\ }\href {https://doi.org/10.1103/PhysRevB.102.054510} {\bibfield
   {journal} {\bibinfo  {journal} {Phys. Rev. B}\ }\textbf {\bibinfo {volume}
  {102}},\ \bibinfo {pages} {054510} (\bibinfo {year}
  {2020}{\natexlab{a}})}\BibitemShut {NoStop}%
\bibitem [{\citenamefont {Papenkort}\ \emph {et~al.}(2007)\citenamefont
  {Papenkort}, \citenamefont {Axt},\ and\ \citenamefont
  {Kuhn}}]{Papenkort2007}%
  \BibitemOpen
  \bibfield  {author} {\bibinfo {author} {\bibfnamefont {T.}~\bibnamefont
  {Papenkort}}, \bibinfo {author} {\bibfnamefont {V.~M.}\ \bibnamefont {Axt}},\
  and\ \bibinfo {author} {\bibfnamefont {T.}~\bibnamefont {Kuhn}},\ }\bibfield
  {title} {\bibinfo {title} {Coherent dynamics and pump-probe spectra of bcs
  superconductors},\ }\href {https://doi.org/10.1103/PhysRevB.76.224522}
  {\bibfield  {journal} {\bibinfo  {journal} {Phys. Rev. B}\ }\textbf {\bibinfo
  {volume} {76}},\ \bibinfo {pages} {224522} (\bibinfo {year}
  {2007})}\BibitemShut {NoStop}%
\bibitem [{\citenamefont {Papenkort}\ \emph {et~al.}(2009)\citenamefont
  {Papenkort}, \citenamefont {Kuhn},\ and\ \citenamefont {Axt}}]{Axt2009}%
  \BibitemOpen
  \bibfield  {author} {\bibinfo {author} {\bibfnamefont {T.}~\bibnamefont
  {Papenkort}}, \bibinfo {author} {\bibfnamefont {T.}~\bibnamefont {Kuhn}},\
  and\ \bibinfo {author} {\bibfnamefont {V.~M.}\ \bibnamefont {Axt}},\
  }\bibfield  {title} {\bibinfo {title} {Nonequilibrium dynamics and coherent
  control of bcs superconductors driven by ultrashort thz pulses},\ }\href@noop
  {} {\bibfield  {journal} {\bibinfo  {journal} {Journal of Physics}\ }\textbf
  {\bibinfo {volume} {193}},\ \bibinfo {pages} {012050} (\bibinfo {year}
  {2009})}\BibitemShut {NoStop}%
\bibitem [{\citenamefont {Podolsky}\ \emph {et~al.}(2011)\citenamefont
  {Podolsky}, \citenamefont {Auerbach},\ and\ \citenamefont
  {Arovas}}]{Assa2011}%
  \BibitemOpen
  \bibfield  {author} {\bibinfo {author} {\bibfnamefont {D.}~\bibnamefont
  {Podolsky}}, \bibinfo {author} {\bibfnamefont {A.}~\bibnamefont {Auerbach}},\
  and\ \bibinfo {author} {\bibfnamefont {D.~P.}\ \bibnamefont {Arovas}},\
  }\bibfield  {title} {\bibinfo {title} {Visibility of the amplitude (higgs)
  mode in condensed matter},\ }\href
  {https://doi.org/10.1103/PhysRevB.84.174522} {\bibfield  {journal} {\bibinfo
  {journal} {Phys. Rev. B}\ }\textbf {\bibinfo {volume} {84}},\ \bibinfo
  {pages} {174522} (\bibinfo {year} {2011})}\BibitemShut {NoStop}%
\bibitem [{\citenamefont {Podolsky}\ and\ \citenamefont
  {Sachdev}(2012)}]{Sachdev2012}%
  \BibitemOpen
  \bibfield  {author} {\bibinfo {author} {\bibfnamefont {D.}~\bibnamefont
  {Podolsky}}\ and\ \bibinfo {author} {\bibfnamefont {S.}~\bibnamefont
  {Sachdev}},\ }\bibfield  {title} {\bibinfo {title} {Spectral functions of the
  higgs mode near two-dimensional quantum critical points},\ }\href
  {https://doi.org/10.1103/PhysRevB.86.054508} {\bibfield  {journal} {\bibinfo
  {journal} {Phys. Rev. B}\ }\textbf {\bibinfo {volume} {86}},\ \bibinfo
  {pages} {054508} (\bibinfo {year} {2012})}\BibitemShut {NoStop}%
\bibitem [{\citenamefont {Gazit}\ \emph {et~al.}(2013)\citenamefont {Gazit},
  \citenamefont {Podolsky},\ and\ \citenamefont {Auerbach}}]{Assa2013}%
  \BibitemOpen
  \bibfield  {author} {\bibinfo {author} {\bibfnamefont {S.}~\bibnamefont
  {Gazit}}, \bibinfo {author} {\bibfnamefont {D.}~\bibnamefont {Podolsky}},\
  and\ \bibinfo {author} {\bibfnamefont {A.}~\bibnamefont {Auerbach}},\
  }\bibfield  {title} {\bibinfo {title} {Fate of the higgs mode near quantum
  criticality},\ }\href {https://doi.org/10.1103/PhysRevLett.110.140401}
  {\bibfield  {journal} {\bibinfo  {journal} {Phys. Rev. Lett.}\ }\textbf
  {\bibinfo {volume} {110}},\ \bibinfo {pages} {140401} (\bibinfo {year}
  {2013})}\BibitemShut {NoStop}%
\bibitem [{\citenamefont {Volovik}\ and\ \citenamefont
  {Zubkov}(2014)}]{Volovik2014}%
  \BibitemOpen
  \bibfield  {author} {\bibinfo {author} {\bibfnamefont {G.~E.}\ \bibnamefont
  {Volovik}}\ and\ \bibinfo {author} {\bibfnamefont {M.~A.}\ \bibnamefont
  {Zubkov}},\ }\bibfield  {title} {\bibinfo {title} {Higgs bosons in particle
  physics and in condensed matter},\ }\href
  {https://doi.org/10.1007/s10909-013-0905-7} {\bibfield  {journal} {\bibinfo
  {journal} {Journal of Low Temperature Physics}\ }\textbf {\bibinfo {volume}
  {175}},\ \bibinfo {pages} {486} (\bibinfo {year} {2014})}\BibitemShut
  {NoStop}%
\bibitem [{\citenamefont {Ran\ifmmode~\mbox{\c{c}}\else \c{c}\fi{}on}\ and\
  \citenamefont {Dupuis}(2014)}]{Dupuis2014}%
  \BibitemOpen
  \bibfield  {author} {\bibinfo {author} {\bibfnamefont {A.}~\bibnamefont
  {Ran\ifmmode~\mbox{\c{c}}\else \c{c}\fi{}on}}\ and\ \bibinfo {author}
  {\bibfnamefont {N.}~\bibnamefont {Dupuis}},\ }\bibfield  {title} {\bibinfo
  {title} {Higgs amplitude mode in the vicinity of a $(2+1)$-dimensional
  quantum critical point},\ }\href {https://doi.org/10.1103/PhysRevB.89.180501}
  {\bibfield  {journal} {\bibinfo  {journal} {Phys. Rev. B}\ }\textbf {\bibinfo
  {volume} {89}},\ \bibinfo {pages} {180501} (\bibinfo {year}
  {2014})}\BibitemShut {NoStop}%
\bibitem [{\citenamefont {Krull}\ \emph {et~al.}(2014)\citenamefont {Krull},
  \citenamefont {Manske}, \citenamefont {Uhrig},\ and\ \citenamefont
  {Schnyder}}]{Manske2014}%
  \BibitemOpen
  \bibfield  {author} {\bibinfo {author} {\bibfnamefont {H.}~\bibnamefont
  {Krull}}, \bibinfo {author} {\bibfnamefont {D.}~\bibnamefont {Manske}},
  \bibinfo {author} {\bibfnamefont {G.~S.}\ \bibnamefont {Uhrig}},\ and\
  \bibinfo {author} {\bibfnamefont {A.~P.}\ \bibnamefont {Schnyder}},\
  }\bibfield  {title} {\bibinfo {title} {Signatures of nonadiabatic bcs state
  dynamics in pump-probe conductivity},\ }\href
  {https://doi.org/10.1103/PhysRevB.90.014515} {\bibfield  {journal} {\bibinfo
  {journal} {Phys. Rev. B}\ }\textbf {\bibinfo {volume} {90}},\ \bibinfo
  {pages} {014515} (\bibinfo {year} {2014})}\BibitemShut {NoStop}%
\bibitem [{\citenamefont {Cea}\ \emph {et~al.}(2016)\citenamefont {Cea},
  \citenamefont {Castellani},\ and\ \citenamefont {Benfatto}}]{Cea2016}%
  \BibitemOpen
  \bibfield  {author} {\bibinfo {author} {\bibfnamefont {T.}~\bibnamefont
  {Cea}}, \bibinfo {author} {\bibfnamefont {C.}~\bibnamefont {Castellani}},\
  and\ \bibinfo {author} {\bibfnamefont {L.}~\bibnamefont {Benfatto}},\
  }\bibfield  {title} {\bibinfo {title} {Nonlinear optical effects and
  third-harmonic generation in superconductors: Cooper pairs versus higgs mode
  contribution},\ }\href {https://doi.org/10.1103/PhysRevB.93.180507}
  {\bibfield  {journal} {\bibinfo  {journal} {Phys. Rev. B}\ }\textbf {\bibinfo
  {volume} {93}},\ \bibinfo {pages} {180507} (\bibinfo {year}
  {2016})}\BibitemShut {NoStop}%
\bibitem [{\citenamefont {Moor}\ \emph {et~al.}(2017)\citenamefont {Moor},
  \citenamefont {Volkov},\ and\ \citenamefont {Efetov}}]{Moore2017}%
  \BibitemOpen
  \bibfield  {author} {\bibinfo {author} {\bibfnamefont {A.}~\bibnamefont
  {Moor}}, \bibinfo {author} {\bibfnamefont {A.~F.}\ \bibnamefont {Volkov}},\
  and\ \bibinfo {author} {\bibfnamefont {K.~B.}\ \bibnamefont {Efetov}},\
  }\bibfield  {title} {\bibinfo {title} {Amplitude higgs mode and admittance in
  superconductors with a moving condensate},\ }\href
  {https://doi.org/10.1103/PhysRevLett.118.047001} {\bibfield  {journal}
  {\bibinfo  {journal} {Phys. Rev. Lett.}\ }\textbf {\bibinfo {volume} {118}},\
  \bibinfo {pages} {047001} (\bibinfo {year} {2017})}\BibitemShut {NoStop}%
\bibitem [{\citenamefont {Fischer}\ \emph {et~al.}(2018)\citenamefont
  {Fischer}, \citenamefont {Hecker}, \citenamefont {Hoyer},\ and\ \citenamefont
  {Schmalian}}]{Joerg2018}%
  \BibitemOpen
  \bibfield  {author} {\bibinfo {author} {\bibfnamefont {S.}~\bibnamefont
  {Fischer}}, \bibinfo {author} {\bibfnamefont {M.}~\bibnamefont {Hecker}},
  \bibinfo {author} {\bibfnamefont {M.}~\bibnamefont {Hoyer}},\ and\ \bibinfo
  {author} {\bibfnamefont {J.}~\bibnamefont {Schmalian}},\ }\bibfield  {title}
  {\bibinfo {title} {Short-distance breakdown of the higgs mechanism and the
  robustness of the bcs theory for charged superconductors},\ }\href
  {https://doi.org/10.1103/PhysRevB.97.054510} {\bibfield  {journal} {\bibinfo
  {journal} {Phys. Rev. B}\ }\textbf {\bibinfo {volume} {97}},\ \bibinfo
  {pages} {054510} (\bibinfo {year} {2018})}\BibitemShut {NoStop}%
\bibitem [{\citenamefont {Sun}\ \emph {et~al.}(2020)\citenamefont {Sun},
  \citenamefont {Fogler}, \citenamefont {Basov},\ and\ \citenamefont
  {Millis}}]{Basov2020}%
  \BibitemOpen
  \bibfield  {author} {\bibinfo {author} {\bibfnamefont {Z.}~\bibnamefont
  {Sun}}, \bibinfo {author} {\bibfnamefont {M.~M.}\ \bibnamefont {Fogler}},
  \bibinfo {author} {\bibfnamefont {D.~N.}\ \bibnamefont {Basov}},\ and\
  \bibinfo {author} {\bibfnamefont {A.~J.}\ \bibnamefont {Millis}},\ }\bibfield
   {title} {\bibinfo {title} {Collective modes and terahertz near-field
  response of superconductors},\ }\href
  {https://doi.org/10.1103/PhysRevResearch.2.023413} {\bibfield  {journal}
  {\bibinfo  {journal} {Phys. Rev. Res.}\ }\textbf {\bibinfo {volume} {2}},\
  \bibinfo {pages} {023413} (\bibinfo {year} {2020})}\BibitemShut {NoStop}%
\bibitem [{\citenamefont {Barresi}\ \emph {et~al.}(2023)\citenamefont
  {Barresi}, \citenamefont {Boulet}, \citenamefont {Wlaz{\l}owski},\ and\
  \citenamefont {Magierski}}]{Barresi2023}%
  \BibitemOpen
  \bibfield  {author} {\bibinfo {author} {\bibfnamefont {A.}~\bibnamefont
  {Barresi}}, \bibinfo {author} {\bibfnamefont {A.}~\bibnamefont {Boulet}},
  \bibinfo {author} {\bibfnamefont {G.}~\bibnamefont {Wlaz{\l}owski}},\ and\
  \bibinfo {author} {\bibfnamefont {P.}~\bibnamefont {Magierski}},\ }\bibfield
  {title} {\bibinfo {title} {Generation and decay of higgs mode in a strongly
  interacting fermi gas},\ }\href {https://doi.org/10.1038/s41598-023-38176-9}
  {\bibfield  {journal} {\bibinfo  {journal} {Scientific Reports}\ }\textbf
  {\bibinfo {volume} {13}},\ \bibinfo {pages} {11285} (\bibinfo {year}
  {2023})}\BibitemShut {NoStop}%
\bibitem [{\citenamefont {Derendorf}\ \emph {et~al.}(2024)\citenamefont
  {Derendorf}, \citenamefont {Volkov},\ and\ \citenamefont
  {Eremin}}]{Eremin2024}%
  \BibitemOpen
  \bibfield  {author} {\bibinfo {author} {\bibfnamefont {P.}~\bibnamefont
  {Derendorf}}, \bibinfo {author} {\bibfnamefont {A.~F.}\ \bibnamefont
  {Volkov}},\ and\ \bibinfo {author} {\bibfnamefont {I.~M.}\ \bibnamefont
  {Eremin}},\ }\bibfield  {title} {\bibinfo {title} {Nonlinear response of
  diffusive superconductors to ac electromagnetic fields},\ }\href
  {https://doi.org/10.1103/PhysRevB.109.024510} {\bibfield  {journal} {\bibinfo
   {journal} {Phys. Rev. B}\ }\textbf {\bibinfo {volume} {109}},\ \bibinfo
  {pages} {024510} (\bibinfo {year} {2024})}\BibitemShut {NoStop}%
\bibitem [{\citenamefont {Collado}\ \emph {et~al.}(2023)\citenamefont
  {Collado}, \citenamefont {Defenu},\ and\ \citenamefont
  {Lorenzana}}]{Lorenzana2023}%
  \BibitemOpen
  \bibfield  {author} {\bibinfo {author} {\bibfnamefont {H.~P.~O.}\
  \bibnamefont {Collado}}, \bibinfo {author} {\bibfnamefont {N.}~\bibnamefont
  {Defenu}},\ and\ \bibinfo {author} {\bibfnamefont {J.}~\bibnamefont
  {Lorenzana}},\ }\bibfield  {title} {\bibinfo {title} {Engineering higgs
  dynamics by spectral singularities},\ }\href
  {https://doi.org/10.1103/PhysRevResearch.5.023011} {\bibfield  {journal}
  {\bibinfo  {journal} {Phys. Rev. Res.}\ }\textbf {\bibinfo {volume} {5}},\
  \bibinfo {pages} {023011} (\bibinfo {year} {2023})}\BibitemShut {NoStop}%
\bibitem [{\citenamefont {Phan}\ and\ \citenamefont
  {Chubukov}(2023)}]{Phan2023}%
  \BibitemOpen
  \bibfield  {author} {\bibinfo {author} {\bibfnamefont {D.}~\bibnamefont
  {Phan}}\ and\ \bibinfo {author} {\bibfnamefont {A.~V.}\ \bibnamefont
  {Chubukov}},\ }\bibfield  {title} {\bibinfo {title} {Following the higgs mode
  across the bcs-bec crossover in two dimensions},\ }\href
  {https://doi.org/10.1103/PhysRevB.107.134519} {\bibfield  {journal} {\bibinfo
   {journal} {Phys. Rev. B}\ }\textbf {\bibinfo {volume} {107}},\ \bibinfo
  {pages} {134519} (\bibinfo {year} {2023})}\BibitemShut {NoStop}%
\bibitem [{\citenamefont {Li}\ and\ \citenamefont {Dzero}(2025)}]{Li2025}%
  \BibitemOpen
  \bibfield  {author} {\bibinfo {author} {\bibfnamefont {Y.}~\bibnamefont
  {Li}}\ and\ \bibinfo {author} {\bibfnamefont {M.}~\bibnamefont {Dzero}},\
  }\bibfield  {title} {\bibinfo {title} {Collective modes in terahertz field
  response of disordered superconductors},\ }\href
  {https://doi.org/10.1088/1361-648X/ada65e} {\bibfield  {journal} {\bibinfo
  {journal} {Journal of Physics: Condensed Matter}\ }\textbf {\bibinfo {volume}
  {37}},\ \bibinfo {pages} {115602} (\bibinfo {year} {2025})}\BibitemShut
  {NoStop}%
\bibitem [{\citenamefont {Krull}\ \emph {et~al.}(2016)\citenamefont {Krull},
  \citenamefont {Bittner}, \citenamefont {Uhrig}, \citenamefont {Manske},\ and\
  \citenamefont {Schnyder}}]{Uhrig2016}%
  \BibitemOpen
  \bibfield  {author} {\bibinfo {author} {\bibfnamefont {H.}~\bibnamefont
  {Krull}}, \bibinfo {author} {\bibfnamefont {N.}~\bibnamefont {Bittner}},
  \bibinfo {author} {\bibfnamefont {G.~S.}\ \bibnamefont {Uhrig}}, \bibinfo
  {author} {\bibfnamefont {D.}~\bibnamefont {Manske}},\ and\ \bibinfo {author}
  {\bibfnamefont {A.~P.}\ \bibnamefont {Schnyder}},\ }\bibfield  {title}
  {\bibinfo {title} {Coupling of higgs and leggett modes in non-equilibrium
  superconductors},\ }\href {https://doi.org/10.1038/ncomms11921} {\bibfield
  {journal} {\bibinfo  {journal} {Nature Communications}\ }\textbf {\bibinfo
  {volume} {7}},\ \bibinfo {pages} {11921} (\bibinfo {year}
  {2016})}\BibitemShut {NoStop}%
\bibitem [{\citenamefont {Dzero}(2024{\natexlab{a}})}]{DzeroIFE2024}%
  \BibitemOpen
  \bibfield  {author} {\bibinfo {author} {\bibfnamefont {M.}~\bibnamefont
  {Dzero}},\ }\bibfield  {title} {\bibinfo {title} {Inverse faraday effect in
  superconductors with potential impurities},\ }\href
  {https://doi.org/10.1103/PhysRevB.110.054506} {\bibfield  {journal} {\bibinfo
   {journal} {Phys. Rev. B}\ }\textbf {\bibinfo {volume} {110}},\ \bibinfo
  {pages} {054506} (\bibinfo {year} {2024}{\natexlab{a}})}\BibitemShut
  {NoStop}%
\bibitem [{\citenamefont {Li}\ and\ \citenamefont {Dzero}(2024)}]{Dzero2024}%
  \BibitemOpen
  \bibfield  {author} {\bibinfo {author} {\bibfnamefont {Y.}~\bibnamefont
  {Li}}\ and\ \bibinfo {author} {\bibfnamefont {M.}~\bibnamefont {Dzero}},\
  }\bibfield  {title} {\bibinfo {title} {Amplitude higgs mode in
  superconductors with magnetic impurities},\ }\href
  {https://doi.org/10.1103/PhysRevB.109.054520} {\bibfield  {journal} {\bibinfo
   {journal} {Phys. Rev. B}\ }\textbf {\bibinfo {volume} {109}},\ \bibinfo
  {pages} {054520} (\bibinfo {year} {2024})}\BibitemShut {NoStop}%
\bibitem [{\citenamefont {Althüser}\ and\ \citenamefont
  {Uhrig}(2025)}]{Uhrig2025}%
  \BibitemOpen
  \bibfield  {author} {\bibinfo {author} {\bibfnamefont {J.}~\bibnamefont
  {Althüser}}\ and\ \bibinfo {author} {\bibfnamefont {G.~S.}\ \bibnamefont
  {Uhrig}},\ }\bibfield  {title} {\bibinfo {title} {{Collective modes in
  superconductors including Coulomb repulsion}},\ }\href
  {https://doi.org/10.21468/SciPostPhys.19.3.067} {\bibfield  {journal}
  {\bibinfo  {journal} {SciPost Phys.}\ }\textbf {\bibinfo {volume} {19}},\
  \bibinfo {pages} {067} (\bibinfo {year} {2025})}\BibitemShut {NoStop}%
\bibitem [{\citenamefont {Dzero}\ and\ \citenamefont
  {Kamenev}(2025)}]{Kamenev2025}%
  \BibitemOpen
  \bibfield  {author} {\bibinfo {author} {\bibfnamefont {M.}~\bibnamefont
  {Dzero}}\ and\ \bibinfo {author} {\bibfnamefont {A.}~\bibnamefont
  {Kamenev}},\ }\bibfield  {title} {\bibinfo {title} {Schmid-higgs mode in the
  presence of pair-breaking interactions},\ }\href
  {https://doi.org/10.1103/PhysRevB.111.174502} {\bibfield  {journal} {\bibinfo
   {journal} {Phys. Rev. B}\ }\textbf {\bibinfo {volume} {111}},\ \bibinfo
  {pages} {174502} (\bibinfo {year} {2025})}\BibitemShut {NoStop}%
\bibitem [{\citenamefont {Nosov}\ \emph {et~al.}(2025)\citenamefont {Nosov},
  \citenamefont {Andriyakhina},\ and\ \citenamefont {Burmistrov}}]{Pasha2025}%
  \BibitemOpen
  \bibfield  {author} {\bibinfo {author} {\bibfnamefont {P.~A.}\ \bibnamefont
  {Nosov}}, \bibinfo {author} {\bibfnamefont {E.~S.}\ \bibnamefont
  {Andriyakhina}},\ and\ \bibinfo {author} {\bibfnamefont {I.~S.}\ \bibnamefont
  {Burmistrov}},\ }\bibfield  {title} {\bibinfo {title} {Spatially resolved
  dynamics of the amplitude schmid-higgs mode in disordered superconductors},\
  }\href {https://doi.org/10.1103/x12p-q7bj} {\bibfield  {journal} {\bibinfo
  {journal} {Phys. Rev. Lett.}\ }\textbf {\bibinfo {volume} {135}},\ \bibinfo
  {pages} {056001} (\bibinfo {year} {2025})}\BibitemShut {NoStop}%
\bibitem [{\citenamefont {Silaev}(2019)}]{Silaev2019-Disorder}%
  \BibitemOpen
  \bibfield  {author} {\bibinfo {author} {\bibfnamefont {M.}~\bibnamefont
  {Silaev}},\ }\bibfield  {title} {\bibinfo {title} {Nonlinear electromagnetic
  response and higgs-mode excitation in bcs superconductors with impurities},\
  }\href {https://doi.org/10.1103/PhysRevB.99.224511} {\bibfield  {journal}
  {\bibinfo  {journal} {Phys. Rev. B}\ }\textbf {\bibinfo {volume} {99}},\
  \bibinfo {pages} {224511} (\bibinfo {year} {2019})}\BibitemShut {NoStop}%
\bibitem [{\citenamefont {Seibold}\ \emph {et~al.}(2021)\citenamefont
  {Seibold}, \citenamefont {Udina}, \citenamefont {Castellani},\ and\
  \citenamefont {Benfatto}}]{Seibold2021-Disorder}%
  \BibitemOpen
  \bibfield  {author} {\bibinfo {author} {\bibfnamefont {G.}~\bibnamefont
  {Seibold}}, \bibinfo {author} {\bibfnamefont {M.}~\bibnamefont {Udina}},
  \bibinfo {author} {\bibfnamefont {C.}~\bibnamefont {Castellani}},\ and\
  \bibinfo {author} {\bibfnamefont {L.}~\bibnamefont {Benfatto}},\ }\bibfield
  {title} {\bibinfo {title} {Third harmonic generation from collective modes in
  disordered superconductors},\ }\href
  {https://doi.org/10.1103/PhysRevB.103.014512} {\bibfield  {journal} {\bibinfo
   {journal} {Phys. Rev. B}\ }\textbf {\bibinfo {volume} {103}},\ \bibinfo
  {pages} {014512} (\bibinfo {year} {2021})}\BibitemShut {NoStop}%
\bibitem [{\citenamefont {Haenel}\ \emph {et~al.}(2021)\citenamefont {Haenel},
  \citenamefont {Froese}, \citenamefont {Manske},\ and\ \citenamefont
  {Schwarz}}]{Haenel2021-Disorder}%
  \BibitemOpen
  \bibfield  {author} {\bibinfo {author} {\bibfnamefont {R.}~\bibnamefont
  {Haenel}}, \bibinfo {author} {\bibfnamefont {P.}~\bibnamefont {Froese}},
  \bibinfo {author} {\bibfnamefont {D.}~\bibnamefont {Manske}},\ and\ \bibinfo
  {author} {\bibfnamefont {L.}~\bibnamefont {Schwarz}},\ }\bibfield  {title}
  {\bibinfo {title} {Time-resolved optical conductivity and higgs oscillations
  in two-band dirty superconductors},\ }\href
  {https://doi.org/10.1103/PhysRevB.104.134504} {\bibfield  {journal} {\bibinfo
   {journal} {Phys. Rev. B}\ }\textbf {\bibinfo {volume} {104}},\ \bibinfo
  {pages} {134504} (\bibinfo {year} {2021})}\BibitemShut {NoStop}%
\bibitem [{\citenamefont {Yang}\ and\ \citenamefont
  {Wu}(2022)}]{Yang2022-Disorder}%
  \BibitemOpen
  \bibfield  {author} {\bibinfo {author} {\bibfnamefont {F.}~\bibnamefont
  {Yang}}\ and\ \bibinfo {author} {\bibfnamefont {M.~W.}\ \bibnamefont {Wu}},\
  }\bibfield  {title} {\bibinfo {title} {Impurity scattering in superconductors
  revisited: Diagrammatic formulation of the supercurrent-supercurrent
  correlation and higgs-mode damping},\ }\href
  {https://doi.org/10.1103/PhysRevB.106.144509} {\bibfield  {journal} {\bibinfo
   {journal} {Phys. Rev. B}\ }\textbf {\bibinfo {volume} {106}},\ \bibinfo
  {pages} {144509} (\bibinfo {year} {2022})}\BibitemShut {NoStop}%
\bibitem [{\citenamefont {Yang}\ and\ \citenamefont
  {Wu}(2020{\natexlab{a}})}]{Yang2020-Disorder2}%
  \BibitemOpen
  \bibfield  {author} {\bibinfo {author} {\bibfnamefont {F.}~\bibnamefont
  {Yang}}\ and\ \bibinfo {author} {\bibfnamefont {M.~W.}\ \bibnamefont {Wu}},\
  }\bibfield  {title} {\bibinfo {title} {Influence of scattering on the optical
  response of superconductors},\ }\href
  {https://doi.org/10.1103/PhysRevB.102.144508} {\bibfield  {journal} {\bibinfo
   {journal} {Phys. Rev. B}\ }\textbf {\bibinfo {volume} {102}},\ \bibinfo
  {pages} {144508} (\bibinfo {year} {2020}{\natexlab{a}})}\BibitemShut
  {NoStop}%
\bibitem [{\citenamefont {Dzero}(2024{\natexlab{b}})}]{IFE-Dzero2024}%
  \BibitemOpen
  \bibfield  {author} {\bibinfo {author} {\bibfnamefont {M.}~\bibnamefont
  {Dzero}},\ }\bibfield  {title} {\bibinfo {title} {Inverse faraday effect in
  superconductors with potential impurities},\ }\href
  {https://doi.org/10.1103/PhysRevB.110.054506} {\bibfield  {journal} {\bibinfo
   {journal} {Phys. Rev. B}\ }\textbf {\bibinfo {volume} {110}},\ \bibinfo
  {pages} {054506} (\bibinfo {year} {2024}{\natexlab{b}})}\BibitemShut
  {NoStop}%
\bibitem [{\citenamefont {Dzero}\ \emph {et~al.}(2009)\citenamefont {Dzero},
  \citenamefont {Yuzbashyan},\ and\ \citenamefont {Altshuler}}]{Dzero2009}%
  \BibitemOpen
  \bibfield  {author} {\bibinfo {author} {\bibfnamefont {M.}~\bibnamefont
  {Dzero}}, \bibinfo {author} {\bibfnamefont {E.~A.}\ \bibnamefont
  {Yuzbashyan}},\ and\ \bibinfo {author} {\bibfnamefont {B.~L.}\ \bibnamefont
  {Altshuler}},\ }\bibfield  {title} {\bibinfo {title} {Cooper pair turbulence
  in atomic fermi gases},\ }\href {https://doi.org/10.1209/0295-5075/85/20004}
  {\bibfield  {journal} {\bibinfo  {journal} {Europhysics Letters}\ }\textbf
  {\bibinfo {volume} {85}},\ \bibinfo {pages} {20004} (\bibinfo {year}
  {2009})}\BibitemShut {NoStop}%
\bibitem [{\citenamefont {Dzero}\ \emph {et~al.}(2018)\citenamefont {Dzero},
  \citenamefont {Yuzbashyan},\ and\ \citenamefont
  {Altshuler}}]{dzero2018comment}%
  \BibitemOpen
  \bibfield  {author} {\bibinfo {author} {\bibfnamefont {M.}~\bibnamefont
  {Dzero}}, \bibinfo {author} {\bibfnamefont {E.~A.}\ \bibnamefont
  {Yuzbashyan}},\ and\ \bibinfo {author} {\bibfnamefont {B.~L.}\ \bibnamefont
  {Altshuler}},\ }\href {https://arxiv.org/abs/1806.03474} {\bibinfo {title}
  {Comment on "nonequilibrium dynamics of superconductivity in the attractive
  hubbard model"}} (\bibinfo {year} {2018}),\ \Eprint
  {https://arxiv.org/abs/1806.03474} {arXiv:1806.03474 [cond-mat.supr-con]}
  \BibitemShut {NoStop}%
\bibitem [{\citenamefont {Chern}\ and\ \citenamefont
  {Barros}(2019)}]{Chern2019}%
  \BibitemOpen
  \bibfield  {author} {\bibinfo {author} {\bibfnamefont {G.-W.}\ \bibnamefont
  {Chern}}\ and\ \bibinfo {author} {\bibfnamefont {K.}~\bibnamefont {Barros}},\
  }\bibfield  {title} {\bibinfo {title} {Nonequilibrium dynamics of
  superconductivity in the attractive hubbard model},\ }\href
  {https://doi.org/10.1103/PhysRevB.99.035162} {\bibfield  {journal} {\bibinfo
  {journal} {Phys. Rev. B}\ }\textbf {\bibinfo {volume} {99}},\ \bibinfo
  {pages} {035162} (\bibinfo {year} {2019})}\BibitemShut {NoStop}%
\bibitem [{\citenamefont {Grankin}\ \emph {et~al.}(2025)\citenamefont
  {Grankin}, \citenamefont {Galitski},\ and\ \citenamefont
  {Oganesyan}}]{Grankin2025}%
  \BibitemOpen
  \bibfield  {author} {\bibinfo {author} {\bibfnamefont {A.}~\bibnamefont
  {Grankin}}, \bibinfo {author} {\bibfnamefont {V.}~\bibnamefont {Galitski}},\
  and\ \bibinfo {author} {\bibfnamefont {V.}~\bibnamefont {Oganesyan}},\ }\href
  {https://arxiv.org/abs/2501.08216} {\bibinfo {title} {Negative superfluid
  density and spatial instabilities in driven superconductors}} (\bibinfo
  {year} {2025}),\ \Eprint {https://arxiv.org/abs/2501.08216} {arXiv:2501.08216
  [cond-mat.supr-con]} \BibitemShut {NoStop}%
\bibitem [{\citenamefont {Katsumi}\ \emph
  {et~al.}(2018{\natexlab{b}})\citenamefont {Katsumi}, \citenamefont {Tsuji},
  \citenamefont {Hamada}, \citenamefont {Matsunaga}, \citenamefont
  {Schneeloch}, \citenamefont {Zhong}, \citenamefont {Gu}, \citenamefont
  {Aoki}, \citenamefont {Gallais},\ and\ \citenamefont
  {Shimano}}]{KotaHighTc1}%
  \BibitemOpen
  \bibfield  {author} {\bibinfo {author} {\bibfnamefont {K.}~\bibnamefont
  {Katsumi}}, \bibinfo {author} {\bibfnamefont {N.}~\bibnamefont {Tsuji}},
  \bibinfo {author} {\bibfnamefont {Y.~I.}\ \bibnamefont {Hamada}}, \bibinfo
  {author} {\bibfnamefont {R.}~\bibnamefont {Matsunaga}}, \bibinfo {author}
  {\bibfnamefont {J.}~\bibnamefont {Schneeloch}}, \bibinfo {author}
  {\bibfnamefont {R.~D.}\ \bibnamefont {Zhong}}, \bibinfo {author}
  {\bibfnamefont {G.~D.}\ \bibnamefont {Gu}}, \bibinfo {author} {\bibfnamefont
  {H.}~\bibnamefont {Aoki}}, \bibinfo {author} {\bibfnamefont {Y.}~\bibnamefont
  {Gallais}},\ and\ \bibinfo {author} {\bibfnamefont {R.}~\bibnamefont
  {Shimano}},\ }\bibfield  {title} {\bibinfo {title} {Higgs mode in the
  $d$-wave superconductor
  ${\mathrm{bi}}_{2}{\mathrm{sr}}_{2}{\mathrm{cacu}}_{2}{\mathrm{o}}_{8+x}$
  driven by an intense terahertz pulse},\ }\href
  {https://doi.org/10.1103/PhysRevLett.120.117001} {\bibfield  {journal}
  {\bibinfo  {journal} {Phys. Rev. Lett.}\ }\textbf {\bibinfo {volume} {120}},\
  \bibinfo {pages} {117001} (\bibinfo {year} {2018}{\natexlab{b}})}\BibitemShut
  {NoStop}%
\bibitem [{\citenamefont {Katsumi}\ \emph
  {et~al.}(2020{\natexlab{b}})\citenamefont {Katsumi}, \citenamefont {Li},
  \citenamefont {Raffy}, \citenamefont {Gallais},\ and\ \citenamefont
  {Shimano}}]{KotaHighTc2}%
  \BibitemOpen
  \bibfield  {author} {\bibinfo {author} {\bibfnamefont {K.}~\bibnamefont
  {Katsumi}}, \bibinfo {author} {\bibfnamefont {Z.~Z.}\ \bibnamefont {Li}},
  \bibinfo {author} {\bibfnamefont {H.}~\bibnamefont {Raffy}}, \bibinfo
  {author} {\bibfnamefont {Y.}~\bibnamefont {Gallais}},\ and\ \bibinfo {author}
  {\bibfnamefont {R.}~\bibnamefont {Shimano}},\ }\bibfield  {title} {\bibinfo
  {title} {Superconducting fluctuations probed by the higgs mode in
  ${\mathrm{bi}}_{2}{\mathrm{sr}}_{2}\mathrm{Ca}{\mathrm{cu}}_{2}{\mathrm{o}}_{8+x}$
  thin films},\ }\href {https://doi.org/10.1103/PhysRevB.102.054510} {\bibfield
   {journal} {\bibinfo  {journal} {Phys. Rev. B}\ }\textbf {\bibinfo {volume}
  {102}},\ \bibinfo {pages} {054510} (\bibinfo {year}
  {2020}{\natexlab{b}})}\BibitemShut {NoStop}%
\bibitem [{\citenamefont {Chu}\ \emph {et~al.}(2020)\citenamefont {Chu},
  \citenamefont {Kim}, \citenamefont {Katsumi}, \citenamefont {Kovalev},
  \citenamefont {Dawson}, \citenamefont {Schwarz}, \citenamefont {Yoshikawa},
  \citenamefont {Kim}, \citenamefont {Putzky}, \citenamefont {Li},
  \citenamefont {Raffy}, \citenamefont {Germanskiy}, \citenamefont {Deinert},
  \citenamefont {Awari}, \citenamefont {Ilyakov}, \citenamefont {Green},
  \citenamefont {Chen}, \citenamefont {Bawatna}, \citenamefont {Cristiani},
  \citenamefont {Logvenov}, \citenamefont {Gallais}, \citenamefont {Boris},
  \citenamefont {Keimer}, \citenamefont {Schnyder}, \citenamefont {Manske},
  \citenamefont {Gensch}, \citenamefont {Wang}, \citenamefont {Shimano},\ and\
  \citenamefont {Kaiser}}]{Cuprates2020}%
  \BibitemOpen
  \bibfield  {author} {\bibinfo {author} {\bibfnamefont {H.}~\bibnamefont
  {Chu}}, \bibinfo {author} {\bibfnamefont {M.-J.}\ \bibnamefont {Kim}},
  \bibinfo {author} {\bibfnamefont {K.}~\bibnamefont {Katsumi}}, \bibinfo
  {author} {\bibfnamefont {S.}~\bibnamefont {Kovalev}}, \bibinfo {author}
  {\bibfnamefont {R.~D.}\ \bibnamefont {Dawson}}, \bibinfo {author}
  {\bibfnamefont {L.}~\bibnamefont {Schwarz}}, \bibinfo {author} {\bibfnamefont
  {N.}~\bibnamefont {Yoshikawa}}, \bibinfo {author} {\bibfnamefont
  {G.}~\bibnamefont {Kim}}, \bibinfo {author} {\bibfnamefont {D.}~\bibnamefont
  {Putzky}}, \bibinfo {author} {\bibfnamefont {Z.~Z.}\ \bibnamefont {Li}},
  \bibinfo {author} {\bibfnamefont {H.}~\bibnamefont {Raffy}}, \bibinfo
  {author} {\bibfnamefont {S.}~\bibnamefont {Germanskiy}}, \bibinfo {author}
  {\bibfnamefont {J.-C.}\ \bibnamefont {Deinert}}, \bibinfo {author}
  {\bibfnamefont {N.}~\bibnamefont {Awari}}, \bibinfo {author} {\bibfnamefont
  {I.}~\bibnamefont {Ilyakov}}, \bibinfo {author} {\bibfnamefont
  {B.}~\bibnamefont {Green}}, \bibinfo {author} {\bibfnamefont
  {M.}~\bibnamefont {Chen}}, \bibinfo {author} {\bibfnamefont {M.}~\bibnamefont
  {Bawatna}}, \bibinfo {author} {\bibfnamefont {G.}~\bibnamefont {Cristiani}},
  \bibinfo {author} {\bibfnamefont {G.}~\bibnamefont {Logvenov}}, \bibinfo
  {author} {\bibfnamefont {Y.}~\bibnamefont {Gallais}}, \bibinfo {author}
  {\bibfnamefont {A.~V.}\ \bibnamefont {Boris}}, \bibinfo {author}
  {\bibfnamefont {B.}~\bibnamefont {Keimer}}, \bibinfo {author} {\bibfnamefont
  {A.~P.}\ \bibnamefont {Schnyder}}, \bibinfo {author} {\bibfnamefont
  {D.}~\bibnamefont {Manske}}, \bibinfo {author} {\bibfnamefont
  {M.}~\bibnamefont {Gensch}}, \bibinfo {author} {\bibfnamefont
  {Z.}~\bibnamefont {Wang}}, \bibinfo {author} {\bibfnamefont {R.}~\bibnamefont
  {Shimano}},\ and\ \bibinfo {author} {\bibfnamefont {S.}~\bibnamefont
  {Kaiser}},\ }\bibfield  {title} {\bibinfo {title} {Phase-resolved higgs
  response in superconducting cuprates},\ }\href
  {https://doi.org/10.1038/s41467-020-15613-1} {\bibfield  {journal} {\bibinfo
  {journal} {Nature Communications}\ }\textbf {\bibinfo {volume} {11}},\
  \bibinfo {pages} {1793} (\bibinfo {year} {2020})}\BibitemShut {NoStop}%
\bibitem [{\citenamefont {Barlas}\ and\ \citenamefont
  {Varma}(2013)}]{BarlasVarma2013}%
  \BibitemOpen
  \bibfield  {author} {\bibinfo {author} {\bibfnamefont {Y.}~\bibnamefont
  {Barlas}}\ and\ \bibinfo {author} {\bibfnamefont {C.~M.}\ \bibnamefont
  {Varma}},\ }\bibfield  {title} {\bibinfo {title} {Amplitude or higgs modes in
  $d$-wave superconductors},\ }\href
  {https://doi.org/10.1103/PhysRevB.87.054503} {\bibfield  {journal} {\bibinfo
  {journal} {Phys. Rev. B}\ }\textbf {\bibinfo {volume} {87}},\ \bibinfo
  {pages} {054503} (\bibinfo {year} {2013})}\BibitemShut {NoStop}%
\bibitem [{\citenamefont {Peronaci}\ \emph {et~al.}(2015)\citenamefont
  {Peronaci}, \citenamefont {Schir\'o},\ and\ \citenamefont
  {Capone}}]{PRL2015}%
  \BibitemOpen
  \bibfield  {author} {\bibinfo {author} {\bibfnamefont {F.}~\bibnamefont
  {Peronaci}}, \bibinfo {author} {\bibfnamefont {M.}~\bibnamefont {Schir\'o}},\
  and\ \bibinfo {author} {\bibfnamefont {M.}~\bibnamefont {Capone}},\
  }\bibfield  {title} {\bibinfo {title} {Transient dynamics of $d$-wave
  superconductors after a sudden excitation},\ }\href
  {https://doi.org/10.1103/PhysRevLett.115.257001} {\bibfield  {journal}
  {\bibinfo  {journal} {Phys. Rev. Lett.}\ }\textbf {\bibinfo {volume} {115}},\
  \bibinfo {pages} {257001} (\bibinfo {year} {2015})}\BibitemShut {NoStop}%
\bibitem [{\citenamefont {Kirmani}\ and\ \citenamefont
  {Dzero}(2019)}]{Kirmani2019}%
  \BibitemOpen
  \bibfield  {author} {\bibinfo {author} {\bibfnamefont {A.~A.}\ \bibnamefont
  {Kirmani}}\ and\ \bibinfo {author} {\bibfnamefont {M.}~\bibnamefont
  {Dzero}},\ }\bibfield  {title} {\bibinfo {title} {Non-adiabatic dynamics in d
  + id-wave fermionic superfluids},\ }\href
  {https://doi.org/10.1007/s10948-019-5133-1} {\bibfield  {journal} {\bibinfo
  {journal} {Journal of Superconductivity and Novel Magnetism}\ }\textbf
  {\bibinfo {volume} {32}},\ \bibinfo {pages} {3473} (\bibinfo {year}
  {2019})}\BibitemShut {NoStop}%
\bibitem [{\citenamefont {Yang}\ and\ \citenamefont
  {Wu}(2020{\natexlab{b}})}]{Wu2020}%
  \BibitemOpen
  \bibfield  {author} {\bibinfo {author} {\bibfnamefont {F.}~\bibnamefont
  {Yang}}\ and\ \bibinfo {author} {\bibfnamefont {M.~W.}\ \bibnamefont {Wu}},\
  }\bibfield  {title} {\bibinfo {title} {Theory of higgs modes in $d$-wave
  superconductors},\ }\href {https://doi.org/10.1103/PhysRevB.102.014511}
  {\bibfield  {journal} {\bibinfo  {journal} {Phys. Rev. B}\ }\textbf {\bibinfo
  {volume} {102}},\ \bibinfo {pages} {014511} (\bibinfo {year}
  {2020}{\natexlab{b}})}\BibitemShut {NoStop}%
\bibitem [{\citenamefont {Awelewa}\ and\ \citenamefont
  {Dzero}(2025)}]{Awelewa2025}%
  \BibitemOpen
  \bibfield  {author} {\bibinfo {author} {\bibfnamefont {S.}~\bibnamefont
  {Awelewa}}\ and\ \bibinfo {author} {\bibfnamefont {M.}~\bibnamefont
  {Dzero}},\ }\bibfield  {title} {\bibinfo {title} {Dynamics of the
  schmid-higgs mode in $d$-wave superconductors},\ }\href@noop {} {\bibfield
  {journal} {\bibinfo  {journal} {pre-print}\ }\textbf {\bibinfo {volume}
  {arXiv:2511.03790}} (\bibinfo {year} {2025})}\BibitemShut {NoStop}%
\bibitem [{\citenamefont {Artemenko}\ and\ \citenamefont
  {Kobelkov}(1997)}]{ArtemenkoSN1997}%
  \BibitemOpen
  \bibfield  {author} {\bibinfo {author} {\bibfnamefont {S.~N.}\ \bibnamefont
  {Artemenko}}\ and\ \bibinfo {author} {\bibfnamefont {A.~G.}\ \bibnamefont
  {Kobelkov}},\ }\bibfield  {title} {\bibinfo {title} {Linear response and
  collective oscillations in superconductors with d-wave pairing},\ }\href
  {https://doi.org/10.1103/PhysRevB.55.9094} {\bibfield  {journal} {\bibinfo
  {journal} {Phys. Rev. B}\ }\textbf {\bibinfo {volume} {55}},\ \bibinfo
  {pages} {9094} (\bibinfo {year} {1997})}\BibitemShut {NoStop}%
\bibitem [{\citenamefont {Artemenko}\ and\ \citenamefont
  {Remizov}(2001)}]{ArtemenkoSN2001}%
  \BibitemOpen
  \bibfield  {author} {\bibinfo {author} {\bibfnamefont {S.~N.}\ \bibnamefont
  {Artemenko}}\ and\ \bibinfo {author} {\bibfnamefont {S.~V.}\ \bibnamefont
  {Remizov}},\ }\bibfield  {title} {\bibinfo {title} {Effect of equilibrium
  fluctuations on superfluid density in layered superconductors},\ }\href
  {https://doi.org/10.1103/PhysRevLett.86.708} {\bibfield  {journal} {\bibinfo
  {journal} {Phys. Rev. Lett.}\ }\textbf {\bibinfo {volume} {86}},\ \bibinfo
  {pages} {708} (\bibinfo {year} {2001})}\BibitemShut {NoStop}%
\bibitem [{\citenamefont {Paramekanti}\ \emph {et~al.}(2000)\citenamefont
  {Paramekanti}, \citenamefont {Randeria}, \citenamefont {Ramakrishnan},\ and\
  \citenamefont {Mandal}}]{Paramekanti2000}%
  \BibitemOpen
  \bibfield  {author} {\bibinfo {author} {\bibfnamefont {A.}~\bibnamefont
  {Paramekanti}}, \bibinfo {author} {\bibfnamefont {M.}~\bibnamefont
  {Randeria}}, \bibinfo {author} {\bibfnamefont {T.~V.}\ \bibnamefont
  {Ramakrishnan}},\ and\ \bibinfo {author} {\bibfnamefont {S.~S.}\ \bibnamefont
  {Mandal}},\ }\bibfield  {title} {\bibinfo {title} {Effective actions and
  phase fluctuations in d-wave superconductors},\ }\href
  {https://doi.org/10.1103/PhysRevB.62.6786} {\bibfield  {journal} {\bibinfo
  {journal} {Phys. Rev. B}\ }\textbf {\bibinfo {volume} {62}},\ \bibinfo
  {pages} {6786} (\bibinfo {year} {2000})}\BibitemShut {NoStop}%
\bibitem [{\citenamefont {Ohashi}\ and\ \citenamefont
  {Takada}(2000)}]{TakadaDwave}%
  \BibitemOpen
  \bibfield  {author} {\bibinfo {author} {\bibfnamefont {Y.}~\bibnamefont
  {Ohashi}}\ and\ \bibinfo {author} {\bibfnamefont {S.}~\bibnamefont
  {Takada}},\ }\bibfield  {title} {\bibinfo {title} {Collective phase
  oscillation in two-dimensional d-wave superconductors},\ }\href
  {https://doi.org/10.1103/PhysRevB.62.5971} {\bibfield  {journal} {\bibinfo
  {journal} {Phys. Rev. B}\ }\textbf {\bibinfo {volume} {62}},\ \bibinfo
  {pages} {5971} (\bibinfo {year} {2000})}\BibitemShut {NoStop}%
\bibitem [{\citenamefont {Sharapov}\ and\ \citenamefont
  {Beck}(2002)}]{Sharapov2002}%
  \BibitemOpen
  \bibfield  {author} {\bibinfo {author} {\bibfnamefont {S.~G.}\ \bibnamefont
  {Sharapov}}\ and\ \bibinfo {author} {\bibfnamefont {H.}~\bibnamefont
  {Beck}},\ }\bibfield  {title} {\bibinfo {title} {Effective action approach
  and carlson-goldman mode in d-wave superconductors},\ }\href
  {https://doi.org/10.1103/PhysRevB.65.134516} {\bibfield  {journal} {\bibinfo
  {journal} {Phys. Rev. B}\ }\textbf {\bibinfo {volume} {65}},\ \bibinfo
  {pages} {134516} (\bibinfo {year} {2002})}\BibitemShut {NoStop}%
\bibitem [{\citenamefont {Karuzin}\ and\ \citenamefont
  {Skvortsov}(2025)}]{karuzin2025plasmon}%
  \BibitemOpen
  \bibfield  {author} {\bibinfo {author} {\bibfnamefont {D.~K.}\ \bibnamefont
  {Karuzin}}\ and\ \bibinfo {author} {\bibfnamefont {M.~A.}\ \bibnamefont
  {Skvortsov}},\ }\href {https://arxiv.org/abs/2511.23431} {\bibinfo {title}
  {Plasmon excitations and their attenuation in dirty superconductors}}
  (\bibinfo {year} {2025}),\ \Eprint {https://arxiv.org/abs/2511.23431}
  {arXiv:2511.23431 [cond-mat.supr-con]} \BibitemShut {NoStop}%
\bibitem [{\citenamefont {Larkin}(1965)}]{Larkin1965}%
  \BibitemOpen
  \bibfield  {author} {\bibinfo {author} {\bibfnamefont {A.~I.}\ \bibnamefont
  {Larkin}},\ }\bibfield  {title} {\bibinfo {title} {Quasiclassical method in
  the theory of superconductivity},\ }\href@noop {} {\bibfield  {journal}
  {\bibinfo  {journal} {Sov. Phys. - JETP}\ }\textbf {\bibinfo {volume} {20}},\
  \bibinfo {pages} {208} (\bibinfo {year} {1965})}\BibitemShut {NoStop}%
\bibitem [{\citenamefont {Larkin}\ and\ \citenamefont
  {Ovchinnikov}(1977)}]{LarkinVertex}%
  \BibitemOpen
  \bibfield  {author} {\bibinfo {author} {\bibfnamefont {A.~I.}\ \bibnamefont
  {Larkin}}\ and\ \bibinfo {author} {\bibfnamefont {Y.~N.}\ \bibnamefont
  {Ovchinnikov}},\ }\bibfield  {title} {\bibinfo {title} {Nonlinear effects
  during the motion of vortices in superconductors},\ }\href@noop {} {\bibfield
   {journal} {\bibinfo  {journal} {Sov. Phys. - JETP}\ }\textbf {\bibinfo
  {volume} {46}},\ \bibinfo {pages} {155} (\bibinfo {year} {1977})}\BibitemShut
  {NoStop}%
\bibitem [{\citenamefont {Eilenberger}(1968)}]{Eilenberger1968}%
  \BibitemOpen
  \bibfield  {author} {\bibinfo {author} {\bibfnamefont {G.}~\bibnamefont
  {Eilenberger}},\ }\bibfield  {title} {\bibinfo {title} {Transformation of
  gorkov's equation for type ii superconductors into transport-like
  equations},\ }\href {https://doi.org/10.1007/BF01379803} {\bibfield
  {journal} {\bibinfo  {journal} {Zeitschrift f{\"u}r Physik A Hadrons and
  nuclei}\ }\textbf {\bibinfo {volume} {214}},\ \bibinfo {pages} {195}
  (\bibinfo {year} {1968})}\BibitemShut {NoStop}%
\bibitem [{\citenamefont {Usadel}(1970)}]{Usadel1970}%
  \BibitemOpen
  \bibfield  {author} {\bibinfo {author} {\bibfnamefont {K.~D.}\ \bibnamefont
  {Usadel}},\ }\bibfield  {title} {\bibinfo {title} {Generalized diffusion
  equation for superconducting alloys},\ }\href
  {https://doi.org/10.1103/PhysRevLett.25.507} {\bibfield  {journal} {\bibinfo
  {journal} {Phys. Rev. Lett.}\ }\textbf {\bibinfo {volume} {25}},\ \bibinfo
  {pages} {507} (\bibinfo {year} {1970})}\BibitemShut {NoStop}%
\bibitem [{\citenamefont {Abrikosov}\ \emph {et~al.}(1977)\citenamefont
  {Abrikosov}, \citenamefont {Gorkov},\ and\ \citenamefont
  {Dzyaloshinski}}]{AGD}%
  \BibitemOpen
  \bibfield  {author} {\bibinfo {author} {\bibfnamefont {A.~A.}\ \bibnamefont
  {Abrikosov}}, \bibinfo {author} {\bibfnamefont {L.~P.}\ \bibnamefont
  {Gorkov}},\ and\ \bibinfo {author} {\bibfnamefont {I.~E.}\ \bibnamefont
  {Dzyaloshinski}},\ }\href@noop {} {\emph {\bibinfo {title} {{\sl Methods of
  Quantum Field Theory in Statistical Physics}}}}\ (\bibinfo  {publisher}
  {Dover},\ \bibinfo {year} {1977})\BibitemShut {NoStop}%
\bibitem [{\citenamefont {Kamenev}\ and\ \citenamefont
  {Levchenko}(2009)}]{Kamenev2009}%
  \BibitemOpen
  \bibfield  {author} {\bibinfo {author} {\bibfnamefont {A.}~\bibnamefont
  {Kamenev}}\ and\ \bibinfo {author} {\bibfnamefont {A.}~\bibnamefont
  {Levchenko}},\ }\bibfield  {title} {\bibinfo {title} {Keldysh technique and
  non-linear $\sigma$-model: basic principles and applications},\ }\href
  {https://doi.org/10.1080/00018730902850504} {\bibfield  {journal} {\bibinfo
  {journal} {Advances in Physics}\ }\textbf {\bibinfo {volume} {58}},\ \bibinfo
  {pages} {197} (\bibinfo {year} {2009})},\ \Eprint
  {https://arxiv.org/abs/https://doi.org/10.1080/00018730902850504}
  {https://doi.org/10.1080/00018730902850504} \BibitemShut {NoStop}%
\bibitem [{\citenamefont {Devereaux}\ and\ \citenamefont
  {Hackl}(2007)}]{Devereaux2007}%
  \BibitemOpen
  \bibfield  {author} {\bibinfo {author} {\bibfnamefont {T.~P.}\ \bibnamefont
  {Devereaux}}\ and\ \bibinfo {author} {\bibfnamefont {R.}~\bibnamefont
  {Hackl}},\ }\bibfield  {title} {\bibinfo {title} {Inelastic light scattering
  from correlated electrons},\ }\href
  {https://doi.org/10.1103/RevModPhys.79.175} {\bibfield  {journal} {\bibinfo
  {journal} {Rev. Mod. Phys.}\ }\textbf {\bibinfo {volume} {79}},\ \bibinfo
  {pages} {175} (\bibinfo {year} {2007})}\BibitemShut {NoStop}%
\bibitem [{\citenamefont {Abramowitz}\ and\ \citenamefont
  {Stegun}(1964)}]{Irene1964}%
  \BibitemOpen
  \bibfield  {author} {\bibinfo {author} {\bibfnamefont {M.}~\bibnamefont
  {Abramowitz}}\ and\ \bibinfo {author} {\bibfnamefont {I.}~\bibnamefont
  {Stegun}},\ }\href@noop {} {\emph {\bibinfo {title} {{\sl Handbook of
  Mathematical Functions with Formulas, Graphs, and Mathematical Tables}}}}\
  (\bibinfo  {publisher} {United States Department of Commerce, National Bureau
  of Standards},\ \bibinfo {year} {1964})\BibitemShut {NoStop}%
\bibitem [{\citenamefont {Petrovic}\ \emph {et~al.}(2001)\citenamefont
  {Petrovic}, \citenamefont {Pagliuso}, \citenamefont {Hundley}, \citenamefont
  {Movshovich}, \citenamefont {Sarrao}, \citenamefont {Thompson}, \citenamefont
  {Fisk},\ and\ \citenamefont {Monthoux}}]{Petrovic2001}%
  \BibitemOpen
  \bibfield  {author} {\bibinfo {author} {\bibfnamefont {C.}~\bibnamefont
  {Petrovic}}, \bibinfo {author} {\bibfnamefont {P.}~\bibnamefont {Pagliuso}},
  \bibinfo {author} {\bibfnamefont {M.}~\bibnamefont {Hundley}}, \bibinfo
  {author} {\bibfnamefont {R.}~\bibnamefont {Movshovich}}, \bibinfo {author}
  {\bibfnamefont {J.}~\bibnamefont {Sarrao}}, \bibinfo {author} {\bibfnamefont
  {J.}~\bibnamefont {Thompson}}, \bibinfo {author} {\bibfnamefont
  {Z.}~\bibnamefont {Fisk}},\ and\ \bibinfo {author} {\bibfnamefont
  {P.}~\bibnamefont {Monthoux}},\ }\bibfield  {title} {\bibinfo {title}
  {Heavy-fermion superconductivity in cecoin5 at 2.3 k},\ }\href@noop {}
  {\bibfield  {journal} {\bibinfo  {journal} {Journal of Physics: Condensed
  Matter}\ }\textbf {\bibinfo {volume} {13}},\ \bibinfo {pages} {L337}
  (\bibinfo {year} {2001})}\BibitemShut {NoStop}%
\bibitem [{\citenamefont {Movshovich}\ \emph {et~al.}(2001)\citenamefont
  {Movshovich}, \citenamefont {Jaime}, \citenamefont {Thompson}, \citenamefont
  {Petrovic}, \citenamefont {Fisk}, \citenamefont {Pagliuso},\ and\
  \citenamefont {Sarrao}}]{Movshovich2001}%
  \BibitemOpen
  \bibfield  {author} {\bibinfo {author} {\bibfnamefont {R.}~\bibnamefont
  {Movshovich}}, \bibinfo {author} {\bibfnamefont {M.}~\bibnamefont {Jaime}},
  \bibinfo {author} {\bibfnamefont {J.}~\bibnamefont {Thompson}}, \bibinfo
  {author} {\bibfnamefont {C.}~\bibnamefont {Petrovic}}, \bibinfo {author}
  {\bibfnamefont {Z.}~\bibnamefont {Fisk}}, \bibinfo {author} {\bibfnamefont
  {P.}~\bibnamefont {Pagliuso}},\ and\ \bibinfo {author} {\bibfnamefont
  {J.}~\bibnamefont {Sarrao}},\ }\bibfield  {title} {\bibinfo {title}
  {Unconventional superconductivity in ceiri n 5 and cecoin 5: Specific heat
  and thermal conductivity studies},\ }\href@noop {} {\bibfield  {journal}
  {\bibinfo  {journal} {Physical review letters}\ }\textbf {\bibinfo {volume}
  {86}},\ \bibinfo {pages} {5152} (\bibinfo {year} {2001})}\BibitemShut
  {NoStop}%
\bibitem [{\citenamefont {Miyake}(2007)}]{Miyake2007}%
  \BibitemOpen
  \bibfield  {author} {\bibinfo {author} {\bibfnamefont {K.}~\bibnamefont
  {Miyake}},\ }\bibfield  {title} {\bibinfo {title} {New trend of
  superconductivity in strongly correlated electron systems},\ }\href
  {https://doi.org/10.1088/0953-8984/19/12/125201} {\bibfield  {journal}
  {\bibinfo  {journal} {Journal of Physics: Condensed Matter}\ }\textbf
  {\bibinfo {volume} {19}},\ \bibinfo {pages} {125201} (\bibinfo {year}
  {2007})}\BibitemShut {NoStop}%
\bibitem [{\citenamefont {Van~Dyke}\ \emph {et~al.}(2014)\citenamefont
  {Van~Dyke}, \citenamefont {Massee}, \citenamefont {Allan}, \citenamefont
  {Davis}, \citenamefont {Petrovic},\ and\ \citenamefont {Morr}}]{Morr2014}%
  \BibitemOpen
  \bibfield  {author} {\bibinfo {author} {\bibfnamefont {J.~S.}\ \bibnamefont
  {Van~Dyke}}, \bibinfo {author} {\bibfnamefont {F.}~\bibnamefont {Massee}},
  \bibinfo {author} {\bibfnamefont {M.~P.}\ \bibnamefont {Allan}}, \bibinfo
  {author} {\bibfnamefont {J.~S.}\ \bibnamefont {Davis}}, \bibinfo {author}
  {\bibfnamefont {C.}~\bibnamefont {Petrovic}},\ and\ \bibinfo {author}
  {\bibfnamefont {D.~K.}\ \bibnamefont {Morr}},\ }\bibfield  {title} {\bibinfo
  {title} {Direct evidence for a magnetic f-electron--mediated pairing
  mechanism of heavy-fermion superconductivity in cecoin5},\ }\href@noop {}
  {\bibfield  {journal} {\bibinfo  {journal} {Proceedings of the National
  Academy of Sciences}\ }\textbf {\bibinfo {volume} {111}},\ \bibinfo {pages}
  {11663} (\bibinfo {year} {2014})}\BibitemShut {NoStop}%
\bibitem [{\citenamefont {Sarrao}\ and\ \citenamefont
  {Thompson}(2007)}]{Sarrao2007}%
  \BibitemOpen
  \bibfield  {author} {\bibinfo {author} {\bibfnamefont {J.}~\bibnamefont
  {Sarrao}}\ and\ \bibinfo {author} {\bibfnamefont {J.}~\bibnamefont
  {Thompson}},\ }\bibfield  {title} {\bibinfo {title} {Superconductivity in
  cerium- and plutonium-based ‘115’ materials},\ }\href
  {https://doi.org/10.1143/JPSJ.76.051013} {\bibfield  {journal} {\bibinfo
  {journal} {Journal of the Physical Society of Japan}\ }\textbf {\bibinfo
  {volume} {76}},\ \bibinfo {pages} {1013} (\bibinfo {year}
  {2007})}\BibitemShut {NoStop}%
\end{thebibliography}%

\end{document}